\documentclass[aps,prd,a4paper,twocolumn]{revtex4}

\usepackage{graphicx}
\usepackage{bm}
\usepackage{amsfonts}
\usepackage{amsmath}

\usepackage[russian,ngerman,english]{babel}

\newcommand{\ve}[1]{\mbox{\boldmath$#1$}}

\let\oldbibitem\bibitem
\renewcommand\bibitem[2][]{\oldbibitem{#2}}

\begin{document}

\title{Time delay in the quadrupole field of a body at rest in 2PN approximation} 
\author{Sven Zschocke}

\affiliation{Lohrmann Observatory,
Dresden Technical University, Helmholtzstrasse 10, D-01069 Dresden, Germany}


\begin{abstract} 
	The time delay of a light signal in the quadrupole field of a body at rest is determined in the second post-Newtonian (2PN) approximation 
	in harmonic coordinates. 
	For grazing light rays at Sun, Jupiter, and Saturn the 2PN quadrupole effect in time delay amounts up to $0.004$, $0.14$, and $0.04$ pico-second, 
	respectively. These values are compared with the time delay in the first post-Newtonian (1PN and 1.5PN) approximation, where it turns out that only 
	the first eight mass-multipoles and the spin-dipole of these massive bodies are required for a given 
	goal accuracy of $0.001$ pico-second in time-delay measurements in the solar system. In addition, the spin-hexapole of Jupiter 
	is required on that scale of accuracy. 
\end{abstract}

\pacs{95.10.Jk, 95.10.Ce, 95.30.Sf, 04.25.Nx, 04.80.Cc}

\maketitle

\begin{center}


\end{center}


\section{Introduction}\label{Section0}

The time delay of a light signal in the gravitational field of a massive body was predicted by Shapiro in 1964 \cite{Shapiro1} 
and belongs to the four classical tests of general relativity: perihelion precession of Mercury, light deflection at the Sun, gravitational redshift of light, 
and light-travel time delay \cite{MTW}. In its original formulation of the Shapiro effect one considers a light signal 
which propagates in the monopole field of one massive body with mass $M$ which is at rest with respect to the coordinate system. 

Assume, the space-time is covered by harmonic coordinates, $\left(t, \ve{x}\right)$ \cite{MTW,Poisson_Will,Brumberg1991,Kopeikin_Efroimsky_Kaplan}  
(cf. Eq.~(5.177) in \cite{Poisson_Will}) and the origin of spatial axes is located at the center-of-mass of the massive body. 
The light signal is emitted by a light source at $\left(t_0, \ve{x}_0\right)$ and then received by an observer 
at $\left(t_1, \ve{x}_1\right)$. The Shapiro time delay is the difference between the light-travel-time, $\left(t_1 - t_0 \right)$, and the Euclidean distance 
between source and observer, $R =  \left|\ve{x}_1 - \ve{x}_0\right|$, divided by the speed of light,  
\begin{eqnarray}
	\Delta \tau &=& \left(t_1 - t_0 \right) - \frac{R}{c}\;.
	\label{Definition_Shapiro_Time_Delay}
\end{eqnarray}

\noindent 
The Newtonian theory predicts no time delay. In General Relativity (GR), however, the light-travel-time $\left(t_1 - t_0 \right)$ differs from $R/c$, 
because the light signal propagates through the gravitational fields of the massive body, which decelerate the speed of the light signal. 
In first post-Newtonian (1PN) approximation for a massive body at rest the time delay is given by \cite{MTW,Poisson_Will,Brumberg1991}  
\begin{eqnarray}
	\Delta \tau^{M}_{\rm 1PN} &=& \frac{2 G M}{c^3} \ln \frac{x_1 + \ve{k} \cdot \ve{x}_1}{x_0 + \ve{k} \cdot \ve{x}_0} \;, 
	\label{Introduction_5} 
\end{eqnarray}

\noindent 
where $\ve{k} = \left(\ve{x}_1 - \ve{x}_0\right)/R$ is the unit-vector pointing from the source towards the observer; superscript label $M$ stands for monopole. 

In the first time-delay measurements, performed in $1968$ \cite{Shapiro2} and $1971$ \cite{Shapiro3}, radar signals were emitted from Earth, which have 
passed nearby the limb of the Sun, then they were reflected by an inner planet, either Mercury or Venus, and finally the radar signals were received back 
on Earth. This round trip of the light signal is called two-way Shapiro effect and yields the double of 
Eq.~(\ref{Introduction_5}) (cf. Eq.~(10.102) in \cite{Poisson_Will}) which gives up to $248$ micro-seconds for the constellation Earth-Sun-Mercury, 
and amounts up to $251$ micro-seconds for the constellation Earth-Sun-Venus. 
In these experiments the time delay predicted by GR has been confirmed up to an error of a few percent, which corresponds 
to a precision in time measurements of a few micro-seconds. Ever since, time delay measurements have been performed with increasing accuracy. In $1977$ the 
Viking1 and Viking2 spacecrafts (Mars landers and orbiters) were used as radar reflectors, 
where an accuracy of about $0.5$ percent in time delay measurements was achieved \cite{Shapiro4}, which was later improved towards an accuracy of 
about $0.1$ percent \cite{Shapiro5}, which corresponds to a precision in time measurements of about $300$ nano-seconds.  
The most accurate time delay measurements in the solar system were achieved in $2003$ by using the Cassini spacecraft (orbiting Saturn) as reflector 
of the radar signals with an error of about $0.001$ percent \cite{Shapiro6}. The two-way Shapiro time delay for a grazing ray at the Sun for the configuration 
Earth-Sun-Saturn amounts up to $288$ micro-seconds, thus that error corresponds to an accuracy of a few nano-seconds in time delay measurements. 

Future time-delay experiments will be performed by optical laser rather than radar signals, as suggested by several mission proposals of the European Space 
Agency (ESA) \cite{Astrod1,Lator1,Odyssey,Sagas,TIPO,EGE}. These missions are designed to significantly improve the test of relativistic gravity of the 
solar system. One aim of these experiments are time-delay measurements at the pico-second and sub-pico-second level of accuracy. 
In these mission proposals it has been suggested that a laser signal is emitted by the observer and then reflected by the spacecraft 
and afterwards received back by the observer. The decisive advantage of this two-way Shapiro effect is, that there is no need for clock-synchronization between 
observer and spacecraft \cite{Book_Clifford_Will}. Thus, besides laser availability and reliability, significant improvements in measurements of the Shapiro 
effect are mainly dependent on advancements in the determination of the proper time at the observer's position, either at ground stations or in space, 
which have made impressive progress during recent decades. 

Today, accuracies on the sub-nano-second scale and even pico-second scale in time measurements are becoming standard in high-precision experiments in space. For 
instance, both Lunar Laser Ranging (LLR) as well as Satellite Laser Ranging (SLR) have reached the sub-nano-second and even the pico-second level of accuracy 
\cite{LLR1,LLR2,LLR3,SLR1,SLR2,Book_Soffel_Han} which implies a standard deviation of the atomic clocks of about $\Delta t / t \sim 10^{-13}$. In these experiments 
a laser signal is sent from a ground station to the Moon or satellite, where it is reflected from retroreflectors, and then the laser signal is received back 
by the ground station; a review of LLR and future developments of SLR are given in \cite{SLR1,LLR4}. 
Meanwhile, there exists a global network of $45$ active ground stations which represent the 
International Laser Ranging Service. The measurement of the round-trip travel time allows one to determine the distance to the Moon or spacecraft, and such 
laser transfer measurements have reached the centimeter and even the millimeter level of accuracy, which corresponds to an accuracy of about $3$ pico-seconds 
in time measurements. 

Furthermore, the two hydrogen maser atomic clocks onboard of each satellite of the European Galileo navigation system are mentioned, which have a standard 
deviation of $\Delta t / t \sim 10^{-14}$ which can be considered as minimal criterion for present-day technology of time measurements in space. The present-day 
most precise atomic clock onboard of a satellite is the Deep Space Atomic Clock (DSAC) \cite{DSAC1} launched in 2019 by National Aeronautics and Space 
Administration (NASA), which has a standard deviation of $\Delta t / t \sim 10^{-15}$. For a light signal in the solar system with a travel-time of about 
$10^4\,{\rm s}$ such a standard deviation of DSAC implies an accuracy of about $\Delta t \sim 10$ pico-second, 
which one may consider as minimal criterion for present-day technology of time measurements for the time-of-flight of such a light signal. In fact, by comparing 
DSAC to the U.S. Naval Observatory's hydrogen maser master clock on the ground, the researchers found that the space clock deviates by about $26$ pico-seconds 
during one day \cite{DSAC_Nature}. A follow-up project, DSAC-2, has recently been selected by NASA for demonstration on the 
upcoming space mission VERITAS (Venus Emissivity Radio Science Insar Topography and Spectroscopy) to Venus \cite{DSAC2}. 

The atmosphere of the Earth has a significant impact on the speed and trajectory of light signals. In view of this fact, the advantage of space-based missions 
is that the atmosphere of Earth cannot disturb the time-of-flight measurements of light signals between spacecrafts. If ground-stations on Earth are involved 
in time-of-flight measurements, then the local meteorological data (i.e. altitude profile of temperature, pressure, humidity)  
need carefully to be determined with high accuracy during the period of time measurements.
The modeling and description of atmospheric corrections of the ground-to-satellite time-transfer of light signals has made important advancements 
during recent years and has reached the pico-second level of accuracy \cite{ground_to_satellite}. Thus, time-delay measurements with ground-stations 
remain an option also for future highly precise experiments on the pico-second and maybe on the sub-pico-second level. 

As example of Earth-bound clocks are the Caesium atomic clocks NIST-F1 and NIST-F2 at the National Institute of Standard and Technology (NIST) 
are mentioned, where a standard deviation of $\Delta t / t \sim 10^{-16}$ has been achieved \cite{NIST}. 
The highest accuracies for Earth-bound atomic clocks have been 
achieved with optical atomic clocks with a standard deviation of $\Delta t / t \sim 10^{-19}$ \cite{Atomic_Clock1}. If one considers a light signal emitted
from Earth towards a spacecraft located in the solar system, for instance, nearby Uranus,
and back, then the light-travel-time would be about $t \sim 10^4\,{\rm s}$. Hence, the standard deviation of such an atomic clock corresponds to a precision 
of about $\Delta t \sim 0.001$ pico-second, which one may consider as maximal criterion for present-day accuracy of time measurements for the time-of-flight
of such a light signal, being aware that in near future the precision of optical atomic clocks will further be improved.

Accordingly, in consideration of these facts and being aware of further rapid progress in the precisions of time measurements in foreseen future 
\cite{Optical_Clocks1}, it seems necessary to develop the theoretical model of Shapiro time delay up to an accuracy of about $\Delta t = 0.001$ pico-second. 
Also regarding the fact that a theoretical model should be at least one order of magnitude more precise than actual real measurements, this magnitude should 
be assumed as most upper accuracy threshold in theoretical considerations for prospective astrometry missions.  

In view of these considerations it becomes apparent that the classical monopole formula (\ref{Introduction_5}) of time delay is by far not sufficient to meet 
near-future accuracies in time measurements and it is clear that the shape and inner structure of the bodies as well as their rotational motions become relevant 
on such scale of accuracy \cite{Zschocke_1PN,Zschocke_15PN}. The expansion of the metric tensor in terms of mass-multipoles, $\hat{M}_L$, 
and spin-multipoles, $\hat{S}_L$, of the massive solar system bodies allows one to account for these effects. The multipole expansion of the metric tensor  
implicates a corresponding multipole expansion of the Shapiro time-delay in terms of mass-multipoles and spin-multipoles. In particular, it is necessary to 
include some post-Newtonian terms (1PN and 1.5PN) in the theory of light propagation,  
\begin{eqnarray}
	\Delta \tau &=& \sum\limits_{l=0}^{\infty} \Delta \tau_{\rm 1PN}^{M_L} + \sum\limits_{l=1}^{\infty} \Delta \tau_{\rm 1.5PN}^{S_L} 
	 + {\cal O}\left(c^{-4}\right),   
	\label{Introduction_20} 
\end{eqnarray}

\noindent
where the first term $\left(l=0\right)$ is just the 1PN mass-monopole term as given by (\ref{Introduction_5}). 
It is clear that some of these higher mass-multipoles $\hat{M}_L$ (describe shape and inner structure of the massive body) and perhaps some spin-multipoles 
$\hat{S}_L$ (describe rotational motions and inner currents of the massive body) are relevant on the sub-pico-second level of accuracy. 
The mathematical expressions for the 1PN mass-multipole and 1.5PN spin-multipole terms in the Shapiro time delay, $\Delta \tau_{\rm 1PN}^{M_L}$ and
$\Delta \tau_{\rm 1.5PN}^{S_L}$, were derived a long time ago \cite{Kopeikin1997}. It is one aim of this investigation to quantify these terms
and to clarify which of these 1PN and 1.5PN terms need to be taken into account for the assumed goal accuracy of about $0.001$ pico-second.

Besides these 1PN and 1.5PN terms in (\ref{Introduction_20}) it might well be that also some 2PN terms are relevant on the sub-pico-second level of accuracy in 
time-delay measurements. For a long time, the knowledge about 2PN effects in the Shapiro time delay was restricted to the case of spherically symmetric bodies, 
that means in 2PN approximation only the mass-monopole term $M$ has been taken into account. The next subsequent term in the multipole decomposition is the 
mass-quadrupole term $M_{ab}$. Clearly, these terms are the most dominant 2PN terms beyond the 2PN mass-monopole.  
Recently, the light trajectory in 2PN approximation in the field of one body at rest with mass-monopole and mass-quadrupole structure was  
determined \cite{Zschocke_Quadrupole_1}. The investigation in \cite{Zschocke_Quadrupole_1} allows us to determine these 2PN mass-quadrupole terms in the 
Shapiro time-delay, that means 
\begin{eqnarray} 
        \Delta \tau &=& \Delta \tau_{\rm 1PN}^{M} + \Delta \tau_{\rm 1PN}^{M_{ab}} 
	\nonumber\\ 
	&& \hspace{-1.0cm} +\,\Delta \tau_{\rm 2PN}^{M \times M} + \Delta \tau_{\rm 2PN}^{M \times M_{ab}} + \Delta \tau_{\rm 2PN}^{M_{ab} \times M_{cd}}
        + {\cal O}\left(c^{-6}\right).  
	\label{Introduction_25} 
\end{eqnarray}
 
\noindent 
In this investigation we will examine the impact of the 2PN monopole-monopole term, $\Delta \tau_{\rm 2PN}^{M \times M}$, 
the monopole-quadrupole term, $\Delta \tau_{\rm 2PN}^{M \times M_{ab}}$, and the quadrupole-quadrupole term, $\Delta \tau_{\rm 2PN}^{M_{ab} \times M_{cd}}$, 
and will compare them with the 1PN and 1.5PN terms 
in (\ref{Introduction_20}). Of course, the 1PN terms in (\ref{Introduction_20}) beyond the mass-quadrupole as well as the 1.5PN terms in (\ref{Introduction_20}) 
can finally be added to (\ref{Introduction_25}) in an appropriate manner. 

The manuscript is organized as follows: In Section~\ref{Section1} the exact geodesic equation and the exact metric tensor for a body at rest 
is discussed. The 1PN and 1.5PN effect on the Shapiro time delay is determined in Section~\ref{Section_Shapiro_1PN}. The initial value problem 
of the 2PN light propagation in the quadrupole field of one body at rest is considered in the Sections~\ref{Section3}.  
The Shapiro time delay in 2PN approximation is examined in Section~\ref{Time_Delay}.  
Finally a summary and outlook are given in Section~\ref{Summary}. The notations as well as details of the 
calculations are relegated to a set of several Appendixes.

\section{Geodesic equation and metric tensor}\label{Section1}

\begin{figure}[!ht]
\begin{center}
\includegraphics[scale=0.15]{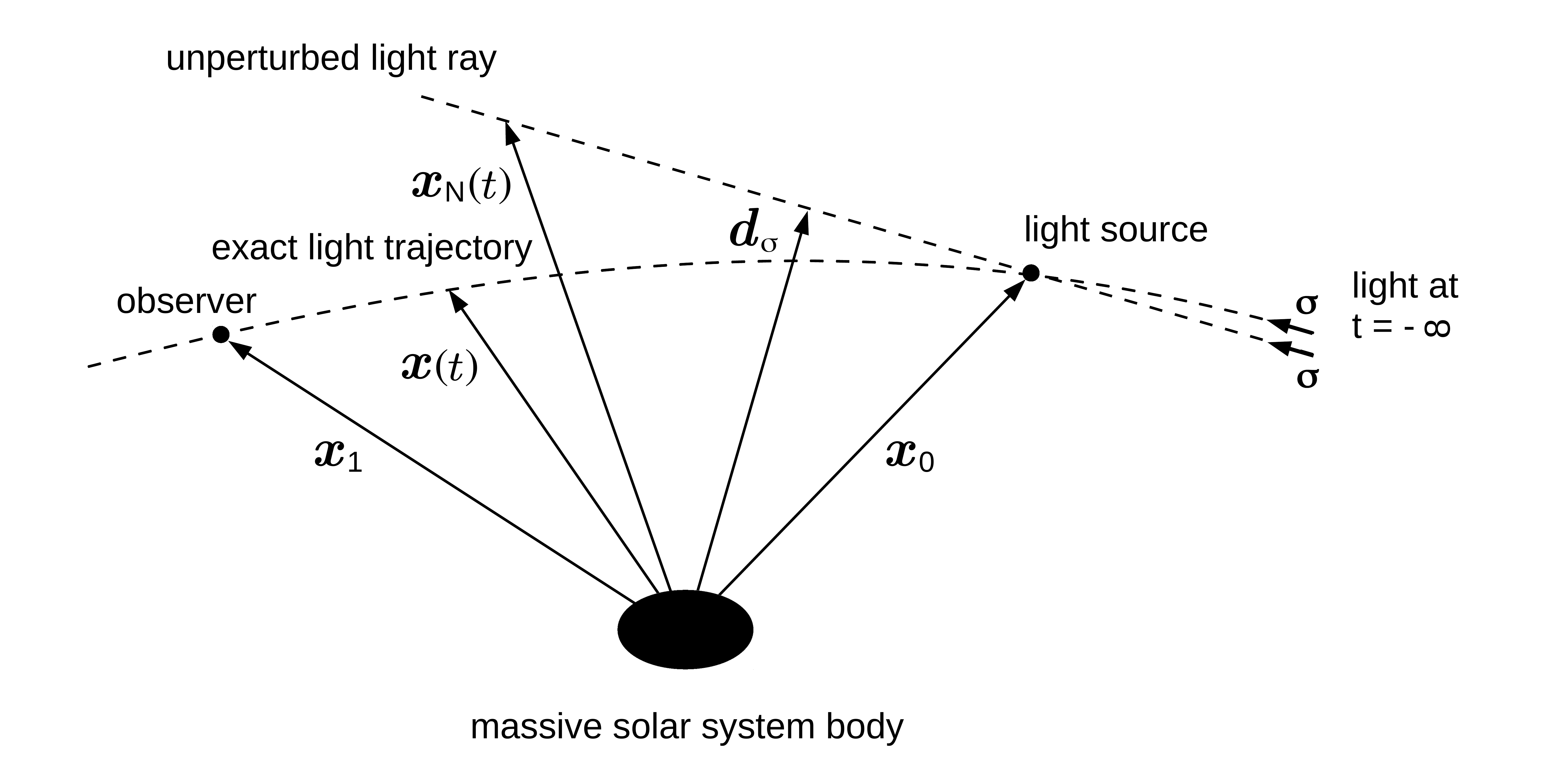}
\end{center}
        \caption{A geometrical representation of the propagation of a light signal through the gravitational field of a massive solar system body
        at rest. The light signal is emitted by the light source at $\ve{x}_0$ and propagates along the exact light trajectory $\ve{x}\left(t\right)$.
        The unit tangent vector along the light trajectory at past null infinity is $\ve{\sigma}$.
        The unperturbed light ray $\ve{x}_{\rm N}\left(t\right)$ is given by Eq.~(\ref{Unperturbed_Lightray_2}) and propagates in the direction
        of $\ve{\sigma}$ along a straight line through the position of the light source at $\ve{x}_0$. The impact vector $\ve{d}_{\sigma}$ of the
	unperturbed light ray is given by Eq.~(\ref{impact_vector_x0}). The impact vector $\hat{\ve{d}_{\sigma}}$ is defined by 
	Eq.~(\ref{appendix_impact_vector_sigma_1}) and is parallel to the impact vector $\ve{d}_{\sigma}$ but a tiny bit smaller and not shown in the diagram.}
\label{Diagram}
\end{figure}

A unique interpretation of astrometric observations, like the time-delay of light signals, requires the determination of light trajectory, $\ve{x}\left(t\right)$, 
as function of coordinate time. 
In Minkowskian space-time, a light signal would travel along a straight trajectory, the so-called unperturbed light ray. 
If the flat space-time is covered by Cartesian coordinates, the components of the 
Minkowskian metric read $\eta_{\alpha\beta} = \left(-1,+1,+1,+1\right)$ and then the trajectory of a light signal is given by 
\begin{eqnarray}
        \ve{x}_{\rm N} &=& \ve{x}_0 + c \left(t - t_0\right) \ve{\sigma}\;,   
\label{Unperturbed_Lightray_2}
\end{eqnarray}

\noindent
where the subindex ${\rm N}$ stands for Newtonian. 
That means a light signal, emitted at the spatial position of the light source, $\ve{x}_0$, would propagate along a straight line in the direction 
of some unit-vector $\ve{\sigma}$. For graphical illustration of the unperturbed light trajectory see Figure~\ref{Diagram}. 

The trajectory of a light signal propagating in curved space-time is determined by the geodesic equation (\ref{Geodetic_Equation2}) and
isotropic condition (\ref{Null_Condition1}), which in terms of coordinate time read as follows \cite{Brumberg1991,Kopeikin_Efroimsky_Kaplan,MTW}
(e.g. Eqs.~(1.2.48) - (1.2.49) in \cite{Brumberg1991} or Eqs.~(7.20) - (7.23) in \cite{Kopeikin_Efroimsky_Kaplan}):
\begin{eqnarray}
        && \frac{\ddot{x}^{i}\left(t\right)}{c^2} + \Gamma^{i}_{\mu\nu} \frac{\dot{x}^{\mu}\left(t\right)}{c} \frac{\dot{x}^{\nu}\left(t\right)}{c}
- \Gamma^{0}_{\mu\nu} \frac{\dot{x}^{\mu}\left(t\right)}{c} \frac{\dot{x}^{\nu}\left(t\right)}{c} \frac{\dot{x}^{i}\left(t\right)}{c} = 0\;,
\nonumber\\ 
        \label{Geodetic_Equation2}
\\
        && g_{\alpha\beta}\,\frac{\dot{x}^{\alpha}\left(t\right)}{c}\,\frac{\dot{x}^{\beta}\left(t\right)}{c} = 0 \;, 
\label{Null_Condition1}
\end{eqnarray}

\noindent
where $g_{\alpha\beta}$ are the covariant components of the metric tensor of space-time; for the signature $\left(-,+,+,+\right)$ has been chosen. The isotropic 
condition (\ref{Null_Condition1}) states that light trajectories are null rays, a condition which must be satisfied at any point along the light trajectory. 
Furthermore, a dot denotes total derivative with respect to coordinate time, and $\Gamma^{\alpha}_{\mu\nu}$ are the Christoffel symbols, given by  
\cite{Kopeikin_Efroimsky_Kaplan,Brumberg1991,MTW} (e.g. Eq.(21.27) in \cite{MTW})  
\begin{eqnarray}
\Gamma^{\alpha}_{\mu\nu} = \frac{1}{2}\,g^{\alpha\beta} 
\left(\frac{\partial g_{\beta\mu}}{\partial x^{\nu}} + \frac{\partial g_{\beta\nu}}{\partial x^{\mu}} 
- \frac{\partial g_{\mu\nu}}{\partial x^{\beta}}\right).   
\label{Christoffel_Symbols2}
\end{eqnarray}

\noindent
The Christoffel symbols are functions of the metric tensor. For weak gravitational fields it is meaningful to separate the metric tensor 
into the flat metric and a metric perturbation, 
\begin{eqnarray}
g_{\alpha \beta} \left(t,\ve{x}\right) &=& \eta_{\alpha \beta} + h_{\alpha \beta}\left(t,\ve{x}\right). 
\label{PM_Expansion_0}
\end{eqnarray}

\noindent
The geodesic equation is a differential equation of second order of one variable, $t$, thus a unique solution of (\ref{Geodetic_Equation2}) 
necessitates two initial-boundary conditions: the spatial position of light source $\ve{x}_0$ and the unit-direction $\ve{\sigma}$ of the 
light signal at past infinity \cite{Brumberg1991,Kopeikin1997,KopeikinSchaefer1999_Gwinn_Eubanks,KlionerKopeikin1992,Zschocke_1PN,Zschocke_15PN}: 
\begin{eqnarray}
        \ve{\sigma} &=& \frac{\dot{\ve{x}}\left(t\right)}{c}\,\bigg|_{t = - \infty} \quad {\rm with} \quad 
        \ve{\sigma} \cdot \ve{\sigma} = 1\;,  
        \label{Initial_A}
        \\ 
        \ve{x}_0 \; &=& \; \ve{x}\left(t\right) \bigg|_{t = t_0}\;.  
        \label{Initial_B}
\end{eqnarray}

\noindent
Then, by inserting the decomposition (\ref{PM_Expansion_0}) into (\ref{Geodetic_Equation2}) and using the initial-boundary conditions (\ref{Initial_A}) 
and (\ref{Initial_B}), the solution of the second integration of geodesic equation (trajectory of light signal) (\ref{Geodetic_Equation2}) is given by
\begin{eqnarray}
        \ve{x}\left(t\right) &=& \ve{x}_0 + c \left(t-t_0\right) \ve{\sigma} + \Delta \ve{x}\left(t,t_0\right), 
        \label{Introduction_2}
\end{eqnarray}

\noindent
where $\Delta \ve{x}$ is the correction to the trajectory of the unperturbed light ray (\ref{Unperturbed_Lightray_2}).
The formal solution of the initial value problem (\ref{Introduction_2}) implies the following limit,
\begin{eqnarray}
        \lim_{t \rightarrow t_0} \Delta \ve{x}\left(t,t_0\right) &=& 0 \;, 
        \label{Introduction_4} 
\end{eqnarray}

\noindent
in order to be consistent with the condition  (\ref{Initial_B}).

For solving the geodesic equation (\ref{Geodetic_Equation2}) one needs the metric tensor (\ref{PM_Expansion_0}) of the specific problem under consideration. 
Usually, the metric tensor (\ref{PM_Expansion_0}) is not known in its exact form and one has to apply for some approximation scheme. If the gravitational fields 
are weak and the speed of matter is slow compared to the speed of light, then one can utilize the post-Newtonian expansion (weak-field slow-motion expansion) 
of the metric tensor, which is an expansion of the metric tensor in inverse powers of the speed of light \cite{Thorne,Blanchet_Damour1},
\begin{eqnarray}
g_{\alpha \beta} \left(t,\ve{x}\right) &=& \eta_{\alpha \beta} + \sum\limits_{n=2}^{\infty} h^{\left({\rm n}\right)}_{\alpha \beta}\left(t,\ve{x}, \ln c\right).
\label{PN_Expansion_1}
\end{eqnarray}

\noindent
In general, the post-Newtonian expansion (\ref{PN_Expansion_1}) is a non-analytic series, because at higher order $n \ge 8$ non-analytic terms involving powers
of logarithms occur \cite{Thorne,Blanchet_Damour1}, while by definition the $n$-th post-Newtonian perturbation, $h^{\left({\rm n}\right)}_{\alpha \beta}$, 
is the factor of $n$-th inverse power of $c$. 

In reality, a solar system body can be of arbitrary shape, inner structure, rotational and oscillating motions and can have inner currents of matter. 
From the Multipolar Post-Minkowskian (MPM) formalism \cite{Thorne,Blanchet_Damour1,Blanchet6} it follows that the post-Newtonian solution for the metric tensor for 
such a body can be given in terms of two kinds of symmetric and trace-free (STF) multipoles: mass-multipoles $\hat{M}_L$ (describing shape, inner structure 
and oscillations of the body) and spin-multipoles $\hat{S}_L$ (describing rotational motions and inner currents of the body) 
\begin{eqnarray}
g_{\alpha \beta} \left(t,\ve{x}\right) &=& \eta_{\alpha \beta} 
	+ \sum\limits_{n=2}^{\infty} h^{\left({\rm n}\right)}_{\alpha \beta} (\hat{M}_L\left(s\right), \hat{S}_L\left(s\right),\ln c) 
\label{PN_Expansion_Multipoles_A}
\end{eqnarray}

\noindent
where the origin of spatial axes of the coordinate system is located somewhere nearby the center of mass of the source of matter (body), and $s = t - x/c$ 
is the retarded time which describes the fact that the metric at field point $\left(t, \ve{x}\right)$ is determined by the multipoles at the earlier time $s$ 
because gravitational action propagates with the finite speed of light. 
In case of a stationary source of matter the multipoles and the metric perturbations are time-independent and then the post-Newtonian expansion of the 
metric tensor reads 
\begin{eqnarray}
g_{\alpha \beta} \left(\ve{x}\right) &=& \eta_{\alpha \beta}
	+ \sum\limits_{n=2}^{\infty} h^{\left({\rm n}\right)}_{\alpha \beta}(\hat{M}_L, \hat{S}_L,\ln c). 
\label{PN_Expansion_Multipoles_B}
\end{eqnarray}

\noindent
These multipoles $\hat{M}_L$ and $\hat{S}_L$ in (\ref{PN_Expansion_Multipoles_A}) and (\ref{PN_Expansion_Multipoles_B}) are integrals over the
stress-energy tensor of the source of matter. They are considered in Appendix \ref{Appendix_Multipoles}.

\section{Shapiro effect in 1.5PN approximation}\label{Section_Shapiro_1PN}

In 1.5PN approximation the expansion (\ref{PN_Expansion_Multipoles_B}) reads 
\begin{eqnarray}
        g_{\alpha\beta} &=& \eta_{\alpha\beta} + h_{\alpha\beta}^{\left(2\right)}(\hat{M}_L)  
        + h_{\alpha\beta}^{\left(3\right)}(\hat{S}_L) 
	\label{PN_Expansion_1PN_15PN}
\end{eqnarray}

\noindent
up to terms of the order ${\cal O}(c^{-4})$, and 
where the non-vanishing metric perturbations $h_{\alpha\beta}^{\left(2\right)}$ and $h_{\alpha\beta}^{\left(3\right)}$ are given by  
\cite{Thorne,Blanchet_Damour1,Multipole_Damour_2,Kopeikin1997,Zschocke_2PM_Metric} 
\begin{eqnarray}
	h_{00}^{\left(2\right)} &=& + \frac{2}{c^2} \sum\limits_{l=0}^{\infty} \frac{\left(-1\right)^l}{l!}\,\hat{M}_L\, 
	\hat{\partial}_L \frac{1}{r}\;,
	\label{Metric_00}
	\\
	h_{0i}^{\left(3\right)} &=& + \frac{4}{c^3} \sum\limits_{l=1}^{\infty} \frac{\left(-1\right)^l\,l}{\left(l+1\right)!} \, 
	\epsilon_{iab}\,\hat{S}_{b L-1}\, \hat{\partial}_{a L-1} \frac{1}{r}\;,  
        \label{Metric_0i}
        \\
	h_{ij}^{\left(2\right)} &=& + \frac{2}{c^2}\,\delta_{ij}  
	\sum\limits_{l=0}^{\infty} \frac{\left(-1\right)^l}{l!}\,\hat{M}_L\,\hat{\partial}_L \frac{1}{r}\;, 
        \label{Metric_ij}
        \end{eqnarray} 

\noindent
where $r = \left|\ve{x}\right|$ and  
\begin{eqnarray}
	\hat{\partial}_L = {\rm STF}_{i_1 \dots i_l}\,\partial_{i_1} \dots \partial_{i_l}\;.
\label{Differential_Operator}
\end{eqnarray}

\noindent 
The mass-multipoles and spin-multipoles in (\ref{Metric_00}) - (\ref{Metric_ij}) in case of stationary source of matter are given by  
\begin{eqnarray}
        \hat{M}_L &=& \int d^3 x \; \hat{x}_L\;\Sigma\,,  
\label{Mass_Multipoles}
\\
        \hat{S}_L &=& \int d^3 x \;\epsilon_{j k < i_l}\,\hat{x}_{L-1 >}\;x^j\;\Sigma^k\,,   
\label{Spin_Multipoles}
\end{eqnarray}

\noindent 
where the notation $\Sigma = \left(T^{00} + T^{kk}\right)/c^2$ and $\Sigma^k = T^{0k}/c$ has been adopted, with $T^{\alpha\beta}$ being the 
stress-energy tensor of the body, and where the integrals run over the three-dimensional volume of the body. 
The geodesic equation in 1.5PN approximation can be deduced from the exact geodesic equation (\ref{Geodetic_Equation2}) 
and is given by Eq.~(2.2.49) in \cite{Brumberg1991} (up to a global sign convention). Inserting the metric perturbations (\ref{Metric_00}) - (\ref{Metric_ij}) 
into the geodesic equation in 1.5PN approximation yields  
\begin{eqnarray}
	\frac{\ddot{\ve{x}}}{c^2} &=&  \sum\limits_{l=0}^{\infty} \frac{\ddot{\ve{x}}_{\rm 1PN}^{M_L}}{c^2} 
	+ \sum\limits_{l=1}^{\infty} \frac{\ddot{\ve{x}}_{\rm 1.5PN}^{S_L}}{c^2} 
	\label{Geodesic_Equation_15PN}
\end{eqnarray}

\noindent 
up to terms of the order ${\cal O}(c^{-4})$, and 
where $\ddot{\ve{x}}_{\rm 1PN}^{M_L}$ and $\ddot{\ve{x}}_{\rm 1.5PN}^{S_L}$ are given by Eq.~(13) in \cite{Kopeikin1997}. 
The solution of (\ref{Geodesic_Equation_15PN}) reads formally as follows:
\begin{eqnarray}
	\ve{x}\left(t\right) &=& \ve{x}_0 + c \left(t - t_0\right) \ve{\sigma} + \sum\limits_{l=0}^{\infty} \Delta\ve{x}^{M_L}_{\rm 1PN} 
        + \sum\limits_{l=1}^{\infty} \Delta\ve{x}^{S_L}_{\rm 1.5PN} 
	\nonumber\\ 
        \label{Second_Interation_1PN}
\end{eqnarray}

\noindent
up to terms of the order ${\cal O}(c^{-4})$, and where $\Delta\ve{x}_{\rm nPN} = {\cal O}\left(c^{-2n}\right)$. 
\begin{table}[h!]
\caption{Numerical parameter for mass $M$, radius $P$,   
	actual zonal harmonic coefficients $J_l$, distance between observer and body $x_1$, of Sun, Jupiter and Saturn. 
	The values for $G M/c^2$ and $P$ are taken from \cite{Ellipticity}. 
	The value for $J_l$ for the Sun are taken from \cite{J_n_Sun} and references therein. 
	The values $J_l$ with $n=2,4,6$ for Jupiter and Saturn are taken from \cite{Book_Zonal_Harmonics},
        while $J_l$ with $n=8,10$ for Jupiter and Saturn are taken from \cite{Zonal_Harmonics_Jupiter} and \cite{Zonal_Harmonics_Saturn}, respectively.
	The angular velocity $\Omega = 2 \pi /T$ (with rotational period $T$) are taken from NASA planetary fact sheets. 
	The dimensionless moment of inertia $\kappa^2$ is defined by Eq.~(\ref{kappa}) and their values are taken from \cite{Ellipticity}.
For the distance between light-source and body we assume $x_0 = 10^{11}\,{\rm m}$
so that the light-source is within the near-zone of the Solar system, while $x_1$ is computed under the assumption that the observer (spacecraft)
is located at Lagrange point $L_2$, i.e. $1.5 \times 10^9\,{\rm m}$ from Earth.}
\begin{tabular}{| c | c | c | c |}
\hline
&&&\\[-12pt]
Parameter
&\hbox to 20mm{\hfill Sun \hfill}
&\hbox to 20mm{\hfill Jupiter \hfill}
&\hbox to 20mm{\hfill Saturn \hfill}\\[3pt]
\hline
&&&\\[-12pt]
$GM/c^2\,[{\rm m}]$ & $1476.8$ & $1.41$ & $0.42$ \\[3pt]
$P\,[{\rm m}]$ & $696 \times 10^6$ & $71.5 \times 10^6$ & $60.3 \times 10^6$ \\[3pt]
$J_2$ & $1.7 \times 10^{-7}$ & $14.696 \times 10^{-3}$ & $16.291 \times 10^{-3}$ \\[3pt]
$J_4$ & $ 9.8 \times 10^{-7} $ & $ - 0.587 \times 10^{-3}$ & $ - 0.936 \times 10^{-3}$ \\[3pt]
$J_6$ & $ 4 \times 10^{-8} $ & $0.034 \times 10^{-3}$ & $0.086 \times 10^{-3}$ \\[3pt]
$J_8$ & $ - 4 \times 10^{-9} $ & $ - 2.5 \times 10^{-6}$ & $ - 10.0 \times 10^{-6}$ \\[3pt]
$J_{10}$ & $ - 2 \times 10^{-10} $ & $0.21 \times 10^{-6}$ & $2.0 \times 10^{-6}$ \\[3pt]
$\Omega\,[{\rm sec}^{-1}]$ & $2.865 \times 10^{-6}$ & $1.758 \times 10^{-4}$ & $1.638 \times 10^{-4}$ \\[3pt]
$\kappa^2$ & $0.059$ & $0.254$ & $0.210$ \\[3pt]
$x_1\,[{\rm m}]$ & $0.150 \times 10^{12}$ & $0.59 \times 10^{12}$ & $1.20 \times 10^{12}$ \\[3pt]
\hline
\end{tabular}
\label{Table1}
\end{table}
In \cite{Kopeikin1997} advanced integration methods have been introduced which allow to integrate (\ref{Geodesic_Equation_15PN}) exactly and which lead to the 
exact expression of (\ref{Second_Interation_1PN}), given by Eqs.~(33), (36) and (38) in \cite{Kopeikin1997}. In that approach two new parameters were introduced,
\begin{eqnarray}
	c \tau &=& \ve{\sigma} \cdot \ve{x}_{\rm N}\;,
	\label{Parameter1}
	\\
	\xi^i &=& P^i_j\,x_{\rm N}^j\;,  
	\label{Parameter2}
\end{eqnarray}
 
\noindent
where $P^{ij} = \delta^{ij} - \sigma^i \sigma^j $ is a projection operator onto the plane perpendicular to vector $\ve{\sigma}$; 
note that $P^{ij} = P_{ij} = P^i_j$. 
Obviously, the unperturbed light ray (\ref{Unperturbed_Lightray_2}) expressed in terms of these new variables takes the form 
\begin{eqnarray}
	\ve{x}_{\rm N} &=& \ve{\xi} + c \tau\,\ve{\sigma}\;. 
        \label{Parameter3}
\end{eqnarray}

\noindent
The three-vector $\ve{\xi}$ is laying in the two-dimensional plane perpendicular to $\ve{\sigma}$, hence only two components are independent, 
which implies $\partial \xi^i / \partial \xi^j = P^i_j$. 
But in practical calculations it is convenient to treat the spatial components of this vector as formally independent,  
which implies $\partial \xi^i / \partial \xi^j = \delta^i_j$. Therefore, a 
subsequent projection onto this two-dimensional plane by means of $P^{ij}$ is necessary (cf. text above Eq.~(31) in \cite{KopeikinSchaefer1999_Gwinn_Eubanks} 
as well as Eqs. (11.2.12) and (11.2.13) in \cite{Book_Soffel_Han}). 
Then, for a spatial derivative expressed in terms of these new variables, one obtains  
\begin{eqnarray}
	\frac{\partial}{\partial x^i} &=& P_i^j\,\frac{\partial}{\partial \xi^j} + \sigma_i\,\frac{\partial}{\partial c \tau}\;. 
	\label{spatial_derivative}
\end{eqnarray}

\noindent
In case of time-independent functions, relation (33) in \cite{KopeikinSchaefer1999_Gwinn_Eubanks} coincides with relation (\ref{spatial_derivative}).
Then, using (\ref{spatial_derivative}) and the binomial theorem, one finds the differential operator in (\ref{Differential_Operator}) 
expressed in terms of these new variables,  
\begin{eqnarray}
	\hat{\partial}_{L} &=& {\rm STF}_{i_1 \dots i_l}\;\sum\limits_{p=0}^{l} \frac{l!}{\left(l-p\right)!\;p!}\;\sigma_{i_1}\,...\,\sigma_{i_p}  
        \nonumber\\ 
	&& \times\, P_{i_{p+1}}^{j_{p+1}}\;...\;P_{i_l}^{j_l} 
	\;\frac{\partial}{\partial \xi^{j_{p+1}}}\;...\;
\frac{\partial}{\partial \xi^{j_{l}}}\;\left(\frac{\partial}{\partial c\tau}\right)^p \,.  
\label{Transformation_Derivative_3}
\end{eqnarray}

\noindent
Here we prefer to use the operator as given by Eq.~(\ref{Transformation_Derivative_3}) where $\partial \xi^i / \partial \xi^j = \delta^i_j$, 
while if one applies the operator as given by Eq.~(24) in \cite{Kopeikin1997} then 
$\partial \xi^i / \partial \xi^j = P^i_j$. The results of either these operations are identical. 
Then, using the basic integral (25) in \cite{Kopeikin1997} one finds for the second integration the formulas given by Eq.~(27) 
in \cite{Kopeikin1997}, which lead to the solution for the second integration of geodesic equation (\ref{Geodesic_Equation_15PN}). 

The approach introduced in \cite{Kopeikin1997} for bodies at rest and time-independent multipoles has further been developed for the case of light propagation in
the gravitational field of a time-dependent source of matter at rest \cite{KopeikinSchaefer1999_Gwinn_Eubanks,KopeikinKorobkovPolnarev2006,KopeikinKorobkov2005}, 
as well as in the gravitational field of $N$ slowly moving bodies with time-dependent multipoles in our investigations in \cite{Zschocke_1PN,Zschocke_15PN}.

According to the solution for the light trajectory as given by Eq.~(31) with (33), (36), (38) in \cite{Kopeikin1997}, the time-of-flight 
in the gravitational field of a body with full mass-multipole and spin-multipole structure is given by the following formula 
(cf. Eq.~(40) in \cite{Kopeikin1997}),  
\begin{eqnarray}
	c \left(t_1 - t_0 \right) &=& R + \sum\limits_{l=0}^{\infty} \Delta c\tau_{\rm 1PN}^{M_L} 
	+ \sum\limits_{l=1}^{\infty} \Delta c\tau_{\rm 1.5PN}^{S_L}  
        \label{Shapiro_15PN} 
\end{eqnarray}

\noindent 
up to terms of the order ${\cal O}(c^{-4})$. 
The mass-multipole (gravitoelectric) term reads (cf. Eqs.~(41) and (42) in \cite{Kopeikin1997}) 
\begin{eqnarray}
\Delta c\tau_{\rm 1PN}^{M_L} &=& + \frac{2 G}{c^2} \frac{\left(-1\right)^l}{l!} \hat{M}_L
\nonumber\\ 
	&& \hspace{-1.5cm} \times \left(\hat{\partial}_{L} \,\ln \left(r_{\rm N} + c \tau\right)\bigg|_{\tau = t_1} 
	\hspace{-0.25cm} - \hat{\partial}_{L} \,\ln \left(r_{\rm N} + c \tau\right)\bigg|_{\tau = t_0}\right),  
\nonumber\\ 
	\label{Shapiro_Mass_Multipole} 
\end{eqnarray}

\noindent
and the spin-multipole (gravitomagnetic) term reads (cf. Eq.~(43) in \cite{Kopeikin1997}; note an overall sign error in Eq.~(43) in \cite{Kopeikin1997}; 
see also footnote $3$ in \cite{Ciufolini} as well as Ref.~$\left[73\right]$ in \cite{Zschocke_15PN}) 
\begin{eqnarray}
	\Delta c\tau_{\rm 1.5PN}^{S_L} &=& + \frac{4 G}{c^3} \frac{\left(-1\right)^l\,l}{\left(l+1\right)!}
	\,\sigma_i\,\epsilon^{iab}\,\hat{S}_{bL-1}
	\nonumber\\
        && \hspace{-1.75cm} \times \left(\!\hat{\partial}_{a L-1} \ln \left(r_{\rm N} + c \tau\right)\bigg|_{\tau = t_1} 
	 \hspace{-0.35cm} - \hat{\partial}_{a L- 1} \ln \left(r_{\rm N} + c \tau\right)\bigg|_{\tau = t_0}\right),
	\nonumber\\ 
\label{Shapiro_Spin_Multipole}
\end{eqnarray}

\noindent
where $r_{\rm N} = \left|\ve{x}_{\rm N}\right|$ with $\ve{x}_{\rm N}$ in (\ref{Parameter3}), that means $r_{\rm N} = \sqrt{\xi^2 + c^2 \tau^2}$. 
These equations were also given by Eqs.~(11.2.34) and (11.2.35) in \cite{Book_Soffel_Han}.  
In (\ref{Shapiro_Mass_Multipole}) and (\ref{Shapiro_Spin_Multipole}) the differentiations have to be performed. 
Afterwards one has to substitute the unperturbed light ray by the standard expression as given by Eq.~(\ref{Unperturbed_Lightray_2}) where 
the coordinate time is either $t_1$ or $t_0$ as indicated by the sub-labels. In particular, at the very end of the calculations one has to replace 
$c \tau$ by $\ve{\sigma} \cdot \ve{x}_{\rm N}$ and $\ve{\xi}$ by $\ve{d}_{\sigma}$. For details about how to perform the differentiations the reader is referred 
to \cite{Kopeikin1997,Book_Soffel_Han}. Because the mass-quadrupole is of specific relevance in our investigation, we consider the application of 
(\ref{Shapiro_Mass_Multipole}) for the mass-quadrupole explicitly in Appendix~\ref{Appendix_Mass_Quadrupole}. 

The largest effect of Shapiro effect is expected from the Sun and the giant planets of the solar system. In order to determine the Shapiro time delay 
one needs the explicit form for mass-multipoles (\ref{Mass_Multipoles}) and for spin-multipoles (\ref{Spin_Multipoles}). 
For an estimation of the individual terms in (\ref{Shapiro_Mass_Multipole}) and (\ref{Shapiro_Spin_Multipole}), one may approximate the Sun and the giant planets 
by a rigid axisymmetric body with radial dependent mass distribution and in uniform rotational motion around the symmetry axis of the body, 
which is aligned with the $x^3$-axis of the coordinate system. Then, the higher mass-multipoles for such a body 
are given by Eqs.~(\ref{M_L_F}) in Appendix \ref{Appendix_Multipoles}, while the spin-dipole and higher spin-multipoles for such a body are given by 
Eq.~(\ref{S_a_B}) and Eq.~(\ref{S_L_F}) in Appendix \ref{Appendix_Multipoles}: 
\begin{eqnarray}
	\hat{M}_0 &=& M\;,
        \label{M_0_Newtonian}
	\\ 
	\hat{M}_L &=& - M \left(P\right)^l\,J_l\;\delta^3_{<{i_1}} \; \dots \; \delta^3_{{i_l}>} 
\label{M_L_Newtonian}
	\\
	&& {\rm with} \quad l = 2, 4, 6, \dots \;, 
	\nonumber\\ 
	\nonumber\\ 
	\hat{S}_{a} &=& \kappa^2\,M\,\Omega\,P^2\,\delta_{3a}\;,
        \label{S_1_Newtonian}
        \\ 
	\hat{S}_L &=& - M\,\Omega\left(P\right)^{l+1}\,J_{l-1}\,\frac{l+1}{l+4}\;\delta^3_{<{i_1}} \; \dots \; \delta^3_{{i_l}>}  
\label{S_L_Newtonian}
	\\ 
	&& {\rm with} \quad l = 3, 5, 7, \dots \;, 
	\nonumber  
\end{eqnarray}

\noindent 
where $M$ is the Newtonian mass of the body, $P$ its equatorial radius, $J_l$ are the actual zonal harmonic coefficients of index $l$, 
$\kappa^2$ is the dimensionless moment of inertia, $\Omega$ is the angular velocity of the rotating body and 
$\delta^3_{<{i_1}} \; \dots \; \delta^3_{{i_l}>} = {\rm STF}_{i_1 \dots i_l}\;\delta_{3 i_1} \dots \delta_{3 i_l}$ 
denotes products of Kronecker symbols which are symmetric and traceless with respect to indices $i_1 \dots i_l$. These multipoles (\ref{M_L_Newtonian}) 
and (\ref{S_L_Newtonian}) are in agreement with the multipoles for an rigid axisymmetric body in uniform rotational motion as given in the resolutions of the 
International Astronomical Union (IAU) \cite{IAU_Resolution1}; that agreement is shown explicitly in Appendix~\ref{Appendix_Multipoles} for the mass-quadrupole 
as well as for the spin-hexapole in case of a rigid axisymmetric body with uniform mass-density. 

The calculations can considerably be simplified by inserting the mass-multipoles and spin-multipoles (\ref{M_L_Newtonian}) and (\ref{S_L_Newtonian}) into 
(\ref{Shapiro_Mass_Multipole}) and (\ref{Shapiro_Spin_Multipole}), respectively, and afterwards one starts with the evaluation of the Shapiro time delay. 
Then, one obtains the following upper limits of the individual terms of Shapiro time delay (cf. text below Eq.~(43) in \cite{Kopeikin1997}): 
\begin{eqnarray}
	\left|\Delta \tau_{\rm 1PN}^{M}\right| &\le& 2\,\frac{G M}{c^3} \,\ln \frac{4\,x_0\,x_1}{\left(d_{\sigma}\right)^2}\;, 
	\label{Shapiro_M_1PN}
	\\
	\left|\Delta \tau_{\rm 1PN}^{M_L}\right| &\le& A_l\,\frac{G M}{c^3}\,\left|J_l\right| \left(\frac{P}{d_{\sigma}}\right)^l
	\label{Shapiro_ML_1PN}
	\\ 
	&& {\rm with} \quad l = 2, 4, 6, \dots\;,
	\nonumber\\
	\left|\Delta \tau_{\rm 1.5PN}^{S}\right| &\le& 4\,\frac{G M}{c^4}\,\kappa^2\,P\,\Omega\;, 
	\label{Shapiro_S_15PN}
	\\
	\nonumber\\ 
	\left|\Delta \tau_{\rm 1.5PN}^{S_L}\right| &\le& B_l\,\frac{G M}{c^4}\,P\,\Omega\,\left|J_{l-1}\right| \left(\frac{P}{d_{\sigma}}\right)^l 
	\label{Shapiro_SL_1PN}
	\\ 
	&& {\rm with} \quad l = 3, 5, 7, \dots\;,
	\nonumber 
\end{eqnarray}

\noindent
where in (\ref{Shapiro_S_15PN}) we have used relation (\ref{S_a_B}). The non-vanishing coefficients for the first few mass-multipoles and spin-multipoles read 
\begin{eqnarray}
	A_2 &=& \frac{11}{5}\;,\; A_4 = \frac{7}{6}\;,\; A_6 = \frac{3}{5}\;,\; A_8 = \frac{3}{10}\;,\; A_{10} = \frac{3}{20}\;,
        \nonumber\\  
	\label{Coefficients_ML}
	\\
	B_3 &=& \frac{7}{6}\;.  
        \label{Coefficients_SL}
\end{eqnarray}

\noindent
The calculation of coefficient $A_2$ is given in some detail in Appendix~\ref{Appendix_Mass_Quadrupole}, while the determination of the 
other coefficients in (\ref{Coefficients_ML}) - (\ref{Coefficients_SL}) proceeds in very similar manner. 
Thus far, to the best of our knowledge, these upper limits have only been determined for mass-monopole, mass-quadrupole and spin-dipole, 
which were given in \cite{Klioner1991}.  
Numerical values of the upper limits in (\ref{Shapiro_M_1PN}) - (\ref{Shapiro_SL_1PN}) are presented in Table~\ref{Table2} for the first mass-multipoles 
and spin-multipoles in case of grazing rays at the Sun and the giant planets of the solar system. 

In Table~\ref{Table2} for the Sun, Jupiter, and Saturn, time delays of mass-quadrupole of $1.8\,{\rm ps}$, $152.1\,{\rm ps}$, and $50.6\,{\rm ps}$ are given. These 
values differ from the values in Table~I in \cite{Klioner1991}, where for Sun, Jupiter, and Saturn, time delays of mass-quadrupole of $16\,{\rm ps}$, 
$240\,{\rm ps}$, and $73\,{\rm ps}$ were given. These differences originate from different upper limits. Here, according to Eq.~(\ref{Shapiro_ML_1PN}), we have 
used $\Delta \tau^{M_2}_{\rm 1PN} \le 2.2\,\frac{GM}{c^3}\,|J_2|$ (which is in line with the statement below Eq.~(47) in \cite{Klioner1991}), 
while in Table~I in \cite{Klioner1991} an upper limit of $\Delta \tau^{M_2}_{\rm 1PN} \le 3.18\,\frac{GM}{c^3}\,|J_2|$ has been used 
(cf. Eq.~(47) in \cite{Klioner1991}). In addition, different values for the second zonal harmonic coefficient $J_2$ have been used for the Sun. 
On the other side, the values for the time delay of spin-dipole presented in Table~\ref{Table2} coincide with the values given in Table~II in \cite{Klioner1991}. 
\begin{table}[h!]
	\caption{The effect of 1PN mass-multipole $\Delta \tau^{M_l}_{\rm 1PN}$ and 1.5PN spin-multipole terms $\Delta \tau^{S_l}_{\rm 1.5PN}$ 
	of (one-way) Shapiro time delay in the gravitational field of the Sun and giant planets of the solar system 
	according to the upper limits presented by Eqs.~(\ref{Shapiro_M_1PN}) - (\ref{Shapiro_SL_1PN}).  
	The time delay is given in units of pico-seconds: $1\,{\rm ps} = 10^{-12}\,{\rm sec}$. 
	The values are given for grazing rays (impact parameter $d_{\sigma}$ equals body's equatorial radius $P$).  
	Values for $\Delta \tau^{M_l}_{\rm 1PN}$ with $\l \ge 10$ and $\Delta \tau^{S_l}_{\rm 1.5PN}$ with $\l \ge 5$ 
	are not shown in the Table, because they are less than a femto-second for any solar system body. 
	The numerical values should be compared with the assumed goal accuracy of $0.001$ pico-seconds in time-delay measurements. 
	A blank entry means a delay of less than a femto-second.} 
\footnotesize
\begin{tabular}{@{}|c|c|c|c|c|c|c|c|c}
\hline
&&&&&&&\\[-7pt]
	Object & $\Delta \tau^{M}_{\rm 1PN}$ & $\Delta \tau^{M_2}_{\rm 1PN}$ & $\Delta \tau^{M_4}_{\rm 1PN}$ & $\Delta \tau^{M_6}_{\rm 1PN}$ & $\Delta \tau^{M_8}_{\rm 1PN}$ & $\Delta \tau^{S_1}_{\rm 1.5PN}$  & $\Delta \tau^{S_3}_{\rm 1.5PN}$\\
&&&&&&&\\[-7pt]
\hline
        Sun & $1.6\!\times\!10^8$ & $1.8$ & $ 5.6 $ & $ 0.1 $ & $ 0.006 $ & $ 7.7 $ & $ - $ \\
	Jupiter & $2.2\!\times\!10^5$ & $152.1$ & $ 3.2 $ & $ 0.1 $ & $ 0.004 $ & $ 0.2 $ & $ 0.001 $ \\
	Saturn & $6.8\!\times\!10^4$ & $50.6$ & $ 1.5 $ & $ 0.07 $ & $ 0.004 $ & $ 0.04 $ & $ - $ \\
\hline
\end{tabular}\\
\label{Table2} 
\end{table}
\normalsize
 
\noindent 
Finally, an important comment should be in order. The solutions for the light trajectory as well as Shapiro time delay in the 1PN and 1.5PN approximation 
are given in terms of the unit vector $\ve{\sigma}$, which can immediately be replaced by the unit vector $\ve{k}$, because they differ by terms beyond 
the 1PN and 1.5PN approximation: $\ve{\sigma} = \ve{k} + {\cal O}\left(c^{-2}\right)$ and $\ve{\sigma} \cdot \ve{k} = 1 +  {\cal O}\left(c^{-4}\right)$. 
However, in 2PN approximation one has carefully to distinguish among these vectors. In addition, in 2PN approximation one must not replace 
$\ve{x}_{\rm N}\left(t_1\right)$ by the spatial position of the observer $\ve{x}_1$, because such a replacement causes an error of the order 
${\cal O}\left(c^{-4}\right)$ which is of second post-Newtonian order. These both aspects make the treatment of the determination of Shapiro time delay 
in 2PN approximation more involved and will be considered in the next sections.

\section{Light propagation in 2PN approximation: Initial value problem \label{Section3}}

A unique solution of geodesic equation (\ref{Geodetic_Equation2}) is given by the initial value problem as defined by Eqs.~(\ref{Initial_A}) and (\ref{Initial_B}). 
In order to get the geodesic equation one needs the metric tensor in Eq.~(\ref{PN_Expansion_Multipoles_B}). 
In 2PN approximation the expansion in Eq.~(\ref{PN_Expansion_Multipoles_B}) reads as follows,
\begin{eqnarray}
        g_{\alpha\beta} &=& \eta_{\alpha\beta} + h_{\alpha\beta}^{\left(2\right)}(\hat{M}_L)
        + h_{\alpha\beta}^{\left(3\right)}(\hat{S}_L)
        + h_{\alpha\beta}^{\left(4\right)}(\hat{M}_L)
        \label{PN_Expansion_2PN}
\end{eqnarray}

\noindent
up to terms of the order ${\cal O}(c^{-5})$, and 
where the mass-multipoles $\hat{M}_L$ and spin-multipoles $\hat{S}_L$ are given by Eqs.~(\ref{M_L}) and (\ref{S_L}), respectively, and they are assumed to be 
time-independent. The 1PN and 1.5PN metric perturbations, $h_{\alpha\beta}^{\left(2\right)}$ and $h_{\alpha\beta}^{\left(3\right)}$, 
were given by Eqs.~(\ref{Metric_00}) - (\ref{Metric_ij}), while the 2PN metric perturbations $h_{\alpha\beta}^{\left(4\right)}$ have been derived from the 
MPM formalism \cite{Thorne,Blanchet_Damour1} and were given by Eqs.~(115) - (117) and Eqs.~(134) - (136) in our article \cite{Zschocke_2PM_Metric} for the 
case of time-independent multipoles. 

For our considerations about the 2PN effect of time-delay in the gravitational field of one body at rest, where only the mass-monopole 
and mass-quadrupole will be taken into account, that means 
\begin{eqnarray}
	\hat{M}_L &=& 0 \quad {\rm for} \quad l > 2\;, 
	\label{M_L_2PN}
	\\
	\hat{S}_L &=& 0 \quad {\rm for} \quad l \ge 1 \;.
	\label{S_L_2PN}
\end{eqnarray}

\noindent 
But we will keep in mind the exact solution of geodesic equation in 1.5PN approximation in (\ref{Second_Interation_1PN}) and the Shapiro time 
delay in 1.5PN approximation in (\ref{Shapiro_15PN}), and we may finally add these terms at the very end of our calculations of the Shapiro time delay in 
2PN approximation. 

Thus far, our knowledge about 2PN effects in the theory of light propagation was restricted to the case of light propagation in the field of monopoles 
\cite{Brumberg1991,KlionerKopeikin1992,Article_Zschocke1}. In our recent article \cite{Zschocke_Quadrupole_1} the initial value problem of 2PN light propagation 
in the field of one body at rest with quadrupole structure has been solved. The metric (\ref{PN_Expansion_2PN}) for one massive solar system body at rest 
with monopole and quadrupole structure takes the form (cf. Eq.~(16) in \cite{Zschocke_Quadrupole_1})
\begin{eqnarray}
        g_{\alpha\beta} &=& \eta_{\alpha\beta} + h_{\alpha\beta}^{\left(2\right)}(M,\hat{M}_{ab}) 
        + h_{\alpha\beta}^{\left(4\right)}(M,\hat{M}_{ab}) 
	\label{PN_Expansion_4}
\end{eqnarray}

\noindent
up to terms of the order ${\cal O}(c^{-6})$ (there are no terms of the order ${\cal O}(c^{-5})$ because the spin-multipoles are neglected), and  
where higher mass-multipoles as well as spin-multipoles have been neglected; the origin of spatial axes of the coordinate system is located 
at the center of mass of the body and, therefore, the mass-dipole vanishes (cf. Eq.~(8.14c) in \cite{Thorne}). 
The explicit expressions for the metric perturbations in (\ref{PN_Expansion_4}) 
have been derived by Eqs.~(145) and (147) as well as Eqs.~(148) - (150) in our article \cite{Zschocke_2PM_Metric}.  
By inserting the 2PN metric tensor (\ref{PN_Expansion_4}) in the geodesic equation (\ref{Geodetic_Equation2}) one obtains the geodesic equation 
in 2PN approximation (cf. Eq.~(74) in \cite{Zschocke_Quadrupole_1})
\begin{eqnarray}
        \frac{\ddot{\ve{x}}}{c^2} &=& \frac{\ddot{\ve{x}}^{M}_{\rm 1PN}}{c^2}
	+ \frac{\ddot{\ve{x}}^{M_{ab}}_{\rm 1PN}}{c^2}
        + \frac{\ddot{\ve{x}}^{{M} \times {M}}_{\rm 2PN}}{c^2}
	+ \frac{\ddot{\ve{x}}^{{M} \times {M_{ab}}}_{\rm 2PN}}{c^2}
	+ \frac{\ddot{\ve{x}}^{{M_{ab}} \times {M_{cd}}}_{\rm 2PN}}{c^2}
\nonumber\\ 
	\label{Geodesic_Equation_2PN}
\end{eqnarray}

\noindent
up to terms of the order ${\cal O}(c^{-6})$. The geodesic equation (\ref{Geodesic_Equation_2PN}) 
can be written in terms of time-independent tensorial coefficients and time-dependent scalar functions. For the explicit form of   
geodesic equation (\ref{Geodesic_Equation_2PN}) we refer to Eqs.~(47) - (49) in \cite{Zschocke_Quadrupole_1} for the 1PN terms as well as  
Eqs.~(75) and (78) - (79) in \cite{Zschocke_Quadrupole_1} for the 2PN terms. The solution of the second integration of the 
geodesic equation (\ref{Geodesic_Equation_2PN}) reads (cf. Eq.~(86) in \cite{Zschocke_Quadrupole_1}) 
\begin{eqnarray}
        \ve{x}\left(t\right) &=& 
        \ve{x}_0 + c \left(t - t_0\right) \ve{\sigma} + \Delta \ve{x}_{\rm 1PN}\left(t,t_0\right) + \Delta \ve{x}_{\rm 2PN}\left(t,t_0\right) 
	\nonumber\\ 
        \label{Second_Integration} 
\end{eqnarray}
 
\noindent
up to terms of the order ${\cal O}(c^{-6})$, and 
where $\Delta\ve{x}_{\rm 1PN} = {\cal O}\left(c^{-2}\right)$ and $\Delta\ve{x}_{\rm 2PN} = {\cal O}\left(c^{-4}\right)$. 
In favor of a simpler notation, the monopole and quadrupole terms in (\ref{Second_Integration}) have been summarized as follows, 
\begin{eqnarray}
        \Delta \ve{x}_{\rm 1PN} &=& \Delta \ve{x}_{\rm 1PN}^{M} + \Delta \ve{x}_{\rm 1PN}^{M_{ab}}\;,
        \label{Notation_1PN_Second}
	\\
	\Delta \ve{x}_{\rm 2PN} &=& \Delta \ve{x}_{\rm 2PN}^{M \times M} 
	+ \Delta \ve{x}_{\rm 2PN}^{M \times M_{ab}} + \Delta \ve{x}_{\rm 2PN}^{M_{ab} \times M_{cd}} \;,
\end{eqnarray}

\noindent 
in obvious meaning: index $M$ means terms proportional to the monopole, index $M_{ab}$ means terms proportional to the quadrupole, 
index $M \times M$ means terms proportional to the monopole times monopole, index $M \times M_{ab}$ means terms proportional to the 
monopole times quadrupole and index $M_{ab} \times M_{cd}$ means terms proportional to the quadrupole times quadrupole. 
In this Section we reconsider the solution of the second integration (\ref{Second_Integration}) 
as it has been obtained in our article \cite{Zschocke_Quadrupole_1}. However, it is necessary to rewrite this solution into a new 
form which is appropriate for subsequent considerations of the Shapiro time delay.

\subsection{Old representation} 

The iterative solution of the second integration of geodesic equation in 2PN approximation (\ref{Geodesic_Equation_2PN}) reads \cite{Zschocke_Quadrupole_1}: 
\begin{eqnarray} 
        \ve{x}_{\rm N}  &=& \,\ve{x}_0 + c \left(t-t_0\right) \ve{\sigma} \;, 
        \label{S_I_N_Old} 
	\\
	\ve{x}_{\rm 1PN} &=& \ve{x}_{\rm N}  
        + \Delta \ve{x}_{\rm 1PN}\left(\ve{x}_{\rm N}\right) - \Delta \ve{x}_{\rm 1PN}\left(\ve{x}_0\right),
        \label{S_I_1PN_Old} 
        \\
        \ve{x}_{\rm 2PN} &=& 
	\ve{x}_{\rm N} + \Delta \ve{x}_{\rm 1PN}\left(\ve{x}_{\rm N}\right) -  \Delta \ve{x}_{\rm 1PN}\left(\ve{x}_0\right) 
	\nonumber\\ 
	&& \hspace{0.55cm} + \, \Delta \ve{x}_{\rm 2PN}\left(\ve{x}_{\rm N}\right) - \Delta \ve{x}_{\rm 2PN}\left(\ve{x}_0\right), 
        \label{Second_Integration_Old} 
\end{eqnarray}

\noindent 
where the spatial components of 1PN terms are given by  
\begin{eqnarray}
        \Delta x^i_{\rm 1PN}\left(\ve{x}_{\rm N}\right) &=& 
        \frac{G M}{c^2} \bigg[{\cal A}^i_{\left(3\right)} \, {\cal W}_{\left(3\right)}\left(t\right)  
        + {\cal B}^i_{\left(3\right)} \, {\cal X}_{\left(3\right)}\left(t\right)\bigg]
        \nonumber\\
	&& \hspace{-2.0cm} + \frac{G \hat{M}_{ab}}{c^2} \sum\limits_{n=5,7} 
        \bigg[{\cal C}^{i\,ab}_{\left(n\right)}\,{\cal W}_{\left(n\right)}\left(t\right)  
        + {\cal D}^{i\,ab}_{\left(n\right)}\,{\cal X}_{\left(n\right)}\left(t\right) \bigg],
        \label{S_I_1PN} 
\end{eqnarray}
 
\noindent
and the spatial components of 2PN terms are given by 
\begin{eqnarray}
        \Delta x^i_{\rm 2PN}\left(\ve{x}_{\rm N}\right) &=& \frac{G^2 M^2}{c^4}
        \bigg[\sum\limits_{n=3}^{6} {\cal E}^{i}_{\left(n\right)} {\cal W}_{\left(n\right)}\!\left(t\right)
	+ \sum\limits_{n=2}^{6} {\cal F}^{i}_{\left(n\right)} {\cal X}_{\left(n\right)}\!\left(t\right) 
	\nonumber\\ 
	&& \hspace{-0.5cm} + {\cal G}^{i}_{\left(5\right)} \; {\cal Y}_{\left(5\right)}\left(t\right)
        + \sum\limits_{n=3,5} {\cal H}^{i}_{\left(n\right)} \; {\cal Z}_{\left(n\right)}\left(t\right)\bigg]
        \nonumber\\
        && \hspace{-2.25cm} + \frac{G^2 M \hat{M}_{ab}}{c^4} 
        \bigg[
        \sum\limits_{n=3}^{10} {\cal K}^{i\,ab}_{\left(n\right)}\,{\cal W}_{\left(n\right)}\left(t\right)
        + \sum\limits_{n=2}^{10} {\cal L}^{i\,ab}_{\left(n\right)}\,{\cal X}_{\left(n\right)}\left(t\right) 
        \nonumber\\
	&& \hspace{-0.5cm} + \sum\limits_{n=7}^{9} {\cal M}^{i\,ab}_{\left(n\right)}\,{\cal Y}_{\left(n\right)}\left(t\right)
        + \sum\limits_{n=5}^{9} {\cal N}^{i\,ab}_{\left(n\right)}\,{\cal Z}_{\left(n\right)}\left(t\right) \bigg] 
        \nonumber\\
        && \hspace{-2.25cm} + \frac{G^2 \hat{M}_{ab} \hat{M}_{cd}}{c^4} \bigg[
        \sum\limits_{n=5}^{14} {\cal P}^{i\,abcd}_{\left(n\right)}\,{\cal W}_{\left(n\right)}\left(t\right)
        + \sum\limits_{n=4}^{14} {\cal Q}^{i\,abcd}_{\left(n\right)}\,{\cal X}_{\left(n\right)}\left(t\right)\bigg]. 
        \nonumber\\
        \label{S_I_2PN} 
\end{eqnarray}

\noindent
In order to get $\Delta \ve{x}_{\rm 1PN}\left(\ve{x}_0\right)$ and $\Delta \ve{x}_{\rm 2PN}\left(\ve{x}_0\right)$ 
we notice that $\ve{x}_0 = \ve{x}_{\rm N}\left(t_0\right)$, that means one has to take the time-argument $t_0$ in 
the scalar functions in (\ref{S_I_1PN}) and (\ref{S_I_2PN}). 

The tensorial coefficients ${\cal A}^{i}_{\left(3\right)}$, ${\cal B}^{i}_{\left(3\right)}$, ${\cal C}^{i\,ab}_{\left(n\right)}$,
${\cal D}^{i\,ab}_{\left(n\right)}$ are given by Eqs.~(52) - (57) in \cite{Zschocke_Quadrupole_1}. In what follows these coefficients 
are essential and have, therefore, been given by Eqs.~(\ref{coefficients_A3_N}) - (\ref{coefficients_D7_N}) in Appendix~\ref{Appendix1PN}.
The tensorial coefficients ${\cal E}^{i}_{\left(n\right)}$, ${\cal F}^{i}_{\left(n\right)}$, ${\cal G}^{i}_{\left(5\right)}$,
${\cal H}^{i}_{\left(n\right)}$, and ${\cal K}^{i\,ab}_{\left(n\right)}$, ${\cal L}^{i\,ab}_{\left(n\right)}$,
${\cal M}^{i\,ab}_{\left(n\right)}$, ${\cal N}^{i\,ab}_{\left(n\right)}$, as well as ${\cal P}^{i\,abcd}_{\left(n\right)}$,
${\cal Q}^{i\,abcd}_{\left(n\right)}$ are given by Eqs.~(E28) - (E39) and Eqs.~(E41) - (E65) as well as Eqs.~(E67) - (E87)
in \cite{Zschocke_Quadrupole_1} (note some corrections 
\footnote{
$\bullet$ In Eq.~(E67) in \cite{Zschocke_Quadrupole_1}:
\newline 
$- \frac{24}{\left(d_{\sigma}\right)^4} \sigma^a \sigma^b d_{\sigma}^c \delta^{di} \rightarrow
 - \frac{24}{\left(d_{\sigma}\right)^4} \sigma^a d_{\sigma}^b \sigma^c \delta^{di}$.
\newline 
$\bullet$ In Eq.~(E69) in \cite{Zschocke_Quadrupole_1}: 
\newline 
$+ \frac{90}{\left(d_{\sigma}\right)^2} \frac{1}{\left(x_0\right)^3}\,\sigma^a d_{\sigma}^b d_{\sigma}^c d_{\sigma}^d d_{\sigma}^i \rightarrow 
 + \frac{60}{\left(d_{\sigma}\right)^2} \frac{1}{\left(x_0\right)^3}\,\sigma^a d_{\sigma}^b d_{\sigma}^c d_{\sigma}^d d_{\sigma}^i$.
\newline  
$\bullet$ In Eq.~(E69) in \cite{Zschocke_Quadrupole_1}:
\newline 
$- \frac{60}{\left(d_{\sigma}\right)^2} \frac{1}{\left(x_0\right)^3}\,d_{\sigma}^a d_{\sigma}^b d_{\sigma}^c d_{\sigma}^d \sigma^i \rightarrow  
- \frac{30}{\left(d_{\sigma}\right)^2} \frac{1}{\left(x_0\right)^3}\,d_{\sigma}^a d_{\sigma}^b d_{\sigma}^c d_{\sigma}^d \sigma^i$.
\newline 
$\bullet$ In Eq.~(E71) in \cite{Zschocke_Quadrupole_1}:
\newline 
$+ \frac{210}{\left(d_{\sigma}\right)^2} \frac{\bm{\sigma} \cdot \bm{x}_0}{x_0}\,\sigma^a d_{\sigma}^b d_{\sigma}^c d_{\sigma}^d \sigma^i \rightarrow 
+ \frac{630}{\left(d_{\sigma}\right)^2} \frac{\bm{\sigma} \cdot \bm{x}_0}{x_0}\,\sigma^a d_{\sigma}^b d_{\sigma}^c d_{\sigma}^d \sigma^i$.}).

The scalar functions ${\cal W}_{\left(n\right)}$, ${\cal X}_{\left(n\right)}$, ${\cal Y}_{\left(n\right)}$, ${\cal Z}_{\left(n\right)}$ are defined
by Eqs.~(D20) - (D23) in \cite{Zschocke_Quadrupole_1} and can be solved in closed form as given by Eqs.~(D25) - (D28) in \cite{Zschocke_Quadrupole_1}.
Some explicit solutions for these functions are provided by Eqs.~(D29) - (D42) in \cite{Zschocke_Quadrupole_1}. In what follows, the scalar functions
${\cal W}_{\left(n\right)}$ and ${\cal X}_{\left(n\right)}$ for $n=3,5,7$ are essential and have been given again 
by Eqs.~(\ref{function_W_3}) - (\ref{function_X_7}) in Appendix~\ref{Appendix1PN}. 

Both the scalar functions as well as the tensorial coefficients in (\ref{S_I_1PN}) - (\ref{S_I_2PN}) are functions of the unperturbed light ray 
$\ve{x}_{\rm N} = \ve{x}_{\rm N}\left(t\right)$ and $\ve{x}_0 = \ve{x}_{\rm N}\left(t_0\right)$. In particular, the tensorial coefficients as well as 
the scalar functions contain the impact vector
\begin{eqnarray}
        \ve{d}_{\sigma} &=& \ve{\sigma} \times \left(\ve{x}_0 \times \ve{\sigma}\right)
\label{impact_vector_x0}
\end{eqnarray}

\noindent
and its absolute value $d_{\sigma} = \left|\ve{d}_{\sigma}\right|$ which called impact parameter $d_{\sigma}$.
The impact vector is perpendicular to the spatial direction of the
unperturbed light ray, that means $\ve{\sigma} \cdot \ve{d}_{\sigma} = 0$, and points from the origin of the coordinate system towards the unperturbed light ray
at the moment of closest approach; see also Figure~\ref{Diagram}. It is noticed that the impact vector
(\ref{impact_vector_x0}) can also be written in terms of the unperturbed light ray (cf. Eq.~(33) in \cite{Zschocke_Quadrupole_1})
\begin{eqnarray}
 \ve{d}_{\sigma} &=& \ve{\sigma} \times \left(\ve{x}_{\rm N} \times \ve{\sigma}\right)
\label{impact_vector_xN}
\end{eqnarray}

\noindent
which is a time-independent quantity as one may see by inserting (\ref{Unperturbed_Lightray_2}) into (\ref{impact_vector_xN}).

\subsection{New representation}

For the solution of the Shapiro time delay it is necessary to rewrite the 2PN solution, given by Eqs.~(\ref{S_I_N_Old}) - (\ref{S_I_2PN}), in the following form,  
\begin{eqnarray} 
        \ve{x}_{\rm N}  &=& \, \ve{x}_0 + c \left(t-t_0\right) \ve{\sigma} \;, 
	\label{S_I_N_New}
        \\
	\ve{x}_{\rm 1PN} &=& \ve{x}_{\rm N} 
        + \Delta \ve{x}_{\rm 1PN}\left(\ve{x}_{\rm N}\right) - \Delta \ve{x}_{\rm 1PN}\left(\ve{x}_0\right), 
        \label{S_I_1PN_New} 
        \\
        \ve{x}_{\rm 2PN} &=& \ve{x}_{\rm N} + \Delta \ve{x}_{\rm 1PN}\left(\ve{x}_{\rm 1PN}\right) -  \Delta \ve{x}_{\rm 1PN}\left(\ve{x}_0\right)
	\nonumber\\ 
	&& \hspace{0.55cm} + \, \Delta \ve{x}_{\rm 2PN}\left(\ve{x}_{\rm N}\right) - \Delta \ve{x}_{\rm 2PN}\left(\ve{x}_0\right), 
        \label{Second_Integration_New} 
\end{eqnarray}
 
\noindent
where the spatial components of 1PN terms are given by  
\begin{eqnarray} 
        \Delta {x}^{i}_{\rm 1PN}\left(\ve{x}\right) &=&
        \frac{G M}{c^2} \sum\limits_{n=1}^2 \left({U}^i_{\left(n\right)}\,{F}_{\left(n\right)}\right)
	\left(\ve{x}\right) 
	\nonumber\\
        && + \,\frac{G \hat{M}_{ab}}{c^2} \sum\limits_{n=1}^8 \left({V}^{i\,ab}_{\left(n\right)}\,{G}_{\left(n\right)}\right)
        \left(\ve{x}\right), 
        \label{Second_Integration_2PN_A}
\end{eqnarray}

\noindent
and the spatial components of 2PN terms are given by 
\begin{eqnarray} 
        \Delta {x}^{i}_{\rm 2PN}\left(\ve{x}\right) &=&
        \frac{G^2 M^2}{c^4} \sum\limits_{n=1}^2 \left(U^i_{\left(n\right)}\,{X}_{\left(n\right)}\right)\left(\ve{x}\right)
        \nonumber\\ 
	&& \hspace{-1.5cm} + \frac{G^2 M \hat{M}_{ab}}{c^4} \sum\limits_{n=1}^{8} \left(V^{i\,ab}_{\left(n\right)}\,{Y}_{\left(n\right)}\right)
        \left(\ve{x}\right)
        \nonumber\\ 
	&& \hspace{-1.5cm} + \frac{G^2 \hat{M}_{ab} \hat{M}_{cd}}{c^4} \sum\limits_{n=1}^{28} \left(W^{i\,abcd}_{\left(n\right)}\,{Z}_{\left(n\right)}\right)
        \left(\ve{x}\right).  
        \label{Second_Integration_2PN_B}
\end{eqnarray}
 
\noindent
The tensorial coefficients $U^i_{\left(n\right)}$, $V^{i\,ab}_{\left(n\right)}$, and $W^{i\,abcd}_{\left(n\right)}$ are given by 
Eqs.~(\ref{coefficient_U1}) - (\ref{coefficient_U2}), Eqs.~(\ref{coefficient_V1}) - (\ref{coefficient_V8}), and 
Eqs.~(\ref{coefficient_W1}) - (\ref{coefficient_W28}) in Appendix~\ref{Tensorial_Coefficients}.  
The scalar functions $F_{\left(n\right)}$ and $G_{\left(n\right)}$ are given by Eqs.~(\ref{F_1}) - (\ref{F_2}) and Eqs.~(\ref{G_1}) - (\ref{G_8}) 
in Appendix~\ref{Scalar_Functions_2}. 
The scalar functions $X_{\left(n\right)}$, $Y_{\left(n\right)}$, and $Z_{\left(n\right)}$ are given by 
Eqs.~(\ref{X_1}) - (\ref{X_2}), Eqs.~(\ref{Y_1}) - (\ref{Y_8}), and Eqs.~(\ref{Z_1}) - (\ref{Z_28}) in Appendix~\ref{Scalar_Functions_2}. 

The difference between the old representation in (\ref{Second_Integration_Old}) and the new representation in (\ref{Second_Integration_New}) 
is the argument of $\Delta \ve{x}_{\rm 1PN}$. In the old representation in (\ref{Second_Integration_Old}) the argument of this term is 
the light trajectory in Newtonian approximation, $\ve{x}_{\rm N}$, while in the new representation in (\ref{Second_Integration_New}) the argument 
of this term is the light trajectory in 1PN approximation, $\ve{x}_{\rm 1PN}$. 
But it is emphasized that the new representation (\ref{S_I_N_New}) - (\ref{Second_Integration_2PN_B}) agrees 
with the old representation (\ref{S_I_N_Old}) - (\ref{S_I_2PN}) up to terms beyond the 2PN approximation. The basic ideas of how to demonstrate the agreement of 
the old representation and the new representation are given in Appendix \ref{Agreement_Second_Integration}. 

The terms proportional to $M$ in (\ref{Second_Integration_2PN_A}) agree with Eq.~(50) in \cite{Article_Zschocke1}, and the terms proportional to $M \times M$ 
in (\ref{Second_Integration_2PN_B}) agree with Eq. (51) in \cite{Article_Zschocke1}. The terms proportional to $M \times \hat{M}_{ab}$ and 
$\hat{M}_{ab} \times \hat{M}_{cd}$ in  (\ref{Second_Integration_2PN_B}) are the new quadrupole terms of the second post-Newtonian approximation. 
In the following we will investigate the influence of these 2PN quadrupole terms within the boundary value problem and in particular their impact on the 
Shapiro time delay.

\section{The Shapiro time delay in 2PN approximation}\label{Time_Delay}

\subsection{The boundary value problem}

The initial value problem has been defined by Eqs.~(\ref{Initial_A}) and (\ref{Initial_B}). The solution of the initial value problem for the 
propagation of a light signal in the monopole and quadrupole field of one body at rest in 2PN approximation has been presented in the previous 
Section. In order to determine the Shapiro time delay one needs the solution of the boundary-value problem, where a unique solution of geodesic equation 
is defined by the space-time point $\left(t_0,\ve{x}_0\right)$ of the light source and by the space-time point $\left(t_1,\ve{x}_1\right)$ of the 
observer \cite{Book_Clifford_Will,Kopeikin_Efroimsky_Kaplan}: 
\begin{eqnarray}
        \ve{x}_0 &=& \ve{x}\left(t\right)\,\,\bigg|_{t = t_0}  \;,
        \label{boundary_0}
        \\
        \ve{x}_1 &=& \ve{x}\left(t\right)\,\,\bigg|_{t = t_1}  \;. 
        \label{boundary_1}
\end{eqnarray}

\noindent
The spatial position of the observer $\left(t_1,\ve{x}_1\right)$ is assumed to be known, while the spatial position of the light source 
$\left(t_0,\ve{x}_0\right)$ has to be determined by a unique interpretation of astronomical observations which is the primary aim of 
astrometric data reduction \cite{Brumberg1991,KlionerKopeikin1992,Kopeikin_Efroimsky_Kaplan,Klioner2003b,Book_Clifford_Will}. 

The solution of the boundary value problem (\ref{boundary_0}) and (\ref{boundary_1}), 
that means a solution of the geodesic equation in terms of the spatial position of source and observer, $\ve{x}_0$ and $\ve{x}_1$, can be obtained from 
the new representation of the initial-boundary solution as given by Eq.~(\ref{Second_Integration_New}) in the following way. 
The spatial coordinates of the unperturbed light ray at the time of observation coincides with the spatial coordinates of the observer up to terms of the order 
${\cal O}\left(c^{-2}\right)$, 
\begin{eqnarray}
        \ve{x}_1 &=& \ve{x}_{\rm N}\left(t_1\right) + {\cal O}\left(c^{-2}\right). 
        \label{Replacement_1}
\end{eqnarray}

\noindent
Therefore, a replacement of $\ve{x}_{\rm N}\left(t_1\right)$ by $\ve{x}_1$ in the expression $\Delta\ve{x}_{\rm 2PN}\left(\ve{x}_{\rm N}\right)$ in 
(\ref{Second_Integration_New}) causes an error of the order ${\cal O}\left(c^{-6}\right)$ which would be in line with the 2PN approximation. 
Furthermore, the spatial coordinates of the light ray in 1PN approximation at the time of observation coincides 
with the spatial coordinates of the observer up to terms of the order ${\cal O}\left(c^{-4}\right)$, 
\begin{eqnarray}
        \ve{x}_1 &=& \ve{x}_{\rm 1PN}\left(t_1\right) + {\cal O}\left(c^{-4}\right). 
        \label{Replacement_2}
\end{eqnarray}

\noindent 
Therefore, a replacement of $\ve{x}_{\rm 1PN}\left(t_1\right)$ by $\ve{x}_1$ in the expression $\Delta\ve{x}_{\rm 1PN}\left(\ve{x}_{\rm 1PN}\right)$ 
in (\ref{Second_Integration_New}) causes also an error of the order ${\cal O}\left(c^{-6}\right)$ which would be in line with the 2PN approximation. 
Finally, the spatial coordinates of the light ray in 2PN approximation at the time of observation coincides
with the spatial coordinates of the observer up to terms of the order ${\cal O}\left(c^{-6}\right)$,
\begin{eqnarray}
        \ve{x}_1 &=& \ve{x}_{\rm 2PN}\left(t_1\right) + {\cal O}\left(c^{-6}\right). 
        \label{Replacement_3}
\end{eqnarray}

\noindent 
Therefore, a replacement of $\ve{x}_{\rm 2PN}\left(t_1\right)$ by $\ve{x}_1$ in the l.h.s. of equation (\ref{Second_Integration_New}) causes an error
of the order ${\cal O}\left(c^{-6}\right)$ which would be in line with the 2PN approximation. 

The sequence of replacements (\ref{Replacement_1}) - (\ref{Replacement_3}) in (\ref{Second_Integration_New}) leads to the following expression
which is valid in 2PN approximation, that means valid up to terms of the order ${\cal O}\left(c^{-6}\right)$: 
\begin{eqnarray}  
        c \left(t_1 - t_0\right) \ve{\sigma} &=& R\,\ve{k}
        - \Delta \ve{x}_{\rm 1PN}\left(\ve{x}_1,\ve{x}_0\right) 
        - \Delta \ve{x}_{\rm 2PN}\left(\ve{x}_1,\ve{x}_0\right),
	\nonumber\\ 
        \label{Second_Integration_Replacement} 
\end{eqnarray}

\noindent
with $R = \left|\ve{x}_1 - \ve{x}_0\right|$ and where
\begin{eqnarray} 
        \Delta \ve{x}_{\rm 1PN}\left(\ve{x}_1, \ve{x}_0\right) &=& \Delta \ve{x}_{\rm 1PN}\left(\ve{x}_1\right) - \Delta \ve{x}_{\rm 1PN}\left(\ve{x}_0\right),
        \label{Delta_1PN}
        \\
\Delta \ve{x}_{\rm 2PN}\left(\ve{x}_1, \ve{x}_0\right) &=& \Delta \ve{x}_{\rm 2PN}\left(\ve{x}_1\right) - \Delta \ve{x}_{\rm 2PN}\left(\ve{x}_0\right),
        \label{Delta_2PN}
\end{eqnarray}

\noindent
with $\Delta \ve{x}_{\rm 1PN}\left(\ve{x}\right)$ and $\Delta \ve{x}_{\rm 2PN}\left(\ve{x}\right)$ given by (\ref{Second_Integration_2PN_A}) and
(\ref{Second_Integration_2PN_B}). It is emphasized that such a replacement would not be possible in the old representation
(\ref{Second_Integration_Old}) because there the corrections $\Delta\ve{x}_{\rm 1PN}$
are given in terms of the unperturbed light ray, but a replacement according to (\ref{Replacement_1}) would cause an error of the order
${\cal O}\left(c^{-4}\right)$ in these terms which would spoil the 2PN approximation.

\subsection{The transformation $\ve{\sigma}$ to $\ve{k}$}

In the boundary value problem the unit-vector $\ve{k}$, pointing from light source towards observer, is of fundamental importance:
\begin{eqnarray}
\ve{k} &=& \frac{\ve{x}_1 - \ve{x}_0}{ \left|\ve{x}_1 - \ve{x}_0\right|}\;.
\label{Tangent_Vector1}
\end{eqnarray}

\noindent
In order to get the expression for the time-delay, one needs the transformation from $\ve{\sigma}$ to $\ve{k}$. In Newtonian approximation
we have
\begin{eqnarray}
        \ve{\sigma} = \ve{k} + {\cal O}\left(c^{-2}\right). 
\label{k_sigma_N}
\end{eqnarray}

\noindent
In 1PN approximation one obtains from (\ref{Second_Integration_Replacement})
\begin{eqnarray}
        \ve{\sigma} &=& \ve{k} - \frac{1}{R}\left[\ve{k}\times \bigg(\Delta\ve{x}_{\rm 1PN}\left(\ve{x}_1,\ve{x}_0\right)\times\ve{k}\bigg)\right] 
        + {\cal O}\left(c^{-4}\right). 
\nonumber\\ 
	\label{k_sigma_1PN}
\end{eqnarray}

\noindent
For later purposes it is noticed here that (\ref{k_sigma_1PN}) implies 
\begin{eqnarray}
	\ve{\sigma} \cdot \ve{k} &=& 1 + {\cal O}\left(c^{-4}\right).
\label{scalar_product_sigma_k}
\end{eqnarray}

\noindent 
Because the three-vector $\ve{\sigma}$ appears in the Newtonian terms in (\ref{Second_Integration_Replacement}), one also needs the transformation
$\ve{\sigma}$ to $\ve{k}$ in 2PN approximation. By iteration, using (\ref{k_sigma_1PN}), one obtains from (\ref{Second_Integration_Replacement})
\begin{eqnarray}
	\ve{\sigma} = \ve{k} && - \frac{1}{R}\left[\ve{k}\times \bigg(\Delta\ve{x}_{\rm 1PN}\left(\ve{x}_1,\ve{x}_0\right)\times\ve{k}\bigg)\right]  
\nonumber\\ 
	&& - \frac{1}{R}\left[\ve{k}\times \bigg(\Delta\ve{x}_{\rm 2PN}\left(\ve{x}_1,\ve{x}_0\right)\times\ve{k}\bigg)\right]  
\nonumber\\ 
&& + \frac{1}{R^2}\,\bigg[\Delta\ve{x}_{\rm 1PN}\left(\ve{x}_1,\ve{x}_0\right) \times 
        \bigg(\ve{k}\times\Delta\ve{x}_{\rm 1PN}\left(\ve{x}_1,\ve{x}_0\right)\bigg)\bigg] 
\nonumber\\ 
	&& - \frac{3}{2}\,\frac{1}{R^2}\,\ve{k}\,\bigg| \ve{k} \times \Delta\ve{x}_{\rm 1PN}\left(\ve{x}_1,\ve{x}_0\right)\bigg|^2  
+ {\cal O}\left(c^{-6}\right)  
\label{k_sigma_2PN}
\end{eqnarray}

\noindent
which generalizes Eq.~(68) in \cite{Article_Zschocke1} which was valid in the field of one monopole at rest.

\subsection{The Shapiro time delay}

Using the expressions for the transformation $\ve{\sigma}$ to $\ve{k}$ in Eqs.~(\ref{k_sigma_N}) - (\ref{k_sigma_2PN}), 
one obtains from (\ref{Second_Integration_Replacement}) the travel time of a light signal in the field of one body
at rest where its monopole and quadrupole structure is taken into account,
\begin{eqnarray}
c \left(t_1 - t_0 \right) &=& R - \ve{k} \cdot \Delta\ve{x}_{{\rm 1PN}}\left(\ve{x}_1,\ve{x}_0\right)
	- \ve{k} \cdot \Delta \ve{x}_{{\rm 2PN}}\left(\ve{x}_1,\ve{x}_0\right)
\nonumber\\ 
	&& +\, \frac{1}{2\,R} \left| \ve{k} \times \Delta \ve{x}_{{\rm 1PN}}\left(\ve{x}_1,\ve{x}_0\right)\right|^2 
	+ {\cal O}\left(c^{-6}\right),
\label{Shapiro_2PN}
\end{eqnarray}

\noindent
which generalizes Eq.~(67) in \cite{Article_Zschocke1} which was valid 
in the field of one monopole at rest. However, formula (\ref{Shapiro_2PN}) is still implicit, because $\Delta \ve{x}_{{\rm 1PN}}$ and 
$\Delta \ve{x}_{{\rm 2PN}}$ are given in terms of $\ve{\sigma}$. Clearly, the last two terms in (\ref{Shapiro_2PN}) are 2PN terms which are 
of the order ${\cal O}\left(c^{-4}\right)$, hence one may immediately replace the vector $\ve{\sigma}$ by the vector $\ve{k}$. But the term
$\ve{k} \cdot \Delta\ve{x}_{\rm 1PN}$ in (\ref{Shapiro_2PN}) is a 1PN term, hence one has to use the transformation $\ve{\sigma}$ to $\ve{k}$ in 1PN approximation
(\ref{k_sigma_1PN}) in order to achieve a formula for $\Delta \ve{x}_{{\rm 1PN}}$ in terms of vector $\ve{k}$ rather than $\ve{\sigma}$. Only in this way one
arrives at a formula for the time delay in 2PN approximation fully in terms of vector $\ve{k}$, which is the central topic of this Section.

The term $\ve{k} \cdot \Delta\ve{x}_{\rm 2PN}$ is calculated in Appendix~\ref{Term_Shapiro_C} and given by Eq.~(\ref{Appendix_Term_Shapiro_C_Final_Form}). 
The term $\ve{k} \cdot \Delta\ve{x}_{\rm 1PN}$ is calculated in Appendix~\ref{Term_Shapiro_A} and given by Eq.~(\ref{Appendix_Term_Shapiro_A_Final_Form}). The term 
$\left|\ve{k} \times \Delta\ve{x}_{\rm 1PN}\right|^2$ is calculated in Appendix~\ref{Term_Shapiro_B} and given by Eq.~(\ref{Appendix_Term_Shapiro_B_Final_Form}). 
According to these results, the light-travel-time in 2PN approximation in the gravitational field of one body at rest with monopole and quadrupole structure is 
given as follows: 
\begin{eqnarray}
c \left(t_1 - t_0 \right) &=& R + \Delta c\tau_{\rm 1PN}^{M} + \Delta c\tau_{\rm 1PN}^{M_{ab}} 
	\nonumber\\ 
	&& \hspace{-1.5cm} + \,\Delta c\tau_{\rm 2PN}^{M \times M} + \Delta c\tau_{\rm 2PN}^{M \times M_{ab}} + \Delta c\tau_{\rm 2PN}^{M_{ab} \times M_{cd}}
    + {\cal O}\left(c^{-6}\right), 
	\nonumber\\ 
\label{Shapiro_2PN_Final}
\end{eqnarray}

\noindent
where the individual terms are given by the following expressions: 
\begin{eqnarray}
    \Delta c\tau_{\rm 1PN}^{M} &=& - \frac{G\,M}{c^2}\,P_{\left(1\right)}\left(\ve{x}_1,\ve{x}_0\right) , 
\label{Shapiro_2PN_Final_M}
    \\
    \Delta c\tau_{\rm 1PN}^{M_{ab}} &=& 
    - \frac{G\,\hat{M}_{ab}}{c^2} \sum\limits_{n=1}^{3} S^{ab}_{\left(n\right)}\,Q_{\left(n\right)}\left(\ve{x}_1,\ve{x}_0\right),
\label{Shapiro_2PN_Final_Mab}
    \\
    \Delta c\tau_{\rm 2PN}^{M \times M} &=& + \frac{G^2 M^2}{c^4}\, R_{\left(1\right)}\left(\ve{x}_1,\ve{x}_0\right), 
\label{Shapiro_2PN_Final_M_M}
    \\
    \Delta c\tau_{\rm 2PN}^{M \times M_{ab}} &=& 
    + \frac{G^2 M\,\hat{M}_{ab}}{c^4} \sum\limits_{n=1}^{3} S^{ab}_{\left(n\right)} \, S_{\left(n\right)}\left(\ve{x}_1,\ve{x}_0\right), 
\label{Shapiro_2PN_Final_M_Mab}
    \\
    \Delta c\tau_{\rm 2PN}^{M_{ab} \times M_{cd}} &=& 
    + \frac{G^2 \hat{M}_{ab} \hat{M}_{cd}}{c^4} \sum\limits_{n=1}^{10} T^{abcd}_{\left(n\right)} \, T_{\left(n\right)}\left(\ve{x}_1,\ve{x}_0\right).
\nonumber\\ 
\label{Shapiro_2PN_Final_Mab_Mcd}
\end{eqnarray}

\noindent 
The tensors $S^{ab}_{\left(n\right)}$ and $T_{\left(n\right)}^{abcd}$ are defined by Eqs.~(\ref{Tensor_S}) and (\ref{Tensor_T}). The scalar functions  
$P_{\left(1\right)}$ and $Q_{\left(n\right)}$ for the 1PN terms are given by Eqs.~(\ref{P_1}) and Eqs.~(\ref{Q_1}) - (\ref{Q_3}), 
while the scalar functions $R_{\left(1\right)}$, $S_{\left(n\right)}$, and $T_{\left(n\right)}$ for the 2PN terms are given by 
Eqs.~(\ref{Shapiro_2PN_Final_Function_R}), (\ref{Shapiro_2PN_Final_Function_S}), and (\ref{Shapiro_2PN_Final_Function_T}). 
 
In order to determine the 2PN effect of the time-delay, higher mass-multipoles beyond the mass-quadrupole as well as spin-multipoles have been neglected, 
as indicated by Eqs.~(\ref{M_L_2PN}) and (\ref{S_L_2PN}). These higher mass-multipoles and spin-multipoles can be taken into account just by adding the other 
1PN mass-multipole terms in (\ref{Shapiro_Mass_Multipole}) (beyond mass-quadrupole) as well as the 1.5PN spin-multipole terms in (\ref{Shapiro_Spin_Multipole}) 
to (\ref{Shapiro_2PN_Final}) in an appropriate manner; cf. text below Eqs.~(\ref{M_L_2PN}) and (\ref{S_L_2PN}) as well as in the introductory Section.  
That means, one has to keep in 
mind that (\ref{Shapiro_2PN_Final}) is given in terms of three-vector $\ve{k}$, while (\ref{Shapiro_Mass_Multipole}) and (\ref{Shapiro_Spin_Multipole}) 
are given in terms of three-vector $\ve{\sigma}$. Therefore, in order to do that consistently, one has to replace the three-vector $\ve{\sigma}$ 
in (\ref{Transformation_Derivative_3}) as well as in (\ref{Shapiro_Mass_Multipole}) and (\ref{Shapiro_Spin_Multipole}) by the three-vector $\ve{k}$. 
In view of relations (\ref{k_sigma_N}) and (\ref{scalar_product_sigma_k}) such a replacement is correct up to higher 2PN multipole terms beyond the mass-quadrupole.

\subsection{The upper limits of 2PN terms in the Shapiro time-delay}

The upper limits for 1PN mass-monopole and mass-quadrupole time delay were given by Eqs.~(\ref{Shapiro_M_1PN}) and (\ref{Shapiro_ML_1PN}), while the upper 
limits for 2PN mass-monopole and mass-quadrupole terms were derived by Eqs.~(\ref{Estimation_Shapiro_2PN_M_M_15}), (\ref{Estimation_Shapiro_2PN_M_Q_10}) and 
(\ref{Estimation_Shapiro_2PN_Q_Q_10}). They read 
\begin{eqnarray}
	\left|\Delta \tau_{\rm 1PN}^{M}\right| &\le& 2\,\frac{GM}{c^3}\,\ln \frac{4 x_0 x_1}{\left(d_k\right)^2}\;, 
\label{Upper_limit_1PN_Final_M}
    \\
        \left|\Delta \tau_{\rm 1PN}^{M_{ab}}\right| &\le& \frac{11}{5}\,\frac{GM}{c^3}\,\left|J_2\right| \left(\frac{P}{d_k}\right)^2\;,
\label{Upper_limit_1PN_Final_Q}
    \\
	\left|\Delta \tau_{\rm 2PN}^{M \times M}\right| &\le& 
	8\,\frac{G^2 M^2}{c^5}\,\frac{x_1}{\left(d_{\sigma}\right)^2} \left(\frac{P}{d_k}\right)^2\;,
\label{Upper_limit_2PN_Final_MM}
    \\
        \left|\Delta \tau_{\rm 2PN}^{M \times M_{ab}}\right| &\le& 
	12\,\frac{G^2 M^2}{c^5}\,\frac{x_1}{\left(d_{\sigma}\right)^2}\,\left|J_2\right| \left(\frac{P}{d_k}\right)^2\;,
\label{Upper_limit_2PN_Final_MQ}
    \\
        \left|\Delta \tau_{\rm 2PN}^{M_{ab} \times M_{cd}}\right| &\le& 8\,\frac{G^2 M^2}{c^5}\,\frac{x_1}{\left(d_{\sigma}\right)^2}\,\left|J_2\right|^2 
	\left(\frac{P}{d_k}\right)^2\;. 
\label{Upper_limit_2PN_Final_QQ}
\end{eqnarray}

\noindent 
The upper limits of the 1PN mass-monopole term (\ref{Upper_limit_1PN_Final_M}) and 1PN mass-quadrupole term (\ref{Upper_limit_1PN_Final_Q}) were already given by 
Eqs.~(\ref{Shapiro_M_1PN}) and (\ref{Shapiro_ML_1PN}) (with coefficient $A_2$ in (\ref{Coefficients_ML})), while their numerical values have been presented in 
Table~\ref{Table2} for grazing light rays at Sun, Jupiter, and Saturn. 
\begin{table}[h!]
        \caption{\label{Table3} 
	The effect of 2PN terms on the (one-way) Shapiro time delay $\Delta \tau$ in the gravitational field of the Sun and giant planets of the solar system 
	according to the upper limits presented by Eqs.~(\ref{Upper_limit_2PN_Final_MM}), (\ref{Upper_limit_2PN_Final_MQ}) and (\ref{Upper_limit_2PN_Final_QQ}).
        The values are given for grazing rays (impact parameter $d_k$ equals body's equatorial radius $P$).
        The time delay is given in units of pico-seconds: ${\rm ps} = 10^{-12}\,{\rm sec}$. 
        The presented numerical values should be compared with the goal accuracy of $0.001$ pico-seconds in time-delay measurements.
        A blank entry means a delay of less than a femto-second.}
\footnotesize
\begin{tabular}{@{}|c|c|c|c|}
\hline
&&&\\[-7pt]
        Object & $\Delta \tau^{M \times M}_{\rm 2PN}$ & $\Delta \tau^{M \times M_{ab}}_{\rm 2PN}$ & $\Delta \tau^{M_{ab} \times M_{cd}}_{\rm 2PN}$ \\
&&&\\[-7pt]
\hline
        Sun & $1.8 \times 10^4$ & $0.004$ & $ - $ \\
        Jupiter & $6.1$ & $0.14$ & $0.001$ \\
        Saturn & $1.6$ & $0.04$ & $ - $ \\
\hline
\end{tabular}\\
\end{table}
\normalsize

\noindent
The numerical values for the 2PN terms (\ref{Upper_limit_2PN_Final_MM}) - (\ref{Upper_limit_2PN_Final_QQ}) are presented in Table~\ref{Table3} for grazing light 
rays at Sun, Jupiter, and Saturn. It is remarkable that the numerical value of the 2PN monopole-quadrupole term (\ref{Upper_limit_2PN_Final_MQ}) for Jupiter and 
Saturn is of similar magnitude than the 1PN spin-dipole term (\ref{Shapiro_S_15PN}) for Jupiter and Saturn. 
Similarly, the numerical value of the 2PN quadrupole-quadrupole term for Jupiter and Saturn (\ref{Upper_limit_2PN_Final_QQ}) is of similar magnitude than 
the 1PN spin-octupole term (\ref{Shapiro_SL_1PN}) (with $B_3 = 7/6$) for Jupiter and Saturn. 

Finally, by comparing the 2PN values presented in Table~\ref{Table3} with the 1PN values given in Table~I in \cite{Klioner1991}, one finds that the 
2PN monopole-quadrupole effects for Jupiter and Saturn are larger than the 1PN quadrupole effects for Earth-like planets of the solar system.

\section{Summary}\label{Summary}  

The Shapiro time delay is the difference between the travel time of a light-signal in the gravitational field of a body and the Euclidean 
distance between source and observer divided by the speed of light, which belongs to the four classical tests of general relativity. For 
a spherically symmetric body with mass $M$, the Shapiro time delay in the 1PN approximation is given by  
\begin{eqnarray}
        \Delta \tau^{M}_{\rm 1PN} &=& \frac{2 G M}{c^3} \ln \frac{x_1 + \ve{k} \cdot \ve{x}_1}{x_0 + \ve{k} \cdot \ve{x}_0} \;. 
        \label{Summary_5} 
\end{eqnarray}

\noindent 
The first measurements of this effect (\ref{Summary_5}) have been performed by radar signals, which were emitted from Earth and which were reflected 
either by the inner planets or by spacecrafts. Since the early days of time-delay measurements in the solar system, the accuracies 
have been improved from a few micro-seconds in $1968$ and $1971$ by radar echos from Mercury and Venus \cite{Shapiro2,Shapiro3} towards a few nano-seconds 
in $2003$ by radar echos from the Cassini spacecraft which orbits Saturn \cite{Shapiro6}. 

Future time-delay measurements in the solar system aim at the pico-second and sub-pico-second level of accuracy, which will be performed by 
optical laser rather than radar signals, as suggested by a series of several ESA mission proposals \cite{Astrod1,Lator1,Odyssey,Sagas,TIPO,EGE}. 
These advancements make it necessary to improve the theoretical models of time delay measurements up to an accuracy of $0.001$ pico-seconds. 
On this level of precision the Shapiro time delay in 1PN monopole approximation (\ref{Summary_5}) is by far not sufficient. 
It is necessary to take into account higher mass-multipoles $\hat{M}_L$ (describe shape and inner structure of the massive body) and 
some spin-multipoles $\hat{S}_L$ (describe rotational motions and inner currents of the massive body) in the post-Newtonian (1PN and 1.5PN) approximation,  
\begin{eqnarray}
        \Delta \tau &=& \sum\limits_{l=0}^{\infty} \Delta \tau_{\rm 1PN}^{M_L} + \sum\limits_{l=1}^{\infty} \Delta \tau_{\rm 1.5PN}^{S_L}\;.  
        \label{Summary_10} 
\end{eqnarray}

\noindent 
The mathematical expressions for the 1PN and 1.5PN terms in the Shapiro time delay were derived a long time ago \cite{Kopeikin1997}. In this investigation 
we have quantified these terms and have clarified that only the first eight mass-multipoles and the spin-dipole and the spin-hexapole terms (for Jupiter) are required 
in order to achieve an assumed accuracy of about $0.001$ pico-seconds. The numerical values for the 1PN mass-multipoles and 1.5PN spin-dipole term were presented 
in Table~\ref{Table2}. It has been shown that higher mass-multipoles $l \ge 10$ as well as spin-multipoles $l \ge 5$ are not relevant for an accuracy of about 
$0.001$ pico-seconds in time delay measurements in the solar system. 

It is clear that on the sub-pico-second level of accuracy in time-delay measurements some 2PN effects need to be taken into account. Thus far, however, 
the knowledge about 2PN effects in the Shapiro time delay was restricted to the case of spherically symmetric bodies. The next term in the multipole decomposition 
is the mass-quadrupole term. In this investigation we have taken into account the monopole and quadrupole structure of a massive body at rest and have 
determined the 2PN quadrupole effects on time delay for a light signal,   
\begin{eqnarray} 
	\Delta \tau &=& \Delta \tau_{\rm 1PN}^{M} + \Delta \tau_{\rm 1PN}^{M_{ab}} 
	\nonumber\\ 
	&& + \,\Delta \tau_{\rm 2PN}^{M \times M} + \Delta \tau_{\rm 2PN}^{M \times M_{ab}} + \Delta \tau_{\rm 2PN}^{M_{ab} \times M_{cd}}\;. 
        \label{Summary_20}
\end{eqnarray}

\noindent
The explicit expression of the 1PN terms in (\ref{Summary_20}) were presented by Eqs.~(\ref{Shapiro_2PN_Final_M}) and (\ref{Shapiro_2PN_Final_Mab}) and 
the 2PN terms in (\ref{Summary_20}) were presented by Eqs.~(\ref{Shapiro_2PN_Final_M_M}) - (\ref{Shapiro_2PN_Final_Mab_Mcd}). 
The 2PN quadrupole effect amounts up to $0.004$, $0.14$, and $0.04$ pico-second for grazing light rays at the Sun, Jupiter, and Saturn, respectively; 
see Table~\ref{Table3}. The values of the 2PN terms are tiny but, nevertheless, they are comparable with the 1PN and 1.5PN terms of some 
higher mass-multipoles and spin-dipoles on time-delay; see Table~\ref{Table2}.  

In the expression for the time delay in 2PN approximation (\ref{Summary_20}) higher multipoles beyond the quadrupole are not taken into account. It is, however, 
not certain whether such higher multipole terms can be neglected in 2PN approximation on the level of $0.001$ pico-second in the accuracy of time delay 
measurements. Namely, the next 2PN term beyond the monopole-quadrupole term, $M \times M_{ab}$, which is proportional to the second zonal harmonic 
coefficient $J_2$, would be the monopole-octupole term, $M \times M_{abcd}$, which is proportional to the fourth zonal harmonic coefficient $J_4$. 
Taking the ratio $J_4/J_2$ and multiplying with the 2PN monopole-quadrupole effect one obtains about 
$0.02$, $0.006$, and $0.002$ pico-second time delay for grazing rays at Sun, Jupiter, and Saturn. These rough estimates show that the  
monopole-octupole term might be relevant for time delay measurements on the level of $0.001$ pico-second. On the other side, these 2PN monopole-octupole terms 
scale with $\displaystyle \left(P/d_k\right)^4$ where $P$ is the equatorial radius of the massive body and $d_k$ is the impact parameter of the light ray. 
Thus, these 2PN effects decrease very rapidly with increasing distance from the massive body. 

Finally, it is also mentioned that the impact of the mass-monopole on a time delay has been determined in the 3PN approximation  
for the case of one body at rest \cite{Linet_Teyssandier}, where it has been shown that on the pico-second level such 3PN effects are relevant, but only 
in case of grazing light ray at the Sun, that means light signals which pass near the limb of the Sun.

\section*{Acknowledgments}

This work was funded by the German Research Foundation (Deutsche Forschungsgemeinschaft DFG) under Grant No. $447922800$. Sincere gratitude is expressed 
to Prof. Sergei A. Klioner for kind encouragement and enduring support. Prof. Michael H. Soffel, Prof. Ralf Sch\"utzhold, Dr. Alexey G. Butkevich, 
Dipl-Inf. Robin Geyer, Priv.-Doz. G\"unter Plunien, Prof. Burkhard K\"ampfer, and Prof. Laszlo P. Csernai, 
are greatly ack\-nowledged for inspiring discussions about astrometry and general theory of relativity.

\appendix

\section{Notation}\label{Appendix_Notation} 

Throughout the investigation the same notation as in Ref.~\cite{Zschocke_Quadrupole_1} is in use:  
\begin{itemize}
\item Lower case Latin indices $i$, $j$, \dots take values $1,2,3$.
\item $\dot{f}$ denotes total time derivative of $f$.
\item $f_{\,,\,i} = \partial f / \partial x^{i}$ denotes partial derivative of $f$ with respect to $x^i$.
\item Kronecker delta: $\delta^i_j\!=\!\delta_{ij}\!=\!\delta^{ij}\!=\!{\rm diag} \left(+1,+1,+1\right)$.
\item $n! = n \left(n-1\right)\left(n-2\right)\cdot\cdot\cdot 2 \cdot 1$ is the factorial for positive integer $\left(0! = 1\right)$.
\item $n!! = n \left(n-2\right) \left(n-4\right)\cdot\cdot\cdot \left(2\;{\rm or}\;1\right)$ is the double factorial for positive integer $\left(0!! = 1\right)$.
\item $\varepsilon_{ijk} = \varepsilon^{ijk}$ with $\varepsilon_{123} = + 1$ is the fully anti-symmetric Levi-Civita symbol.
\item Triplet of three-vectors are in boldface, e.g. $\ve{a}$, $\ve{b}$, $\ve{\sigma}$, $\ve{x}$.
\item Contravariant components of three-vectors: $a^{i} = \left(a^{\,1},a^2,a^3\right)$.
\item Absolute value of a three-vector: $a = |\ve{a}| = \sqrt{a^{\,1}\,a^{\,1}+a^2\,a^2+a^3\,a^3}$.
\item Scalar product of three-vectors: $\ve{a}\,\cdot\,\ve{b}=\delta_{ij}\,a^i\,b^j$.
\item Vector product of two three-vectors: $\left(\ve{a}\times\ve{b}\right)^i=\varepsilon_{ijk}\,a^j\,b^k$.
\item Angle between three-vectors $\ve{a}$ and $\ve{b}$ is denoted by $\delta\left(\ve{a},\ve{b}\right)$.
\item Lowercase Greek indices take values 0,1,2,3.
\item $f_{\,,\,\mu} = \partial f / \partial x^{\mu}$ denotes partial derivative of $f$ with respect to $x^{\mu}$.
\item $\eta_{\alpha\beta} = \eta^{\alpha \beta} = {\rm diag}\left(-1,+1,+1,+1\right)$ is the metric tensor of flat space-time.
\item $g_{\alpha\beta}$ and $g^{\alpha\beta}$ are the covariant and contravariant components of the metric tensor.
\item Contravariant components of four-vectors: $a^{\mu} = \left(a^{\,0},a^{\,1},a^2,a^3\right)$.
\item Repeated indices are implicitly summed over ({\it Einstein's} sum convention).
\end{itemize}

\section{Mass and Spin multipoles}\label{Appendix_Multipoles}

\subsection{STF tensors} 

Here we will present only those few standard notations about symmetric tracefree (STF) tensors, 
which are really necessary for our considerations, while further STF relations can be found 
in \cite{Thorne,Blanchet_Damour1,Multipole_Damour_2,Hartmann_Soffel}. 
\begin{itemize}
\item $L=i_1 i_2 ...i_l$ is a Cartesian multi-index of a given tensor $T$, that means
$T_L \equiv T_{i_1 i_2 \,.\,.\,.\,i_l}$.
\item two identical multi-indices imply summation:
$A_L\,B_L \equiv \sum\limits_{i_1\,.\,.\,.\,i_l}\,A_{i_1\,.\,.\,.\,i_l}\,B_{i_1\,.\,.\,.\,i_l}$.
\item The symmetric part of a Cartesian tensor $T_L$ is (cf. Eq.~(2.1) in \cite{Thorne}):
\begin{eqnarray} 
T_{\left(L\right)} &=& T_{\left(i_1 ... i_l \right)} = \frac{1}{l!} \sum\limits_{\sigma} 
A_{i_{\sigma\left(1\right)} ... i_{\sigma\left(l\right)}}\,,
\end{eqnarray}

\noindent
where $\sigma$ is running over all permutations of $\left(1,2,...,l\right)$.
\item The symmetric tracefree part of a Cartesian tensor $T_L$ (notation:
	$\hat{T}_L \equiv {\rm STF}_L\,T_L \equiv T_{<i_1 \dots i_l>}$) is (cf. Eq.~(2.2) in \cite{Thorne}):
\begin{eqnarray}
	\hspace{1.0cm} \hat{T}_L = \sum_{k=0}^{\left[l/2\right]} a_{l k}\,\delta_{(i_1 i_2 ...} \delta_{i_{2k-1} i_{2k}}\,
S_{i_{2k+1 ... i_l) \,a_1 a_1 ... a_k a_k}}\,,
\nonumber\\ 
	\label{anti_symmetric_1}
\end{eqnarray}

\noindent
where $\left[l/2\right]$ means the largest integer less than or equal to $l/2$, and $S_L \equiv T_{\left(L\right)}$
abbreviates the symmetric part of tensor $T_L$. The coefficient in (\ref{anti_symmetric_1}) is given by
\begin{eqnarray}
a_{l k} &=& \left(-1\right)^k \frac{l!}{\left(l - 2 k\right)!}\,
\frac{\left(2 l - 2 k - 1\right)!!}{\left(2 l - 1\right)!! \left(2k\right)!!}\,.
\label{coefficient_anti_symmetric}
\end{eqnarray}
\end{itemize}

\noindent
Three comments are in order about STF. First of all, the Kronecker delta has no symmetric tracefree part, 
\begin{eqnarray}
	{\rm STF}_{ab} \,\delta^{ab} &=& 0\;. 
	\label{STF_comment_1}
\end{eqnarray}

\noindent
Second, the symmetric tracefree part of any tensor which contains Kronecker delta is zero, if the Kronecker delta 
has not any summation (dummy) index, for instance, 
\begin{eqnarray}
	{\rm STF}_{abc} \,\delta^{ab}\,d_{\sigma}^c &=& 0\;, 
	\label{STF_comment_2a}
	\\
	{\rm STF}_{abc} \,\delta^{ab}\,\sigma^c &=& 0\;. 
	\label{STF_comment_2b}
\end{eqnarray}
 
\noindent
And third, the following relation is very useful (cf. Eq.~(A1) in \cite{Hartmann_Soffel}), 
\begin{eqnarray}
	\hat{A}_{L}\,\hat{B}_{L} &=& A_{L}\,\hat{B}_{L} = \hat{A}_{L}\,B_{L} 
	\label{STF_comment_3}
\end{eqnarray}

\noindent
which often simplifies the analytical evaluations, because the STF structure can be determined at the very end of the calculations. 
In this Appendix the normalizations and definitions as used in \cite{Poisson_Will} will be applied. 
In particular, we need the following Cartesian STF tensor,
\begin{eqnarray}
	\hat{n}_L = \frac{x_{<\,i_1}}{r}\,\dots\,\frac{x_{i_l\,>}}{r} \;, 
\label{Appendix_Cartesian_Tensor}
\end{eqnarray}

\noindent
where $x_i$ are the spatial coordinates of some arbitrary field point and $r = \left|\ve{x}\right|$; we note that $x_i = x^i$ and $\hat{n}_L = \hat{n}^L$.

A basis in the $\left(2 l +1\right)$-dimensional space of STF tensors with $L$ indices is provided by the tensors $\hat{Y}_L^{lm}$. They are given by 
(cf. Eqs.~(A6.a) - (A6.c) in \cite{Blanchet_Damour1}; a few examples of these basis tensors are provided in Box $1.5$ p.~$33$ in \cite{Poisson_Will}) 
\begin{eqnarray}
	\hat{Y}_L^{lm} &=& A^{lm}\;E^{lm}_{< L >}\;,
	\label{STF_Basis}
\end{eqnarray}

\noindent
where $E^{lm}_{< L >} = {\rm STF}_{i_1 \dots i_l}\;E^{lm}_{i_1 \dots i_l}$ with 
\begin{eqnarray}
	E^{lm}_{L} &=&  
	\left(\delta^1_{i_1} + i\,\delta^2_{i_1}\right) \dots \left(\delta^1_{i_m} + i\,\delta^2_{i_m}\right) 
	\delta^3_{i_{m+1}} \dots \delta^3_{i_l}  
	\nonumber\\ 
	\label{Basis_SFT_E}
\end{eqnarray}

\noindent
and 
\begin{eqnarray}
	A^{lm} &=& \left(-1\right)^m \!\left(2 l - 1\right)!! \sqrt{\frac{2 l + 1}{4 \pi\left(l-m\right)! \left(l+m\right)!}} . 
\end{eqnarray}

\noindent 
These basis tensors are normalized by (cf. Eq.~(2.26a) in \cite{Thorne} or cf. Eq.~(A7) in \cite{Blanchet_Damour1})
\begin{eqnarray}
	\hat{Y}^{l m}_L\;\hat{Y}_L^{\ast\,l m^{\prime}} &=& \delta_{m m^{\prime}}\,\frac{\left(2 l + 1\right)!!}{4 \pi\,l!} 
	\label{normalization_basis_tensor}
\end{eqnarray}

\noindent
where $\hat{Y}_L^{\ast\,lm}$ are the complex conjugate of the basis tensors. 
Using the transformation between Cartesian coordinates $\left(x^1, x^2, x^3\right)$ and spherical coordinates $\left(r,\theta,\phi\right)$, 
\begin{eqnarray}
	x^1 &=& r\,\sin \theta\,\cos \phi\;,\; x^2 = r\,\sin \theta\,\sin \phi\;,\; x^3 = r\,\cos \theta\;, 
	\nonumber\\ 
	\label{spherical_coordinates}
\end{eqnarray}

\noindent
one may show that the STF basis tensors $\hat{Y}^{lm}_L$ are related to the spherical harmonics 
$Y_{lm}$ as follows (cf. Eq.~(2.11) in \cite{Thorne} or Eq.(A8) in \cite{Blanchet_Damour1}) 
\begin{eqnarray}
	\hat{Y}^{lm}_L\,\hat{n}_L &=& Y_{lm}\;, 
	\label{Relation_Y_lm}
\end{eqnarray}

\noindent 
which are normalized by (cf. Eq.~(1.117) in \cite{Poisson_Will})
\begin{eqnarray}
        \int Y_{lm}\,Y_{l^{\prime} m^{\prime}}^{\ast}\;d \Omega &=& \delta_{m m^{\prime}}\;\delta_{l l^{\prime}}\;,  
\label{Spherical_Harmonics}
\end{eqnarray}

\noindent
where $Y_{lm}^{\ast}$ are the complex conjugate of spherical harmonics and 
$d \Omega = \sin \theta\,d \theta\,d \phi$ is the infinitesimal solid angle in the direction $\left(\theta, \phi\right)$.

Any STF tensor $\hat{T}_L$ can be expanded in terms of these basis tensors 
\begin{eqnarray}
	\hat{T}_L &=& \frac{4 \pi\,l!}{\left(2 l + 1\right)!!}\,\sum \limits_{m = - l}^{l} T_{lm}\;\hat{Y}_L^{lm}\;. 
	\label{Expansion_STF_Tensor}
\end{eqnarray}

\noindent
The expansion coefficients $T_{lm}$ are called moments of the STF tensor $\hat{T}_L$ and are obtained by the inverse of (\ref{Expansion_STF_Tensor}). 
That means, if both sides of (\ref{Expansion_STF_Tensor}) are multiplied with $\hat{Y}_L^{\ast\,lm^{\prime}}$, then one obtains 
\begin{eqnarray}
	T_{lm} &=& \hat{T}_L\;\hat{Y}_L^{\ast\,lm}\;, 
        \label{Expansion_STF_Moments}
\end{eqnarray}

\noindent
where the normalization (\ref{normalization_basis_tensor}) of the STF basis tensors has been used. Let us notice that the normalization prefactor 
$\displaystyle \frac{4 \pi\,l!}{\left(2 l + 1\right)!!}$ is convention and appears either in front of (\ref{Expansion_STF_Tensor}) or 
(\ref{Expansion_STF_Moments}). Only the combination of (\ref{Expansion_STF_Tensor}) and (\ref{Expansion_STF_Moments}) is relevant, which 
agrees with the combinations of Eqs.~(2.13a) and (2.13b) in \cite{Thorne}. Here we follow the convention as used, for instance, 
in \cite{Poisson_Will,Hartmann_Soffel}. 

In particular, we need the expansion of the STF part $\hat{x}_L = r^l\,\hat{n}_L$ in terms of these basis tensors, which reads   
\begin{eqnarray}
	\hat{x}_L &=& \frac{4 \pi\,l!}{\left(2 l + 1\right)!!} \sum\limits_{m = - l}^{l} x_{lm}\;\hat{Y}_L^{lm}\;. 
\label{Expansion_xL_5}
\end{eqnarray}

\noindent
According to (\ref{Expansion_STF_Moments}), the moments are given by 
\begin{eqnarray}
x_{lm} = \hat{x}_L\;\hat{Y}_L^{\ast\,lm} = r^l\;Y_{lm}^{\ast}\;, 
\label{Expansion_xL_15}
\end{eqnarray}

\noindent 
where the relation between the STF basis tensors (\ref{Relation_Y_lm}) has been used. 
Hence, one obtains for the expansion of the STF tensor $\hat{x}_L$ the following expression (cf. Eq.~(2.23) in \cite{Hartmann_Soffel}): 
\begin{eqnarray}
	\hat{x}_L &=& \frac{4 \pi\,l!}{\left(2 l + 1\right)!!}\; r^l \sum\limits_{m = - l}^{l} Y_{lm}^{\ast}\;\hat{Y}_L^{lm}\;,  
\label{F_L_A}
\end{eqnarray}

\noindent
which will be used in order to determine the mass-multipole moments and spin-multipole moments.

\subsection{Mass multipoles}

The mass-multipoles $\hat{M}_L$ have been obtained in \cite{Multipole_Damour_2}. In case of time-independent multipoles, they simplify 
to the following form, up to terms of the order ${\cal O}\left(c^{-4}\right)$ (cf. Eq.~(5.38) in \cite{Multipole_Damour_2})
\begin{eqnarray}
        \hat{M}_L &=& \int d^3 x \; \hat{x}_L\;\Sigma\;, 
\label{M_L}
\end{eqnarray}

\noindent
where $\Sigma = \left(T^{00} + T^{kk}\right)/c^2$ is the gravitational mass-energy density of the body 
with $T^{\alpha\beta}$ being the stress-energy tensor of the body. The integration runs over the 
three-dimensional volume of the body.  The zeroth term $l=0$ is the mass of the body: $\hat{M}_0 = M$. The first term $l=1$ is the mass-dipole moment 
which defines the spatial position of the center of mass of the body. In case the origin of the coordinate system coincides with the center of mass 
of the body the mass-dipole moment would vanish \cite{Thorne,Poisson_Will,Kopeikin_Efroimsky_Kaplan} (cf. Eq.~(8.14c) in \cite{Thorne}). 
According to Eq.~(\ref{Expansion_STF_Tensor}) the expansion of the STF mass-multipole (\ref{M_L}) in terms of basis tensors $\hat{Y}_L^{lm}$ 
reads 
\begin{eqnarray}
        \hat{M}_L &=& \frac{4 \pi\,l!}{\left(2 l + 1\right)!!}\,\sum\limits_{m = - l}^{l} M_{lm}\;\hat{Y}_L^{lm}\;.  
\label{M_L_A}
\end{eqnarray}

\noindent
The mass-moments $M_{lm}$ are obtained from the inverse of (\ref{M_L_A}) and read (cf. Eq.~(\ref{Expansion_STF_Moments}))
\begin{eqnarray}
	M_{lm} &=& \hat{M}_L \,\hat{Y}_L^{\ast\,lm} \;. 
\label{Mass_Moment}
\end{eqnarray}

\noindent
Let us notice that the combination of relations (\ref{M_L_A}) and (\ref{Mass_Moment}) coincides with the combination of equations (4.6a) and (4.7a) 
in \cite{Thorne} in case of time-independent multipoles. By inserting (\ref{M_L}) into (\ref{Mass_Moment}) one obtains, with virtue of (\ref{F_L_A}) 
and (\ref{normalization_basis_tensor}), the following expression for the mass-moments (cf. Eq.(1.139) in \cite{Poisson_Will})
\begin{eqnarray}
	M_{lm} &=&  \int d^3 x \,r^l\,\Sigma\, Y_{lm}^{\ast}\;,
\label{M_L_B}
\end{eqnarray}

\noindent
where the integration runs over the volume of the body. The giant planets can be described by a rigid axisymmetric body.  
Accordingly, in order to determine the impact of mass-multipoles on the Shapiro time delay we consider a Newtonian rigid axisymmetric body, having the shape 
\begin{eqnarray}
	\frac{\left(x^1\right)^2}{A^2} + \frac{\left(x^2\right)^2}{B^2} + \frac{\left(x^3\right)^2}{C^2} &=& 1\;, 
\label{Shape}
\end{eqnarray}

\noindent
where $A = B$ is the semi-major axis (i.e. equatorial radius $P$) and $C$ is the semi-minor axis of the body. The oblateness of the axisymmetric body 
is parameterized by the ellipticity parameter $\epsilon^2 = \left(A^2 - C^2\right)/A^2$ which is also used in the IAU resolutions 
(p.~$2698$ in \cite{IAU_Resolution1}).
It is assumed that the unit-vector ${\ve e}_3$ is the symmetry axis of the massive body and the $x^3$-direction of the coordinate system 
is aligned with the symmetry axis of the body. Then, the multipole moments (\ref{M_L_B}) vanish for $m \neq 0$, that means we need 
\begin{eqnarray}
        M_{l0} &=&  \int d^3 x \,r^l\,\Sigma\, Y_{l0}^{\ast} \;. 
\label{M_L_C}
\end{eqnarray}

\noindent 
The spherical harmonics for $m=0$ are real valued functions, $Y_{l0}^{\ast} = Y_{l0}$, 
and they are related to the Legendre polynomials $P_l$ (cf. Eq.~(1.112) in \cite{Poisson_Will}) 
\begin{eqnarray}
	Y_{l0} &=& \sqrt{\frac{2 l + 1}{4 \pi}}\,P_l \left(\cos \theta\right),  
\label{Legende_Polynoms}
\end{eqnarray}

\noindent
where $\theta$ is the angle between integration variable $\ve{x}$ and the $x^3$-direction of the coordinate system (azimuth angle). 
Performing these integrals in (\ref{M_L_C}) one finds that they are proportional to the mass $M$ of the body and the $l$-th power 
of the equatorial radius of the body, $\left(P\right)^l$ (which should not be confused with Legendre polynomial $P_l$) 
and they are non-vanishing only for even $l$, 
\begin{eqnarray}
	\hspace{-0.75cm} M_{l0} = - \sqrt{\frac{2 l + 1}{4 \pi}}\,M \left(P\right)^l J^{\rm el}_l 
\label{M_L_D}
\end{eqnarray}

\noindent
for $l = 0, 2, 4, 6, \dots$. 
Eq.~(\ref{M_L_D}) coincides with Eq.~(1.143) in \cite{Poisson_Will}. The dimensionless parameter $J^{\rm el}_l$ in (\ref{M_L_D}) 
are the gravitoelectric zonal harmonic coefficients, and follow from inserting (\ref{M_L_D}) into (\ref{M_L_C}) (cf. Eq.~(17) in \cite{Teyssandier1})
\begin{eqnarray}
	J^{\rm el}_{l} = - \frac{1}{M \left(P\right)^l} \int d^3 x \,r^l\,\Sigma\,P_l \left(\cos \theta\right)  
\label{zonal_harmonic_coefficient}
\end{eqnarray}

\noindent 
for $l = 0, 2, 4, 6 \dots$.
For an axisymmetric body ((\ref{Shape}) with $A=B$) with uniform density one obtains (cf. Eq.~(56) in \cite{Klioner2003b}) 
\begin{eqnarray}
	J^{\rm el}_l = \left(-1\right)^{l/2 + 1} \frac{3}{\left(l+1\right) \left(l+3\right)}\;\epsilon^l  
\label{zonal_harmonics_el}
\end{eqnarray}

\noindent 
for $l = 0, 2, 4, 6 \dots$.
Obviously, higher mass-moments $\left(l > 0 \right)$ vanish for $\epsilon = 0$, that means for spherically symmetric bodies only the mass-monopole is non-zero.
By inserting (\ref{M_L_D}) into (\ref{M_L_A}) one obtains for the mass-multipole (\ref{M_L})
\begin{eqnarray}
	\hat{M}_L &=& - \sqrt{\frac{2l + 1}{4 \pi}}\,\frac{4 \pi\,l!}{\left(2 l + 1\right)!!}\,M\,P^l\,J^{\rm el}_l\,\hat{Y}_L^{l0}\,,
\label{M_L_E}
\end{eqnarray}

\noindent
where $P^l$ means the $l$-th power of the equatorial radius, while the suffix $l$ in $J^{\rm el}_l$ is an index and denotes the $l$-th zonal harmonic coefficient. 
The basis tensors $\hat{Y}_L^{lm}$ for $m=0$ are given by (cf. Eqs.~(A6.a) - (A6.c) in \cite{Blanchet_Damour1}) 
\begin{eqnarray}
	\hat{Y}_L^{l0} &=& \left(2 l - 1 \right)!! \; \sqrt{\frac{2 l + 1}{4 \pi\,l!\;l!}}\;  
	\;\delta^{3}_{< {i_1}} \; \dots \; \delta^{3}_{{i_l} >} \;.
\label{Y_L}
\end{eqnarray}

\noindent
Finally, inserting (\ref{Y_L}) into (\ref{M_L_E}) yields for the mass-multipoles for the case of an axisymmetric rigid body with uniform density 
the following expression: 
\begin{eqnarray}
	\hat{M}_L &=& - M\,P^l\,J^{\rm el}_l\;\delta^3_{<{i_1}} \; \dots \; \delta^3_{{i_l}>} 
\label{M_L_G}
\end{eqnarray}

\noindent
for $l = 2, 4, 6, \dots$. 
The STF terms are products of Kronecker symbols which are symmetric and traceless with respect to indices $i_1 \dots i_l$.
They are given by the formula (cf. Eq.~(1.155) in \cite{Poisson_Will}):
\begin{eqnarray}
	\delta^3_{<{i_1}} \; \dots \; \delta^3_{{i_l}>} &=& \sum\limits_{p=0}^{[l/2]} \left(-1\right)^p\,
        \frac{\left(2 l - 2 p - 1\right)!!}{\left(2 l - 1 \right)!!}  
	\nonumber\\ 
	&& \times \left[\delta_{2P}\,\delta^3_{L - 2P} + {\rm sym.}\left(q\right) \right], 
        \label{STF_Formula}
\end{eqnarray}

\noindent
where $[l/2]$ is equal to $l/2$ for even $l$ and equal to $(l-1)/2$ for odd $l$. The symbol $\delta_{2P}$ stands for the product of $p$ Kronecker deltas with 
indices running from $\delta_{i_1 i_2} \times \dots \times \delta_{i_{2p-1} i_{2p}}$. The symbol $\delta^3_{L - 2P}$ stands for the product of $l - 2 p$ Kronecker 
deltas with indices running from $\delta^3_{i_{2p+1}} \times \dots \times \delta^3_{i_{l}}$. The notation ${\rm sym.}\left(q\right)$ in (\ref{STF_Formula}) means 
symmetrization with respect to the $2 p$ indices $i_1 \dots i_{2p}$, where the total number of these symmetrized terms 
is $q = l! \,/ \left[ (l - 2 p)! \, (2p)!! \right]$.
The terminology of the first mass-multipoles reads:   
\begin{enumerate}
        \item[$\bullet$] $l=0$: mass-monopole,
        \item[$\bullet$] $l=2$: mass-quadrupole,
        \item[$\bullet$] $l=4$: mass-octupole,
        \item[$\bullet$] $l=6$: mass-dodecapole,
        \item[$\bullet$] $l=8$: mass-hexadecapole,
        \item[$\bullet$] $l=10$: mass-icosadecapole.
\end{enumerate}

\noindent 
Let us show that expression (\ref{M_L_G}) coincides with the IAU resolutions \cite{IAU_Resolution1} for the case of mass-quadrupole. The equation (48) 
in \cite{IAU_Resolution1} states $\hat{M}_L = - \hat{C}_L$ where $\hat{C}_L = {\rm STF}_{i_1 \dots i_l}\,C_{i_1 \dots i_l}$ with 
the tensor $C_{i_1 \dots i_l}$ given by Eq.~(46) in \cite{IAU_Resolution1}. In case of an axisymmetric 
rigid body with uniform density the explicit values $C_{XX} = C_{YY} = M \left(A^2 + C^2\right)/5$ and $C_{ZZ} = 2 M A^2 / 5$ were presented 
(see text below Eq.~(48) in \cite{IAU_Resolution1}). Using (\ref{anti_symmetric_1}) one may determine their STF expressions, which, using Eq.~(48) 
in \cite{IAU_Resolution1}, results in $\hat{M}_{XX} = \hat{M}_{YY} = M \left(A^2 - C^2\right)/15$ and $\hat{M}_{ZZ} = - 2 M \left(A^2 - C^2\right)/15$,
which is in agreement with our expression given by Eq.~(\ref{M_L_G}) for $l=2$. 

In reality the mass distribution $\Sigma$ of the Sun and the giant planets is not uniform but depends on the radial distance. 
Therefore, the theoretical values of the zonal harmonic coefficients, $J^{\rm el}_l$, as calculated for a axisymmmetric body with uniform density by 
Eq.~(\ref{zonal_harmonics_el}), are a bit larger than their actual values. Instead to calculate these actual values by relation (\ref{zonal_harmonic_coefficient}) 
with a model-dependent assumption for the mass-density, the actual zonal harmonic coefficients are deduced from real measurements of the gravitational fields 
of the giant planets and are denoted by $J_l$. These values are given in Table~\ref{Table1}. If one replaces in (\ref{M_L_G}) the theoretical values of the  
zonal harmonic coefficients, $J^{\rm el}_l$, by these actual values from real measurements, $J_l$, then one obtains the mass-multipoles for the case of an 
axisymmetric rigid body with radial-dependent mass-density:  
\begin{eqnarray}
	\hat{M}_L &=& - M\,P^l\,J_l\;\delta^3_{<{i_1}} \; \dots \; \delta^3_{{i_l}>} 
\label{M_L_F}
\end{eqnarray}

\noindent 
for $l = 2, 4, 6, \dots$. 
For estimations of the Shapiro time delay only the first eight terms of the mass-multipoles (\ref{M_L_F}) are needed, even on the 
sub-pico-second level. The mass-quadrupole and the mass-octupole are given in their explicit form as follows: 
\begin{eqnarray}
	\hat{M}_{ab} &=& + M\,P^2\,J_2 \left[\frac{1}{3}\,\delta_{ab} - \delta^3_{a}\,\delta^3_{b}\right], 
\label{M_ab}
\\
	\hat{M}_{abcd} &=& - M\,P^4\,J_4 \,\bigg[\frac{1}{35}\left(\delta_{ab}\,\delta_{cd} + \delta_{ac}\,\delta_{bd} + \delta_{ad}\,\delta_{bc}\right) 
	\nonumber\\ 
	&& \hspace{-0.75cm} + \, \delta^3_{a}\,\delta^3_{b}\,\delta^3_{c}\,\delta^3_{d} 
	- \frac{1}{7} \left(\delta_{ab}\,\delta^3_{c}\,\delta^3_{d} + \delta_{ac}\,\delta^3_{b}\,\delta^3_{d} + \delta_{ad}\,\delta^3_{b}\,\delta^3_{c} \right) 
	\nonumber\\ 
	&& \hspace{-0.75cm} 
	- \frac{1}{7} \left(\delta_{bc}\,\delta^3_{a}\,\delta^3_{d} + \delta_{bd}\,\delta^3_{a}\,\delta^3_{c} + \delta_{cd}\,\delta^3_{a}\,\delta^3_{b}\right) 
	\bigg].  
\label{M_abcd}
\end{eqnarray}

\subsection{Spin multipoles} 

The spin-multipoles $\hat{S}_L$ have been obtained in \cite{Multipole_Damour_2}. In case of time-independent multipoles, they simplify
to the following form, up to terms of the order ${\cal O}\left(c^{-4}\right)$ (cf. Eq.~(5.40) in \cite{Multipole_Damour_2})
\begin{eqnarray}
        \hat{S}_L &=& \int d^3 x \;\epsilon_{j k < i_l}\, \hat{x}_{L-1 >}\;x^j\;\Sigma^k 
\label{S_L}
\end{eqnarray}

\noindent
where the notation $\Sigma^k = T^{0k}/c$ has been adopted, with $T^{\alpha\beta}$ being the stress-energy tensor of the body and the integration 
runs over the three-dimensional volume of the body. The first term $l=1$ is the spin-dipole and describes the rotational motion of the body as a whole.  
In case the body is rigid and spherically symmetric, then the higher spin-multipoles would vanish. However, in case the body is not 
spherically symmetric, then these higher spin-multipoles $l \ge 3$ account for the rotational motion of the body as a whole. 
In addition, if there are inner currents of the body, then the higher spin-multipoles account also for these inner circulations. 

According to Eq.~(\ref{Expansion_STF_Tensor}) the expansion of the STF spin-multipole (\ref{S_L}) in terms of basis tensors $\hat{Y}_L^{lm}$ reads 
\begin{eqnarray}
        \hat{S}_L &=& \frac{4 \pi\,l!}{\left(2 l + 1\right)!!}\,\sum\limits_{m = - l}^{l} S_{lm}\;\hat{Y}_L^{lm}\;. 
\label{S_L_A}
\end{eqnarray}

\noindent
The spin-moments $S_{lm}$ are obtained from the inverse of (\ref{S_L_A}) and read (cf. Eq.~(\ref{Expansion_STF_Moments})) 
\begin{eqnarray}
        S_{lm} &=& \hat{S}_L \,\hat{Y}_L^{\ast\,lm} \;. 
\label{Spin_Moment}
\end{eqnarray}

\noindent
Let us notice that the combination of relations (\ref{S_L_A}) and (\ref{Spin_Moment}) coincides with the combination of equations (4.6b) and (4.7b) 
in \cite{Thorne} in case of time-independent multipoles. By inserting (\ref{S_L}) into (\ref{Spin_Moment}) one obtains, with virtue of (\ref{F_L_A}), 
the following expression for the spin-moments 
\begin{eqnarray}
	S_{lm} &=& \frac{4 \pi\,\left(l-1\right)!}{\left(2 l - 1 \right)!!} \int d^3 x\;r^l\;n^j \,\Sigma^k 
	\nonumber\\ 
	&& \times \!\! \sum \limits_{m^{\prime}= - l + 1}^{l - 1} 
	\epsilon_{j k \,< i_l}\;\hat{Y}^{l-1\,m^{\prime}}_{L-1\,>}\;Y_{l - 1\,m^{\prime}}^{\ast}\;\hat{Y}^{\ast\,lm}_{L}  
\label{S_L_5}
\end{eqnarray}

\noindent
where the integration runs over the volume of the body; note that $n^j = x^j/r$ and $\hat{Y}^{\ast\,lm}_{L} = \hat{Y}^{\ast\,lm}_{i_l L-1}$. 
Now we make use of the following relation (cf. Eq.~(2.26b) in \cite{Thorne}): 
\begin{eqnarray}
	\hat{Y}^{l-1\,m^{\prime}}_{L-1}\,\hat{Y}^{\ast\,lm}_{i_l L - 1} = \frac{\left(2 l + 1 \right)!!}{4\,\pi\;l!}\,\sqrt{\frac{l}{2 l + 1}}  
	\left(1\;\;l - 1\;\;0\;m^{\prime}| l\;m\right) e_3^{i_l} 
	\nonumber\\ 
\label{S_L_20}
\end{eqnarray}

\noindent
where $\left(1\;\;l - 1\;\;0\;m^{\prime}| l\;m\right)$ are the Clebsch-Gordan coefficients \cite{Arfken_Weber} and $e_3^{i_l}$ is the $i_l$-component of 
unit three-vector $\ve{e}_3$. By inserting (\ref{S_L_20}) into (\ref{S_L_5}) one encounters the vector spherical harmonics \cite{Thorne,Arfken_Weber} 
(cf. Eq.~(2.16) in \cite{Thorne} or Eq.~(2.221) in \cite{Arfken_Weber}) 
\begin{eqnarray}
        Y^{\ast\,l-1\,,\,lm}_{i_l} &=& 
	\sum\limits_{m^{\prime} = - l + 1}^{l - 1} \left(1\;\;l - 1\;\;0\;m^{\prime}| l\;m\right) \;Y_{l - 1\,m^{\prime}}^{\ast}\;e_3^{i_l}\;.
	\nonumber\\ 
\label{S_L_25}
\end{eqnarray}

\noindent
Thus, in view of (\ref{S_L_20}) and (\ref{S_L_25}) one obtains for the spin moments (\ref{S_L_5}) 
\begin{eqnarray}
	S_{lm} &=& \frac{2l + 1}{l} \sqrt{\frac{l}{2 l + 1}} \int d^3 x\;r^l\;\epsilon_{i j k}\;n^j\;\Sigma^k\;Y^{\ast\,l-1\,,\,lm}_{i}\;,   
	\nonumber\\ 
\label{S_L_30}
\end{eqnarray}

\noindent
where the spatial dummy index $i_l$ has been designated into the new spatial dummy index $i$. 
Now we use a relation between vector spherical harmonics and STF harmonics (cf. Eq.~(2.24a) in \cite{Thorne}) 
\begin{eqnarray}
	Y^{\ast\,l-1\,,\,lm}_{i} &=& \sqrt{\frac{l}{2 l + 1}} \;\hat{Y}^{\ast\,lm}_{i L-1}\;\hat{n}_{L-1}\;,
	\label{S_L_35}
\end{eqnarray}

\noindent 
as well as (cf. Eq.~(2.23b) in \cite{Thorne}) 
\begin{eqnarray}
	\epsilon_{i j k}\;n^j\;\hat{Y}^{\ast\,lm}_{i L-1}\;\hat{n}_{L-1} &=& - \sqrt{\frac{l + 1}{l}} \;Y_k^{\ast\,B\,,\,lm}\;,
\label{S_L_40}
\end{eqnarray}

\noindent
where $Y_k^{\ast\,B\,,\,lm}$ is the complex conjugate of one of the pure spin-vector harmonics (cf. Eq.~(2.18b) in \cite{Thorne}) and obtain 
\begin{eqnarray}
        S_{lm} &=& - \sqrt{\frac{l + 1}{l}} \int d^3 x\;r^l\;\Sigma^k\;Y_k^{\ast\,B\,,\,lm}\;.  
\label{S_L_45}
\end{eqnarray}

\noindent
Finally, we use the definition of the pure spin-vector harmonics (cf. Eq.~(2.18b) in \cite{Thorne}) 
\begin{eqnarray}
	Y_k^{\ast\,B\,,\,lm} &=& \sqrt{\frac{1}{l \left(l+1\right)}}\;\left(\ve{x} \times \ve{\nabla}\right)^k\;Y_{lm}^{\ast}\;,
\label{S_L_50}
\end{eqnarray}

\noindent
where $\ve{\nabla} = \ve{e}_r\,\partial_r + \ve{e}_{\theta}\,r^{-1}\,\partial_{\theta} + \ve{e}_{\phi}\,\left(r \sin \theta\right)^{-1}\,\partial_{\phi}$ 
is the gradient operator of Euclidean three-space in spherical coordinates which acts on the complex conjugate of 
spherical harmonics $Y_{lm}^{\ast}$ and the position vector in spherical coordinates reads $\ve{x} = r\,\ve{e}_r$. 
Inserting (\ref{S_L_50}) into (\ref{S_L_45}) yields the following expression for the spin-moments 
\begin{eqnarray}
	S_{lm} &=& \frac{1}{l} \int d^3 x \,r^l\left(\ve{x} \times \ve{\Sigma}\right) \cdot \ve{\nabla}\, Y_{lm}^{\ast}  \;, 
\label{S_L_B}
\end{eqnarray}

\noindent
where the integration runs over the three-dimensional volume of the body. The steps from (\ref{S_L_A}) until (\ref{S_L_B}) coincide with the steps from 
Eq.~(5.17b) to Eq.~(5.18b) in \cite{Thorne} for the case of time-independent multipoles. Below we will show, for the case of axisymmetric bodies, that 
(\ref{S_L_B}) coincides with the IAU resolutions \cite{IAU_Resolution1}. Let us also notice that the combination of expressions (\ref{S_L_A}) and (\ref{S_L_B}) 
coincides with the combination of Eqs.~(10) and (11) in \cite{Pannhans_Soffel}. 

In order to determine the impact of spin-multipoles on the Shapiro time delay we consider a rigid Newtonian body in uniform rotational motion and having 
axisymmetric shape (\ref{Shape}), where the unit-vector ${\ve e}_3$ is the symmetry axis of the massive body and the $x^3$-direction of the coordinate system 
is aligned with the rotational axis of the body. Then, the rotational angular velocity $\ve{\Omega}$ is independent of time and for the momentum-density of 
the body one may write (cf. Eq.~(12) in \cite{Pannhans_Soffel} and IAU resolutions (p. $2698$ in \cite{IAU_Resolution1}) where spin-moments for the model of 
a rigidly rotating Earth have been considered):  
\begin{eqnarray}
	\ve{\Sigma} = \Sigma\,\left(\ve{\Omega} \times \ve{x} \right) = \Sigma \,\Omega\,r\,\sin \theta\;\ve{e}_{\phi}\;.  
\label{angular_velocity_1}
\end{eqnarray}

\noindent 
It has been shown in \cite{Pannhans_Soffel} that the only non vanishing spin-moments (\ref{S_L_B}) are those for $m=0$ and odd $l$ 
(cf. Eqs.~(20) in \cite{Pannhans_Soffel}): 
\begin{eqnarray}
	S_{l0} &=& + \frac{1}{l} \int d^3 x \,r^l\,\left(\ve{x} \times \ve{\Sigma}\right) \cdot \ve{\nabla}\,Y_{l0}^{\ast} \;,  
	\nonumber\\
	&=& - \frac{1}{l}\,\sqrt{\frac{2 l + 1}{4 \pi}}\,\Omega 
	\int d^3 x \,\Sigma\,r^{l+1} \sin \theta\,\frac{\partial\,P_l \left(\cos \theta\right)}{\partial \theta}\,,  
	\nonumber\\ 
\label{S_L_C}
\end{eqnarray}

\noindent
where the spherical harmonics for $m=0$ are related to the Legendre polynomials as given by Eq.~(\ref{Legende_Polynoms}) and where $\theta$ is again 
the angle between integration variable $\ve{x} = r\,\ve{e}_r$ and the $x^3$-direction of the coordinate system (azimuth angle) and 
$\ve{e}_r \times \ve{e}_{\phi} = - \ve{e}_{\theta}$ has been used. 
Performing these integrals in (\ref{S_L_C}) one finds that they are proportional to the angular velocity $\Omega$, 
to the mass $M$ of the body and the $\left(l+1\right)$-th power of the equatorial radius $P$ of the body and they are non-vanishing only for odd $l$,
\begin{eqnarray}
	S_{l0} &=& \sqrt{\frac{2 l + 1}{4 \pi}} \left(l + 1\right) M \, \Omega \left(P\right)^{l+1} J^{\rm gm}_l 
\label{S_L_D}
\end{eqnarray}

\noindent 
for $l = 1, 3, 5, \dots$.
The parameter $J^{\rm gm}_l$ in (\ref{S_L_D}) are the gravitomagnetic zonal harmonic coefficients and follow from inserting (\ref{S_L_C}) into (\ref{S_L_D}),  
\begin{eqnarray}
	J^{\rm gm}_l &=& - \frac{1}{M \left(P\right)^{l+1}}\,\frac{1}{l \left(l+1\right)} 
	\int d^3 x\, r^{l+1}\,\Sigma\,\sin \theta\,\frac{\partial\,P_l \left(\cos \theta\right)}{\partial \theta} 
\nonumber\\ 
	\label{zonal_harmonics_gravito_magnetic}
\end{eqnarray}

\noindent
for $l = 1, 3, 5, \dots$.
For an axisymmetric body ((\ref{Shape}) with $A=B$) with uniform mass density they are given by (cf. Eq.~(25) in \cite{Pannhans_Soffel})  
\begin{eqnarray}
	J^{\rm gm}_l = \left(-1\right)^{(l - 1)/2} \frac{3}{l\,\left(l+2\right) \left(l+4\right)}\,\epsilon^{l - 1}  
\label{zonal_harmonics_gm}
\end{eqnarray}

\noindent
for $l = 1, 3, 5, \dots$.
where the ellipticity parameter $\epsilon^2 = \left(A^2 - C^2\right)/A^2$ has already been defined above. 
The combinations of the equations (\ref{S_L_A}) with (\ref{S_L_D}) and (\ref{zonal_harmonics_gm}) agrees with the combination of the equations 
(10) with (22) and (25) in \cite{Pannhans_Soffel}. Obviously, higher spin-moments $\left(l > 1 \right)$ vanish for $\epsilon = 0$, that means 
for spherically symmetric bodies only the spin-dipole is non-zero. 
A comparison between (\ref{zonal_harmonics_gm}) and (\ref{zonal_harmonics_el}) leads to the 
following remarkable relation between the gravitomagnetic and gravitoelectric zonal harmonic coefficients for an axisymmetric body with uniform mass density 
and in uniform rotational motion (cf. Eq.~(28) in \cite{Pannhans_Soffel}): 
\begin{eqnarray}
	J^{\rm gm}_l = - \frac{J^{\rm el}_{l-1}}{l + 4} \;.
\label{relation_zonal_harmonics}
\end{eqnarray}

\noindent 
Finally, in view of relation (\ref{relation_zonal_harmonics}) and by inserting (\ref{zonal_harmonics_gm}) and (\ref{S_L_D}) into (\ref{S_L_A}) one obtains 
for the spin-multipoles for the case of an axisymmetric rigid body with uniform mass-density and in uniform rotational motion the following expression:
\begin{eqnarray}
  \hat{S}_L &=& - M\,\Omega\,P^{l+1}\,J^{\rm el}_{l-1}\,\frac{l+1}{l+4}\;\delta^3_{<{i_1}} \; \dots \; \delta^3_{{i_l}>} 
\label{S_L_G}
\end{eqnarray}

\noindent
for $l = 1, 3, 5, \dots$. The STF terms are products of Kronecker symbols which are symmetric and traceless with respect to indices $i_1 \dots i_l$.
They are given by the formula (\ref{STF_Formula}). The terminology of the first spin-multipoles reads: 
\begin{enumerate}
        \item[$\bullet$] $l=1$: spin-dipole,
        \item[$\bullet$] $l=3$: spin-hexapole,
        \item[$\bullet$] $l=5$: spin-decapole,
        \item[$\bullet$] $l=7$: spin-quattuordecapole,
        \item[$\bullet$] $l=9$: spin-octodecapole.
\end{enumerate}

\noindent 
Let us show that expression (\ref{S_L_G}) coincides with the IAU resolutions \cite{IAU_Resolution1} for the case of spin-hexapole. The equation (45)
in \cite{IAU_Resolution1} states $\hat{S}_L = \hat{C}_{Ld} \Omega^d$ where $\hat{C}_{Ld}$ is STF with respect to indices $L$ but not with respect to 
index $d$, that means $\hat{C}_{Ld} = {\rm STF}_{i_1 \dots i_l}\,C_{i_1 \dots i_l d}$ 
which is given by Eq.~(46) in \cite{IAU_Resolution1}. 
Assuming $\Omega^d = \left(0,0,\Omega\right)$ the non-vanishing terms are $\hat{S}_{XXZ} = \hat{S}_{YYZ} = 3 \eta \Omega$ and $\hat{S}_{ZZZ} = - 6 \eta \Omega$ 
with $\eta = 4 M A^4 \epsilon^2 / 525$, which is in agreement with our expression given by Eq.~(\ref{S_L_G}) for $l=3$.

The spin-multipoles in (\ref{S_L_G}) are valid for a rigid axi\-symmetric body with uniform mass-density and in uniform rotation with angular velocity 
$\Omega = 2 \pi/T$ where $T$ is the rotational period around the spin axis of the body. However, in reality the mass distribution of the Sun and the 
giant planets is not uniform, but increasing towards the center of the massive body. In case of mass-multipoles this fact has been taken into account in the step 
from (\ref{M_L_G}) to (\ref{M_L_F}), where the gravitoelectric zonal harmonic coefficients $J_l^{\rm el}$, for an axisymmetric body with uniform 
mass-density given by (\ref{zonal_harmonics_el}), have been replaced by the actual zonal harmonic coefficients $J_l$ which are determined by 
real measurements of the gravitational fields of these bodies by space missions. 
Here, in similar manner, the gravitoelectric zonal harmonic coefficients $J_l^{\rm em}$ for an axisymmetric body with uniform 
density in (\ref{S_L_G}) are replaced by their actual gravitoelectric zonal harmonic coefficients $J_l$, as they are given in Table~\ref{Table1}. 
In this way, one obtains for the spin-multipoles for the case of a axisymmetric 
rigid body in uniform rotational motion and with radial-dependent mass-density the following expression:  
\begin{eqnarray}
	\hat{S}_L &=& - M\,\Omega\,P^{l+1}\,J_{l-1}\,\frac{l+1}{l+4}\;\delta^3_{<{i_1}} \; \dots \; \delta^3_{{i_l}>} 
\label{S_L_F}
\end{eqnarray}

\noindent 
for $l = 1, 3, 5, \dots$. 
Actually, for estimations of the Shapiro time delay only the first two terms of the spin-multipoles (\ref{S_L_F}) are needed, even on the
sub-pico-second level: spin-dipole and spin-hexapole. They are given in their explicit form as follows:
\begin{eqnarray}
	&& \hspace{-0.75cm} \hat{S}_{a} = + \frac{2}{5}\,M\,\Omega\,P^2\,\delta_{3a}\;, 
\label{S_a}
\\
	&& \hspace{-0.75cm} \hat{S}_{abc} = + \frac{4}{7}\,M\,\Omega\,P^4\,J_2
	\nonumber\\ 
	&& \hspace{-0.25cm} \times \bigg[\frac{1}{5} \left(\delta_{ab}\,\delta_{3c} + \delta_{ac}\,\delta_{3b} + \delta_{bc}\,\delta_{3a}\right) 
	- \delta_{a3}\,\delta_{b3}\,\delta_{c3}\bigg].  
\label{S_abc}
\end{eqnarray}

\noindent
In (\ref{S_a}) we have used $J^{\rm el}_0 = J_0 = - 1$, that means for $l=1$ the theoretical gravitoelectric zonal harmonic coefficient 
for a body with uniform mass-density and the actual zonal harmonic coefficient for a body with radius-dependent mass-density 
are equal. Thus, a replacement of either these terms from (\ref{S_L_G}) to (\ref{S_L_F}) has no impact on the spin-dipole in (\ref{S_a}). 
Therefore, in order to account for the fact that the density of the massive bodies 
is not uniform, one considers the following reasoning for the spin-dipole. In general, the absolute value of the exact spin-dipole $\left|S_a\right|$ 
(i.e. $l=1$ in Eq.~(\ref{S_L})) is the body's spin angular momentum, which is related to the body's moment of inertia $I$ as follows,
\begin{eqnarray}
         \left|S_a\right| &=& I\,\Omega\;.
        \label{Moment_of_Inertia_1}
\end{eqnarray}

\noindent 
For a solid sphere with uniform density the moment of inertia is $\displaystyle I = \frac{2}{5}\,M\,P^2$ (cf. Eq.~(1.20) in \cite{Ellipticity}), 
hence $\displaystyle \left|S_a\right| = \frac{2}{5}\,M\,P^2\,\Omega$ in agreement with absolute value of the spin-dipole (\ref{S_a}). 
In order to take into account also for the spin-dipole the fact that in reality the mass density is increasing towards the 
center of these massive solar system bodies, we implement the so-called dimensionless moment of inertia $\kappa^2$, which is defined as follows 
\cite{Ellipticity}
\begin{eqnarray}
         \kappa^2 &=& \frac{I}{M\,P^2}\;. 
	\label{kappa}
\end{eqnarray}

\noindent 
Then, the spin angular momentum of the body (\ref{Moment_of_Inertia_1}) is given by \cite{Ellipticity,Moment_of_Inertia_Saturn}
\begin{eqnarray}
         \left|S_a\right| &=& \kappa^2\,M\,P^2\,\Omega\;.
        \label{Moment_of_Inertia_2}
\end{eqnarray}

\noindent
For $\kappa^2 = 0.4$ one recovers the case of a solid sphere with uniform density (cf.~(\ref{S_a})), while for real solar system bodies 
$\kappa^2 < 0.4$ because their mass-density increases towards the center of the bodies. 
These realistic values for $\kappa^2$ have been determined for several solar system bodies in \cite{Ellipticity} using 
the Darwin-Radau relation (e.g.. Eq.~(18) in \cite{Darwin_Radau}). Similar values are given in the planetary fact sheets. 
For the Sun the value of $\kappa^2$ fairly coincides with helioseismology data of the Sun's spin angular momentum \cite{Angular_Momentum_Sun}.  
Accordingly, instead of (\ref{S_a}) we will adopt the following expression for the spin-dipole:
\begin{eqnarray}
        \hat{S}_{a} &=& + \kappa^2\,M\,P^2\,\Omega\,\delta_{3a}\;, 
\label{S_a_B}
\end{eqnarray}

\noindent
where $\kappa^2$ is given in Table~\ref{Table1} for Sun, Jupiter, and Saturn.

\section{The 1PN Shapiro effect of mass-quadrupole}\label{Appendix_Mass_Quadrupole}  

From (\ref{Shapiro_Mass_Multipole}) one obtains the following expression for the impact of the 1PN mass-quadrupole on Shapiro time delay: 
\begin{eqnarray}
	&& \hspace{-0.5cm} \Delta c\tau_{\rm 1PN}^{M_{ab}} = + \frac{G\,\hat{M}_{ab}}{c^2} 
	\nonumber\\ 
	&& \hspace{-0.5cm} \times \left(\hat{\partial}_{ab} \,\ln \left(r_{\rm N} + c \tau\right)\bigg|_{\tau = t_1} 
	\hspace{-0.25cm} - \hat{\partial}_{ab} \,\ln \left(r_{\rm N} + c \tau\right)\bigg|_{\tau = t_0}\right).   
	\nonumber\\ 
\label{Appendix_Shapiro_Mass_Quadrupole_5} 
\end{eqnarray}

\noindent 
The application of the differential operator (\ref{Transformation_Derivative_3}), without the STF procedure, yields  
\begin{eqnarray}
	\partial_{ab}\,\ln \left(r_{\rm N} + c \tau\right) &=& 
	P_{a}^{j_1}\,P_{b}^{j_2}\, \frac{\partial}{\partial \xi^{j_1}}\,\frac{\partial}{\partial \xi^{j_2}}\,\ln \left(r_{\rm N} + c \tau\right) 
	\nonumber\\ 
	&& \hspace{-1.0cm} + \,2\,\sigma_a\,P_{b}^{j_2}\,\frac{\partial}{\partial c \tau}\,\frac{\partial}{\partial \xi^{j_2}}\,\ln \left(r_{\rm N} + c \tau\right) 
	\nonumber\\ 
	&& \hspace{-1.0cm} + \, \sigma_a \,\sigma_b \,\frac{\partial}{\partial c \tau}\,\frac{\partial}{\partial c \tau}\, \ln \left(r_{\rm N} + c \tau\right). 
\label{Appendix_Shapiro_Mass_Quadrupole_10}
\end{eqnarray}

\noindent 
where the STF operation with respect to the indices $ab$ has been omitted in view of relation (\ref{STF_comment_3}). 
With $r_{\rm N} =\sqrt{\xi^2 + c^2 \tau^2}$ one gets 
\begin{eqnarray}
	\partial_{ab}\,\ln \left(r_{\rm N} + c \tau\right) &=& 
	+ P_{a}^{j_1}\,P_{b}^{j_2}\,\delta_{j_1 j_2}\,\frac{1}{r_{\rm N}}\,\frac{1}{r_{\rm N} + c \tau} 
	\nonumber\\ 
	&& \hspace{-1.0cm}  -\, P_{a}^{j_1}\,P_{b}^{j_2}\,\xi_{j_1}\,\xi_{j_2}\,\frac{1}{\left(r_{\rm N}\right)^3} \, \frac{1}{r_{\rm N} + c \tau}
	\nonumber\\ 
	&& \hspace{-1.0cm} -\, P_{a}^{j_1}\,P_{b}^{j_2}\,\xi_{j_1}\,\xi_{j_2}\,\frac{1}{\left(r_{\rm N}\right)^2} \, \frac{1}{\left(r_{\rm N} + c \tau\right)^2} 
	\nonumber\\ 
	&& \hspace{-1.0cm} -\, 2\,\sigma_a\,P_{b}^{j_2}\,\xi_{j_2}\,\frac{1}{\left(r_{\rm N}\right)^3} 
	- \sigma^a\,\sigma^b\,\frac{c \tau}{\left(r_{\rm N}\right)^3}\;. 
\label{Appendix_Shapiro_Mass_Quadrupole_15}
\end{eqnarray}

\noindent
Here we have used $\partial \xi^i / \partial \xi^j = \delta^i_j$, because we treat the spatial components of vector $\ve{\xi}$ as 
formally independent. Therefore, a subsequent projection onto the two-dimensional plane perpendicular to the three-vector $\ve{\sigma}$ is performed 
(cf. text above Eq.~(31) in \cite{KopeikinSchaefer1999_Gwinn_Eubanks}). It is emphasized that this projection is automatically included here, 
namely in the differential operator, which has been introduced in the form given by Eq.~(\ref{Transformation_Derivative_3}). 
Using $P_a^{j_1}\,\xi_{j_1} = \xi_a$ (cf. Eq.~(29) in \cite{KopeikinSchaefer1999_Gwinn_Eubanks}) and finally replacing $c \tau = \ve{\sigma} \cdot \ve{x}$ 
as well as $\xi^a = d_{\sigma}^a$, one obtains for the 1PN quadrupole Shapiro effect (\ref{Appendix_Shapiro_Mass_Quadrupole_5}):
\begin{eqnarray}
        \Delta c\tau_{\rm 1PN}^{M_{ab}} &=& + \frac{G\,\hat{M}_{ab}}{c^2}
	\nonumber\\ 
	&& \hspace{-1.5cm} 
	\times \!\left[\!\frac{1}{\left(d_{\sigma}\right)^2}\!\left(\frac{\ve{\sigma} \cdot \ve{x}_1}{x_1} - \frac{\ve{\sigma} \cdot \ve{x}_0}{x_0}\right) 
	- \left(\frac{\ve{\sigma} \cdot \ve{x}_1}{\left(x_1\right)^3} - \frac{\ve{\sigma} \cdot \ve{x}_0}{\left(x_0\right)^3}\right)\!\right]\!\sigma^a \sigma^b 
	\nonumber\\
	&& + \frac{G\,\hat{M}_{ab}}{c^2} 
	\nonumber\\
        && \hspace{-1.5cm} 
        \times \!\left[\!\frac{2}{\left(d_{\sigma}\right)^2}\!\left(\frac{\ve{\sigma} \cdot \ve{x}_1}{x_1} - \frac{\ve{\sigma} \cdot \ve{x}_0}{x_0}\right) 
        + \left(\frac{\ve{\sigma} \cdot \ve{x}_1}{\left(x_1\right)^3} - \frac{\ve{\sigma} \cdot \ve{x}_0}{\left(x_0\right)^3}\right) 
        \!\right]\!\frac{d_{\sigma}^a d_{\sigma}^b}{\left(d_{\sigma}\right)^2} 
        \nonumber\\ 
	&& - \frac{G\,\hat{M}_{ab}}{c^2} \left[\frac{2}{\left(x_1\right)^3} - \frac{2}{\left(x_0\right)^3}\right] \sigma^a\,d_{\sigma}^b\;,  
\label{Appendix_Shapiro_Mass_Quadrupole_20A} 
\end{eqnarray}

\noindent
where $\hat{M}_{ab}\,\delta_{ab} = 0$ has been used. In order to determine the upper limit of (\ref{Appendix_Shapiro_Mass_Quadrupole_20A}) the mass-quadrupole for 
an axisymmetric body (\ref{M_ab}) is inserted, which yields (cf. Eq.~(46) in \cite{Klioner1991}) 
\begin{eqnarray}
	\Delta c\tau_{\rm 1PN}^{M_{ab}} &=& + \frac{G\,M}{c^2}\,J_2 \left(\frac{P}{d_{\sigma}}\right)^2 
        \nonumber\\ 
        && \hspace{-1.5cm} 
	\times \Bigg[\left(\frac{\ve{\sigma} \cdot \ve{x}_1}{x_1} - \frac{\ve{\sigma} \cdot \ve{x}_0}{x_0}\right) 
	\left(1 - \left(\ve{\sigma} \cdot \ve{e}_3\right)^2 - 2 \left(\frac{\ve{d}_{\sigma} \cdot \ve{e}_3}{d_{\sigma}}\right)^2\right) 
        \nonumber\\
	&& \hspace{-1.5cm}
        + \left(\frac{\ve{\sigma} \cdot \ve{x}_1}{x_1}\left(\frac{d_{\sigma}}{x_1}\right)^2 
	- \frac{\ve{\sigma} \cdot \ve{x}_0}{x_0} \left(\frac{d_{\sigma}}{x_0}\right)^2\right) 
	\left(\ve{\sigma} \cdot \ve{e}_3\right)^2
        \nonumber\\
	&& \hspace{-1.5cm}
	- \left(\frac{\ve{\sigma} \cdot \ve{x}_1}{x_1}\left(\frac{d_{\sigma}}{x_1}\right)^2 
        - \frac{\ve{\sigma} \cdot \ve{x}_0}{x_0} \left(\frac{d_{\sigma}}{x_0}\right)^2\right) 
	\left(\frac{\ve{d}_{\sigma} \cdot \ve{e}_3}{d_{\sigma}}\right)^2 
        \nonumber\\
	&& \hspace{-1.5cm} 
	+ 2 \left(\left(\frac{d_{\sigma}}{x_1}\right)^3 - \left(\frac{d_{\sigma}}{x_0}\right)^3 \right)  
	\left(\ve{\sigma} \cdot \ve{e}_3\right)  \left(\frac{\ve{d}_{\sigma} \cdot \ve{e}_3}{d_{\sigma}}\right)\Bigg]\,,  
\label{Appendix_Shapiro_Mass_Quadrupole_20B} 
\end{eqnarray}

\noindent 
where $\ve{\sigma} \cdot \ve{e}_3 = \sigma^3$ and $\ve{d}_{\sigma} \cdot \ve{e}_3 = d_{\sigma}^3$ are the $x^3$-components of these vectors, 
because the symmetry axis of the body $\ve{e}_3$ is aligned with the $x^3$-axis of the coordinate system. 
In order to determine the upper limit of (\ref{Appendix_Shapiro_Mass_Quadrupole_20B}), 
the relations for the angles $\alpha_0 = \delta\left(\ve{\sigma}, \ve{x}_0\right)$ and $\alpha_1 = \delta\left(\ve{\sigma}, \ve{x}_1\right)$ are very useful:  
\begin{eqnarray}
	\cos \alpha_0 &=& \frac{\ve{\sigma} \cdot \ve{x}_0}{x_0} = \frac{\left(x_1\right)^2 - \left(x_0\right)^2 - R^2}{2 R x_0}\;, 
        \label{alpha0}
        \\
	\cos \alpha_1 &=& \frac{\ve{\sigma} \cdot \ve{x}_1}{x_1} = \frac{\left(x_1\right)^2 - \left(x_0\right)^2 + R^2}{2 R x_1}\;. 
        \label{alpha1}
\end{eqnarray}

\noindent
These relations can be shown by using (\ref{Tangent_Vector1}) and (\ref{k_sigma_N}) and they are valid up to terms of the order ${\cal O}\left(c^{-2}\right)$. 
Let us note that for the impact vectors one gets $d_{\sigma} = x_0\,\sin \alpha_0 = x_1\,\sin \alpha_1$.  
It is also meaningful to introduce a further variable
\begin{eqnarray}
        z &=& \frac{x_1}{x_0} \quad {\rm with} \quad 0 \le z \le \infty\;, 
        \label{parameter_z}
\end{eqnarray}

\noindent
as well as the angle
\begin{eqnarray}
	\alpha &=& \delta\left(\ve{x}_0,\ve{x}_1\right) \quad {\rm with} \quad 0 \le \alpha \le 2\,\pi\;.
        \label{parameter_beta0_beta1}
\end{eqnarray}

\noindent
Then one may rewrite (\ref{Appendix_Shapiro_Mass_Quadrupole_20B}) in terms of these two independent variables, $z$ and $\alpha$. 
By using the computer algebra system {\it Maple} \cite{Maple}, one obtains for the upper limit of the 1PN quadrupole term in the Shapiro time delay:
\begin{eqnarray}
	\left| \Delta \tau_{\rm 1PN}^{M_{ab}}\right| &\le& + \frac{11}{5}\,\frac{G M}{c^3} \left|J_2\right| \left(\frac{P}{d_{\sigma}}\right)^2 \,, 
\label{Appendix_Shapiro_Mass_Quadrupole_25} 
\end{eqnarray}

\noindent
which coincides with coefficient $A_2$ asserted by Eq.~(\ref{Coefficients_ML}). For a correct determination of the upper limit given by 
(\ref{Appendix_Shapiro_Mass_Quadrupole_25}) one has to take care about the fact that the three-vectors $\ve{\sigma}$ and $\ve{d}_{\sigma}$ 
are perpendicular to each other, which restricts their possible angles with rotational vector $\ve{e}_3$. 
That means, one may rotate the coordinate system such that $\ve{\sigma}$ is aligned with the $x$-axis and $\ve{d}_{\sigma}$ is aligned with the 
$y$-axis, while $\ve{e}_3 = \left(e_3^x, e_3^y, e_3^z\right)$ has three components now (see also endnote [99] in \cite{Zschocke_2PM_Metric}).
Taking into account that $\ve{e}_3$ is a unit-vector one obtains the upper limit asserted in (\ref{Appendix_Shapiro_Mass_Quadrupole_25}).

\section{The tensorial coefficients and scalar functions of the 1PN solution\label{Appendix1PN}}

The tensorial coefficients in Eqs.~(\ref{S_I_1PN}) and (\ref{S_I_2PN})  
are given by (cf. Eqs.~(52) - (57) in \cite{Zschocke_Quadrupole_1}) 
\begin{eqnarray}
	{\cal A}^i_{\left(3\right)}\left(\ve{x}_{\rm N}\right) &=& + 2\,\sigma^i \;,
        \label{coefficients_A3_N}
        \\
	{\cal B}^i_{\left(3\right)}\left(\ve{x}_{\rm N}\right) &=& - 2\,d_{\sigma}^{\,i} \;,
        \label{coefficients_B3_N}
        \\
        {\cal C}^{i\,ab}_{\left(5\right)}\left(\ve{x}_{\rm N}\right) &=& + \,6\,\sigma^a\,\delta^{bi}
        + 3\,\sigma^a\,\sigma^b\,\sigma^i \;,
        \label{coefficients_C5_N}
        \\
        {\cal C}^{i\,ab}_{\left(7\right)}\left(\ve{x}_{\rm N}\right) &=& - 15 (d_{\sigma})^2 \sigma^a \sigma^b \sigma^i
        + 15 \,d^{\,a}_{\sigma}\,d^{\,b}_{\sigma} \sigma^i
	\nonumber\\ 
	&& - 30 \,\sigma^a d_{\sigma}^{\,b}\,d_{\sigma}^{\,i} \;,
        \label{coefficients_C7_N}
        \\
	{\cal D}^{i\,ab}_{\left(5\right)}\left(\ve{x}_{\rm N}\right)
        &=& + 6\,d_{\sigma}^{\,a}\,\delta^{bi} - 15\,\sigma^a \sigma^b d_{\sigma}^i
        + 18\,\sigma^a d_{\sigma}^{\,b} \sigma^i ,
        \label{coefficients_D5_N}
        \\
        {\cal D}^{i\,ab}_{\left(7\right)}\left(\ve{x}_{\rm N}\right) &=& - 15\,d_{\sigma}^{\,a} \,d_{\sigma}^{\,b}\, d_{\sigma}^{\,i}
        + 15\, (d_{\sigma})^2\, \sigma^a \sigma^b d_{\sigma}^{\,i}
	\nonumber\\ 
	&& - 30 \, (d_{\sigma})^2 \,\sigma^a d_{\sigma}^{\,b} \,\sigma^i \,,
        \label{coefficients_D7_N}
\end{eqnarray}

\noindent
where (cf. Eqs.~(\ref{impact_vector_x0}) and (\ref{impact_vector_xN})) 
\begin{eqnarray}
	\ve{d}_{\sigma} &=& \ve{\sigma} \times \left(\ve{x}_0 \times \ve{\sigma} \right) 
	= \ve{\sigma} \times \left(\ve{x}_{\rm N}\left(t\right) \times \ve{\sigma} \right).
\end{eqnarray}

\noindent 
Actually, the tensorial coefficients in (\ref{coefficients_A3_N}) and (\ref{coefficients_C5_N}) do not depend on $\ve{x}_{\rm N}$ but 
only on $\ve{\sigma}$. Nevertheless, we will keep their arguments as is, in favor of a unique notation for these 
tensorial coefficients (\ref{coefficients_A3_N}) - (\ref{coefficients_D7_N}). 
We note that the tensorial coefficients ${\cal A}^i_{\left(3\right)}\left(\ve{x}_{\rm N}\right) = {\cal A}^i_{\left(3\right)}\left(\ve{x}_0\right) , \dots \,, 
{\cal D}^{i\,ab}_{\left(7\right)}\left(\ve{x}_{\rm N}\right) = {\cal D}^{i\,ab}_{\left(7\right)}\left(\ve{x}_0\right)$.

The scalar functions in Eq.~(\ref{S_I_2PN}) are given by (cf. Eqs.~(D29), (D31), (D33), (D35), (D37), (D39) in \cite{Zschocke_Quadrupole_1})
\begin{eqnarray}
        {\cal W}_{\left(3\right)}\left(t\right) &=& \ln \left(x_{\rm N} - \ve{\sigma} \cdot \ve{x}_{\rm N}\right),
        \label{function_W_3}
        \\
        {\cal W}_{\left(5\right)}\left(t\right) &=& - \frac{1}{3}\,
        \frac{1}{\left(d_{\sigma}\right)^2}\,\frac{\ve{\sigma} \cdot \ve{x}_{\rm N}}{x_{\rm N}}\;,  
        \label{function_W_5}
        \\
        {\cal W}_{\left(7\right)}\left(t\right) &=& - \frac{2}{15}\,\frac{1}{\left(d_{\sigma}\right)^2}  
        \left(\frac{\ve{\sigma} \cdot \ve{x}_{\rm N}}{x_{\rm N}}\,\frac{1}{\left(d_{\sigma}\right)^2} 
        + \frac{1}{2}\,\frac{\ve{\sigma} \cdot \ve{x}_{\rm N}}{\left(x_{\rm N}\right)^3} \right) ,
	\nonumber\\ 
        \label{function_W_7}
        \\
        {\cal X}_{\left(3\right)}\left(t\right) &=& \frac{1}{\left(d_{\sigma}\right)^2}  
         \left(x_{\rm N} + \ve{\sigma} \cdot \ve{x}_{\rm N} \right) , 
        \label{function_X_3}
        \\
        {\cal X}_{\left(5\right)}\left(t\right) &=& \frac{2}{3}\,\frac{1}{\left(d_{\sigma}\right)^2} 
        \left(\frac{x_{\rm N} + \ve{\sigma} \cdot \ve{x}_{\rm N}}{\left(d_{\sigma}\right)^2} - \frac{1}{2}\,\frac{1}{x_{\rm N}}\right) , 
        \label{function_X_5}
        \\
        {\cal X}_{\left(7\right)}\left(t\right) &=& \frac{8}{15} \frac{1}{\left(d_{\sigma}\right)^2}
	\nonumber\\ 
	&& \hspace{-1.0cm} \times 
	\left(\frac{x_{\rm N} + \ve{\sigma} \cdot \ve{x}_{\rm N}}{\left(d_{\sigma}\right)^4} - \frac{1}{2} \frac{1}{x_{\rm N}} \frac{1}{\left(d_{\sigma}\right)^2} 
        - \frac{1}{8} \frac{1}{\left(x_{\rm N}\right)^3}\right), 
        \label{function_X_7}
\end{eqnarray}

\noindent
where $\ve{x}_{\rm N} = \ve{x}_{\rm N}\left(t\right)$ and $x_{\rm N} = x_{\rm N}\left(t\right)$. 
One also needs the scalar functions ${\cal W}_{\left(3\right)}\left(t_0\right)\,,\, \dots \,,\, {\cal X}_{\left(7\right)}\left(t_0\right)$ 
which one obtains from (\ref{function_W_3}) - (\ref{function_X_7}) by replacing $\ve{x}_{\rm N}$ 
and $x_{\rm N}$ by $\ve{x}_0$ and $x_0$, respectively, because $x_{\rm N}\left(t_0\right) = \ve{x}_0$ 
and $x_{\rm N}\left(t_0\right) = \ve{x}_0$; note that $\ve{d}_{\sigma}$ is time-independent.

\section{Tensorial coefficients in (\ref{Second_Integration_2PN_A}) and (\ref{Second_Integration_2PN_B})}\label{Tensorial_Coefficients} 

It is convenient to introduce the impact vector, 
\begin{eqnarray}
	\ve{d} &=& \ve{\sigma} \times \left(\ve{x} \times \ve{\sigma}\right), 
	\label{impact_vector} 
\end{eqnarray}

\noindent 
where the spatial variable $\ve{x}$ can either be the unperturbed light ray $\ve{x}_{\rm N}$ in (\ref{S_I_N_New})  
or the light ray in 1PN approximation $\ve{x}_{\rm 1PN}$ in (\ref{S_I_1PN_New}); the spatial components of this impact vector are $d^i$. 

The tensorial coefficients of monopole-monopole term of the new representation of light trajectory 
in (\ref{Second_Integration_2PN_A}) and (\ref{Second_Integration_2PN_B}) are
\begin{eqnarray}
	U^{i}_{\left(1\right)}\left(\ve{x}\right) &=& \sigma^i\;.
        \label{coefficient_U1}
        \\
        U^{i}_{\left(2\right)}\left(\ve{x}\right) &=& d^i \;.
        \label{coefficient_U2}
\end{eqnarray}

\noindent
The tensorial coefficients of monopole-quadrupole term of the new representation of light trajectory 
in (\ref{Second_Integration_2PN_A}) and (\ref{Second_Integration_2PN_B}) are
\begin{eqnarray}
        V^{i\,ab}_{\left(1\right)}\left(\ve{x}\right) &=& \sigma^a \delta^{bi}\;.
        \label{coefficient_V1} 
        \\
        V^{i\,ab}_{\left(2\right)}\left(\ve{x}\right) &=& d^a \delta^{bi}\;.
        \label{coefficient_V2}
        \\
        V^{i\,ab}_{\left(3\right)}\left(\ve{x}\right) &=& \sigma^a \sigma^b \sigma^i\;.
        \label{coefficient_V3} 
        \\
        V^{i\,ab}_{\left(4\right)}\left(\ve{x}\right) &=& \sigma^a d^b \sigma^i\;.
        \label{coefficient_V4} 
        \\
        V^{i\,ab}_{\left(5\right)}\left(\ve{x}\right) &=& d^a d^b \sigma^i\;.
        \label{coefficient_V5} 
        \\
        V^{i\,ab}_{\left(6\right)}\left(\ve{x}\right) &=& d^a d^b d^i\;.
        \label{coefficient_V6} 
        \\
        V^{i\,ab}_{\left(7\right)}\left(\ve{x}\right) &=& \sigma^a \sigma^b d^i\;.
        \label{coefficient_V7} 
        \\
        V^{i\,ab}_{\left(8\right)}\left(\ve{x}\right) &=& \sigma^a d^b d^i\;.
        \label{coefficient_V8} 
\end{eqnarray}

\noindent
The tensorial coefficients of quadrupole-quadrupole term of the new representation of light trajectory in (\ref{Second_Integration_2PN_B}) are
\begin{eqnarray}
        W^{i\,abcd}_{\left(1\right)}\left(\ve{x}\right) &=& \delta^{ac} \sigma^b \delta^{di} \;.
        \label{coefficient_W1}
        \\
        W^{i\,abcd}_{\left(2\right)}\left(\ve{x}\right) &=& \delta^{ac} d^b \delta^{di} \;.
        \label{coefficient_W2}
        \\
        W^{i\,abcd}_{\left(3\right)}\left(\ve{x}\right) &=& \sigma^a \sigma^b \sigma^c \delta^{di} \;.
        \label{coefficient_W3}
        \\
        W^{i\,abcd}_{\left(4\right)}\left(\ve{x}\right) &=& \sigma^a \sigma^b d^c \delta^{di} \;.
        \label{coefficient_W4}
        \\
        W^{i\,abcd}_{\left(5\right)}\left(\ve{x}\right) &=& \sigma^a d^b \sigma^c \delta^{di} \;.
        \label{coefficient_W5}
        \\
        W^{i\,abcd}_{\left(6\right)}\left(\ve{x}\right) &=& \sigma^a d^b d^c \delta^{di} \;.
        \label{coefficient_W6}
        \\
        W^{i\,abcd}_{\left(7\right)}\left(\ve{x}\right) &=& d^a d^b \sigma^c \delta^{di} \;.
        \label{coefficient_W7}
        \\
        W^{i\,abcd}_{\left(8\right)}\left(\ve{x}\right) &=& d^a d^b d^c \delta^{di} \;.
        \label{coefficient_W8}
	\\
        W^{i\,abcd}_{\left(9\right)}\left(\ve{x}\right) &=& \delta^{ac} \delta^{bd} \sigma^i \;.
        \label{coefficient_W9} 
        \\
        W^{i\,abcd}_{\left(10\right)}\left(\ve{x}\right) &=& \delta^{ac} \sigma^b \sigma^d \sigma^i \;.
        \label{coefficient_W10} 
        \\
        W^{i\,abcd}_{\left(11\right)}\left(\ve{x}\right) &=& \delta^{ac} \sigma^b d^d \sigma^i \;.
        \label{coefficient_W11}
        \\
        W^{i\,abcd}_{\left(12\right)}\left(\ve{x}\right) &=& \delta^{ac} d^b d^d \sigma^i \;.
        \label{coefficient_W12}
        \\
        W^{i\,abcd}_{\left(13\right)}\left(\ve{x}\right) &=& \sigma^a \sigma^b \sigma^c \sigma^d \sigma^i \;.
        \label{coefficient_W13}
        \\
        W^{i\,abcd}_{\left(14\right)}\left(\ve{x}\right) &=& \sigma^a \sigma^b \sigma^c d^d \sigma^i \;.
        \label{coefficient_W14}
        \\ 
        W^{i\,abcd}_{\left(15\right)}\left(\ve{x}\right) &=& \sigma^a \sigma^b d^c d^d \sigma^i \;.
        \label{coefficient_W15}
        \\
        W^{i\,abcd}_{\left(16\right)}\left(\ve{x}\right) &=& \sigma^a d^b \sigma^c d^d \sigma^i \;.
        \label{coefficient_W16}
        \\ 
        W^{i\,abcd}_{\left(17\right)}\left(\ve{x}\right) &=& \sigma^a d^b d^c d^d \sigma^i \;.
        \label{coefficient_W17}
        \\
        W^{i\,abcd}_{\left(18\right)}\left(\ve{x}\right) &=& d^a d^b d^c d^d \sigma^i \;.
	\label{coefficient_W18}
	\\ 
        W^{i\,abcd}_{\left(19\right)}\left(\ve{x}\right) &=& \delta^{ac} \delta^{bd} d^i \;.
        \label{coefficient_W19} 
        \\
        W^{i\,abcd}_{\left(20\right)}\left(\ve{x}\right) &=& \delta^{ac} \sigma^b \sigma^d d^i \;.
        \label{coefficient_W20} 
        \\
        W^{i\,abcd}_{\left(21\right)}\left(\ve{x}\right) &=& \delta^{ac} \sigma^b d^d d^i \;.
        \label{coefficient_W21}
        \\
        W^{i\,abcd}_{\left(22\right)}\left(\ve{x}\right) &=& \delta^{ac} d^b d^d d^i \;.
        \label{coefficient_W22}
        \\
        W^{i\,abcd}_{\left(23\right)}\left(\ve{x}\right) &=& \sigma^a \sigma^b \sigma^c \sigma^d d^i \;.
        \label{coefficient_W23}
        \\
        W^{i\,abcd}_{\left(24\right)}\left(\ve{x}\right) &=& \sigma^a \sigma^b \sigma^c d^d d^i \;.
        \label{coefficient_W24}
        \\
        W^{i\,abcd}_{\left(25\right)}\left(\ve{x}\right) &=& \sigma^a \sigma^b d^c d^d d^i \;.
        \label{coefficient_W25}
        \\
        W^{i\,abcd}_{\left(26\right)}\left(\ve{x}\right) &=& \sigma^a d^b \sigma^c d^d d^i \;.
        \label{coefficient_W26}
        \\
        W^{i\,abcd}_{\left(27\right)}\left(\ve{x}\right) &=& \sigma^a d^b d^c d^d d^i \;.
        \label{coefficient_W27}
        \\
        W^{i\,abcd}_{\left(28\right)}\left(\ve{x}\right) &=& d^a d^b d^c d^d d^i \;.
        \label{coefficient_W28}
\end{eqnarray}

\section{Scalar functions in (\ref{Second_Integration_2PN_A}) and (\ref{Second_Integration_2PN_B})}\label{Scalar_Functions_2}

To simplify the notation, it is appropriate to introduce the following scalar functions, 
\begin{eqnarray}
        a_{\left(n\right)}\left(\ve{x}\right) &=& \left(x + \ve{\sigma} \cdot \ve{x}\right)^n \;,
        \label{a_n}
        \\
        b_{\left(n\right)}\left(\ve{x}\right) &=& \frac{1}{\left(x\right)^n}\;,
        \label{b_n}
        \\
        c_{\left(n\right)}\left(\ve{x}\right) &=& \frac{\ve{\sigma} \cdot \ve{x}}{\left(x\right)^n} \;,
        \label{c_n}
        \\
        d_{\left(1\right)}\left(\ve{x}\right) &=& \ln \left(x - \ve{\sigma} \cdot \ve{x} \right),
        \label{d_1}
        \\
        d_{\left(2\right)}\left(\ve{x}\right) &=& \arctan \frac{\ve{\sigma} \cdot \ve{x}}{d} + \frac{\pi}{2}\;,
        \label{d_2}
        \\
        d_{\left(3\right)}\left(\ve{x}\right) &=& \arctan \frac{\ve{\sigma} \cdot \ve{x}}{d}\;,
        \label{d_3}
        \\
        d_{\left(4\right)}\left(\ve{x}\right) &=& \frac{\ve{\sigma} \cdot \ve{x}}{d} \left(\arctan \frac{\ve{\sigma} \cdot \ve{x}}{d} + \frac{\pi}{2}\right). 
        \label{d_4}
\end{eqnarray}

\noindent
Then, the scalar functions in the new representation in (\ref{Second_Integration_2PN_A}) and (\ref{Second_Integration_2PN_B}) can be expressed in terms 
of these functions (\ref{a_n}) - (\ref{d_4}). 

The scalar functions of the monopole term of the new representation in (\ref{Second_Integration_2PN_A}) are given by
\begin{eqnarray}
	{F}_{\left(1\right)}\left(\ve{x}\right) &=& + 2\,d_{\left(1\right)}\,.
        \label{F_1}
        \\
	{F}_{\left(2\right)}\left(\ve{x}\right) &=& - 2\,\frac{a_{\left(1\right)}}{\left(d\right)^2} \,.
        \label{F_2}
        \end{eqnarray}

\noindent
The scalar functions of the quadrupole term of the new representation in (\ref{Second_Integration_2PN_A}) are given by
\begin{eqnarray}
	{G}_{\left(1\right)}\left(\ve{x}\right) &=& - 2\,\frac{c_{\left(1\right)}}{\left(d\right)^2}\,. 
        \label{G_1}
        \\
	{G}_{\left(2\right)}\left(\ve{x}\right) &=& + 4\,\frac{a_{\left(1\right)}}{\left(d\right)^4} 
	- 2\,\frac{b_{\left(1\right)}}{\left(d\right)^2}\,. 
        \label{G_2}
        \\
	{G}_{\left(3\right)}\left(\ve{x}\right) &=& + \frac{c_{\left(1\right)}}{\left(d\right)^2} + c_{\left(3\right)} \;.
        \label{G_3}
        \\
	{G}_{\left(4\right)}\left(\ve{x}\right) &=& - 4\,\frac{a_{\left(1\right)}}{\left(d\right)^4} 
	+ 2\,\frac{b_{\left(1\right)}}{\left(d\right)^2} + 2\,b_{\left(3\right)}\;.
	\label{G_4}
        \end{eqnarray}
        
\begin{eqnarray}
	{G}_{\left(5\right)}\left(\ve{x}\right) &=& - 2\,\frac{c_{\left(1\right)}}{\left(d\right)^4} 
	- \frac{c_{\left(3\right)}}{\left(d\right)^2}\,. 
        \label{G_5}
        \\
	{G}_{\left(6\right)}\left(\ve{x}\right) &=& - \frac{8}{\left(d\right)^6} a_{\left(1\right)} 
	+ 4\,\frac{b_{\left(1\right)}}{\left(d\right)^4} + \frac{b_{\left(3\right)}}{\left(d\right)^2}\,. 
        \label{G_6}
        \\
	{G}_{\left(7\right)}\left(\ve{x}\right) &=& - \frac{2}{\left(d\right)^4} a_{\left(1\right)} 
	+ \frac{b_{\left(1\right)}}{\left(d\right)^2} - b_{\left(3\right)}\,.
        \label{G_7}
        \\
	{G}_{\left(8\right)}\left(\ve{x}\right) &=& + 4\,\frac{c_{\left(1\right)}}{\left(d\right)^4} 
	+ 2\,\frac{c_{\left(3\right)}}{\left(d\right)^2}\,. 
        \label{G_8}
        \end{eqnarray}

\noindent 
The scalar functions of the monopole-monopole term of the new representation in (\ref{Second_Integration_2PN_B}) are given by
\begin{eqnarray}
	{X}_{\left(1\right)}\left(\ve{x}\right) &=& + 4\,\frac{a_{\left(1\right)}}{\left(d_{\sigma}\right)^2} 
	+ \frac{c_{\left(2\right)}}{4} - \frac{15}{4}\,\frac{d_{\left(3\right)}}{d}\,.
        \label{X_1}
        \\
        {X}_{\left(2\right)}\left(\ve{x}\right) &=& + 4\,\frac{a_{\left(2\right)}}{\left(d_{\sigma}\right)^4} 
	+ \frac{b_{\left(2\right)}}{4} 
	-  \frac{15}{4} \frac{d_{\left(4\right)}}{\left(d\right)^2}\,.  
        \label{X_2}
        \end{eqnarray}

\noindent 
These functions in combination with the coefficients (\ref{coefficient_U1}) and (\ref{coefficient_U2}) are in agreement with
Eq.~(51) in \cite{Article_Zschocke1}. 

The scalar functions of the monopole-quadrupole term of new representation in (\ref{Second_Integration_2PN_B}) are given by 
\begin{eqnarray}
	{Y}_{\left(1\right)}\left(\ve{x}\right) &=& + 12\,\frac{a_{\left(1\right)}}{\left(d\right)^4}
	- 4\,\frac{b_{\left(1\right)}}{\left(d\right)^2} 
	- \frac{93}{32}\,\frac{c_{\left(2\right)}}{\left(d\right)^2} 
	- \frac{7}{16}\,c_{\left(4\right)}
	\nonumber\\ 
	&& - \frac{285}{32}\,\frac{d_{\left(3\right)}}{\left(d\right)^3}\,. 
        \label{Y_1}
        \\
	{Y}_{\left(2\right)}\left(\ve{x}\right) &=& - 16\,\frac{a_{\left(2\right)}}{\left(d\right)^6} 
	- \frac{91}{32}\,\frac{b_{\left(2\right)}}{\left(d\right)^2} 
	- \frac{7}{16}\,b_{\left(4\right)}
	+ 4\,\frac{c_{\left(1\right)}}{\left(d\right)^4}  
	\nonumber\\ 
	&& + \frac{465}{32}\,\frac{d_{\left(4\right)}}{\left(d\right)^4}\,.  
        \label{Y_2}
        \\
	{Y}_{\left(3\right)}\left(\ve{x}\right) &=& - 8\,\frac{a_{\left(1\right)}}{\left(d\right)^4} 
	+ 2\,\frac{b_{\left(1\right)}}{\left(d\right)^2} 
	+ 2\,b_{\left(3\right)} 
	+ \frac{29}{64}\,\frac{c_{\left(2\right)}}{\left(d\right)^2} 
	\nonumber\\ 
	&& + \frac{111}{32}\,c_{\left(4\right)} 
	- \frac{5}{8} \left(d\right)^2 c_{\left(6\right)}  
	+ \frac{285}{64}\,\frac{d_{\left(3\right)}}{\left(d\right)^3}\, .
	\label{Y_3}
	\\ 
	{Y}_{\left(4\right)}\left(\ve{x}\right) &=& + 16\,\frac{a_{\left(2\right)}}{\left(d\right)^6} 
	+ \frac{155}{32}\,\frac{b_{\left(2\right)}}{\left(d\right)^2} 
	+ \frac{199}{16}\,b_{\left(4\right)} 
	- \frac{5}{4}\,\left(d\right)^2 b_{\left(6\right)} 
	\nonumber\\ 
	&& - 8\,\frac{c_{\left(1\right)}}{\left(d\right)^4} 
	- 4\,\frac{c_{\left(3\right)}}{\left(d\right)^2} 
	- \frac{465}{32}\,\frac{d_{\left(4\right)}}{\left(d\right)^4}\,.  
	\label{Y_4}
        \\ 	
	{Y}_{\left(5\right)}\left(\ve{x}\right) &=& + 8\,\frac{a_{\left(1\right)}}{\left(d\right)^6} 
	- 4\,\frac{b_{\left(1\right)}}{\left(d\right)^4} 
	- 2\,\frac{b_{\left(3\right)}}{\left(d\right)^2} 
	- \frac{209}{64}\,\frac{c_{\left(2\right)}}{\left(d\right)^4} 
	\nonumber\\ 
	&& - \frac{91}{32}\,\frac{c_{\left(4\right)}}{\left(d\right)^2} 
	+ \frac{5}{8}\,c_{\left(6\right)} 
	- \frac{465}{64}\,\frac{d_{\left(3\right)}}{\left(d\right)^5}\,. 
        \label{Y_5}
        \\
	{Y}_{\left(6\right)}\left(\ve{x}\right) &=& + 48\,\frac{a_{\left(2\right)}}{\left(d\right)^8} 
	+ \frac{263}{64}\,\frac{b_{\left(2\right)}}{\left(d\right)^4} 
	+ \frac{883}{32}\,\frac{b_{\left(4\right)}}{\left(d\right)^2} 
	+ \frac{5}{8}\,b_{\left(6\right)} 
	\nonumber\\ 
	&& - 16\,\frac{c_{\left(1\right)}}{\left(d\right)^6}\,
	- 4\,\frac{c_{\left(3\right)}}{\left(d\right)^4} 
	- \frac{2325}{64}\,\frac{d_{\left(4\right)}}{\left(d\right)^6}\,. 
        \label{Y_6}
	\end{eqnarray} 
	\\
	\begin{eqnarray} 
	{Y}_{\left(7\right)}\left(\ve{x}\right) &=& + 16\,\frac{a_{\left(2\right)}}{\left(d\right)^6} 
	+ \frac{235}{64}\,\frac{b_{\left(2\right)}}{\left(d\right)^2} 
	- \frac{71}{32}\,b_{\left(4\right)} 
	- \frac{5}{8} \left(d\right)^2 b_{\left(6\right)}  
	\nonumber\\ 
	&& + 4\,\frac{c_{\left(3\right)}}{\left(d\right)^2} 
	- \frac{855}{64}\,\frac{d_{\left(4\right)}}{\left(d\right)^4}\,.  
        \label{Y_7}
        \\
	{Y}_{\left(8\right)}\left(\ve{x}\right) &=& - 32\,\frac{a_{\left(1\right)}}{\left(d\right)^6} 
	+ 12\,\frac{b_{\left(1\right)}}{\left(d\right)^4} 
	+ 8\,\frac{b_{\left(3\right)}}{\left(d\right)^2} 
	+ \frac{81}{32}\,\frac{c_{\left(2\right)}}{\left(d\right)^4} 
	\nonumber\\ 
	&& + \frac{91}{16}\,\frac{c_{\left(4\right)}}{\left(d\right)^2} 
	+ \frac{5}{4}\,c_{\left(6\right)} 
	+ \frac{465}{32}\,\frac{d_{\left(3\right)}}{\left(d\right)^5}\,.
        \label{Y_8}
\end{eqnarray}

\noindent 
The scalar functions of the quadrupole-quadrupole term of the new representation in (\ref{Second_Integration_2PN_B}) are given by 
\begin{widetext} 
\begin{eqnarray}
	{Z}_{\left(1\right)}\left(\ve{x}\right) &=& + 8\,\frac{a_{\left(1\right)}}{\left(d\right)^6} 
	- 8\,\frac{b_{\left(1\right)}}{\left(d\right)^4}  
	- \frac{327}{128}\,\frac{c_{\left(2\right)}}{\left(d\right)^4}
	- \frac{7}{192}\,\frac{c_{\left(4\right)}}{\left(d\right)^2} 
	+ \frac{13}{48}\,c_{\left(6\right)} 
	+ \frac{185}{128} \frac{d_{\left(3\right)}}{\left(d\right)^5} \, . 
        \label{Z_1}
        \\
	{Z}_{\left(2\right)}\left(\ve{x}\right) &=& - 16\,\frac{a_{\left(2\right)}}{\left(d\right)^8} 
	- \frac{985}{384}\, \frac{b_{\left(2\right)}}{\left(d\right)^4} 
	- \frac{5}{192}\, \frac{b_{\left(4\right)}}{\left(d\right)^2} 
	+ \frac{13}{48}\,b_{\left(6\right)} 
	+ 8\,\frac{c_{\left(1\right)}}{\left(d\right)^6} 
	+ \frac{985}{128}\,\frac{d_{\left(4\right)}}{\left(d\right)^6} \,.  
        \label{Z_2}
        \\
	{Z}_{\left(3\right)}\left(\ve{x}\right) &=& + 4\,\frac{a_{\left(1\right)}}{\left(d\right)^6} 
	+ 4\,\frac{b_{\left(1\right)}}{\left(d\right)^4} 
	- \frac{2103}{512} \,\frac{c_{\left(2\right)}}{\left(d\right)^4} 
	+ \frac{451}{256} \,\frac{c_{\left(4\right)}}{\left(d\right)^2} 
	+ \frac{23}{64}\, c_{\left(6\right)} 
	+ \frac{9}{32} \left(d\right)^2 c_{\left(8\right)}  
	- \frac{5175}{512}\,\frac{d_{\left(3\right)}}{\left(d\right)^5} \,. 
        \label{Z_3}
        \\
	{Z}_{\left(4\right)}\left(\ve{x}\right) &=& - 16\,\frac{a_{\left(2\right)}}{\left(d\right)^8} 
	- \frac{27019}{1536}\,\frac{b_{\left(2\right)}}{\left(d\right)^4}
	+ \frac{1585}{768}\,\frac{b_{\left(4\right)}}{\left(d\right)^2} 
	+ \frac{5}{96}\,b_{\left(6\right)} 
	+ \frac{9}{32} \left(d\right)^2 b_{\left(8\right)} 
	+ 20\,\frac{c_{\left(1\right)}}{\left(d\right)^6} 
	- 8\,\frac{c_{\left(3\right)}}{\left(d\right)^4} 
	+ \frac{5515}{512}\,\frac{d_{\left(4\right)}}{\left(d\right)^6} \,.  
        \label{Z_4}
        \\
	{Z}_{\left(5\right)}\left(\ve{x}\right) &=& + 16\,\frac{a_{\left(2\right)}}{\left(d\right)^8} 
	- \frac{3859}{768}\,\frac{b_{\left(2\right)}}{\left(d\right)^4} 
        + \frac{1609}{384}\,\frac{b_{\left(4\right)}}{\left(d\right)^2} 
        + \frac{79}{96}\,b_{\left(6\right)} 
        + \frac{9}{16}\,\left(d\right)^2 b_{\left(8\right)} 
        - 8\,\frac{c_{\left(1\right)}}{\left(d\right)^6} 
        - \frac{2285}{256}\,\frac{d_{\left(4\right)}}{\left(d\right)^6}\,.  
        \label{Z_5}
        \\
	{Z}_{\left(6\right)}\left(\ve{x}\right) &=& - 16\,\frac{a_{\left(1\right)}}{\left(d\right)^8} 
	+ 24\,\frac{b_{\left(1\right)}}{\left(d\right)^6} 
	- 16\,\frac{b_{\left(3\right)}}{\left(d\right)^4} 
	+ \frac{6381}{256}\,\frac{c_{\left(2\right)}}{\left(d\right)^6} 
	- \frac{2323}{384}\,\frac{c_{\left(4\right)}}{\left(d\right)^4} 
	- \frac{119}{96}\,\frac{c_{\left(6\right)}}{\left(d\right)^2} 
	- \frac{9}{16}\,c_{\left(8\right)}  
	+ \frac{2285}{256}\,\frac{d_{\left(3\right)}}{\left(d\right)^7}\,.
        \label{Z_6}
        \\
	{Z}_{\left(7\right)}\left(\ve{x}\right) &=& + 16\,\frac{a_{\left(1\right)}}{\left(d\right)^8} 
	+ 16\,\frac{b_{\left(1\right)}}{\left(d\right)^6} 
	- \frac{1419}{512}\,\frac{c_{\left(2\right)}}{\left(d\right)^6} 
	- \frac{2443}{768}\,\frac{c_{\left(4\right)}}{\left(d\right)^4} 
	- \frac{143}{192}\,\frac{c_{\left(6\right)}}{\left(d\right)^2}  
	- \frac{9}{32}\,c_{\left(8\right)} 
	- \frac{5515}{512}\,\frac{d_{\left(3\right)}}{\left(d\right)^7}\,.
        \label{Z_7}
        \\
	{Z}_{\left(8\right)}\left(\ve{x}\right) &=& + \frac{4831}{512}\,\frac{b_{\left(2\right)}}{\left(d\right)^6} 
	-  \frac{877}{256}\,\frac{b_{\left(4\right)}}{\left(d\right)^4} 
	-  \frac{43}{64}\,\frac{b_{\left(6\right)}}{\left(d\right)^2} 
	-  \frac{9}{32}\,b_{\left(8\right)} 
	+ 8\,\frac{c_{\left(3\right)}}{\left(d\right)^6}
	- \frac{2205}{512}\,\frac{d_{\left(4\right)}}{\left(d\right)^8}\,.  
        \label{Z_8}
        \\
	{Z}_{\left(9\right)}\left(\ve{x}\right) &=& + \frac{1}{128}\,\frac{c_{\left(2\right)}}{\left(d\right)^4} 
	+ \frac{1}{192}\,\frac{c_{\left(4\right)}}{\left(d\right)^2} 
	+ \frac{5}{48}\,c_{\left(6\right)} 
	+ \frac{1}{128}\,\frac{d_{\left(3\right)}}{\left(d\right)^5}\,.
        \label{Z_9}
        \\
	{Z}_{\left(10\right)}\left(\ve{x}\right) &=& - 8\,\frac{a_{\left(1\right)}}{\left(d\right)^6} 
	+ 8\,\frac{b_{\left(1\right)}}{\left(d\right)^4} 
	+ \frac{839}{256}\,\frac{c_{\left(2\right)}}{\left(d\right)^4} 
	+ \frac{199}{384}\,\frac{c_{\left(4\right)}}{\left(d\right)^2}
	- \frac{85}{96}\,c_{\left(6\right)} 
	+ \frac{15}{16} \left(d\right)^2 c_{\left(8\right)} 
	- \frac{185}{256}\,\frac{d_{\left(3\right)}}{\left(d\right)^5}\,. 
	\label{Z_10} 
	\\
	{Z}_{\left(11\right)}\left(\ve{x}\right) &=& + 16\,\frac{a_{\left(2\right)}}{\left(d\right)^8} 
	+ \frac{2521}{384}\,\frac{b_{\left(2\right)}}{\left(d\right)^4} 
	+ \frac{197}{192}\,\frac{b_{\left(4\right)}}{\left(d\right)^2} 
	- \frac{85}{48}\,b_{\left(6\right)}
	+ \frac{15}{8} \left(d\right)^2 b_{\left(8\right)} 
	- 8\,\frac{c_{\left(1\right)}}{\left(d\right)^6} 
	- \frac{985}{128}\,\frac{d_{\left(4\right)}}{\left(d\right)^6}\,.  
        \label{Z_11}
        \\
	{Z}_{\left(12\right)}\left(\ve{x}\right) &=& - \frac{985}{256}\,\frac{c_{\left(2\right)}}{\left(d\right)^6}  
	- \frac{217}{384}\,\frac{c_{\left(4\right)}}{\left(d\right)^4} 
	- \frac{5}{96}\,\frac{c_{\left(6\right)}}{\left(d\right)^2}
	- \frac{15}{16}\,c_{\left(8\right)} 
	- \frac{985}{256}\,\frac{d_{\left(3\right)}}{\left(d\right)^7}\,. 
        \label{Z_12}
        \\
	{Z}_{\left(13\right)}\left(\ve{x}\right) &=& - 4\,\frac{a_{\left(1\right)}}{\left(d\right)^6}  
	- 4\,\frac{b_{\left(1\right)}}{\left(d\right)^4} 
	+ 14\,\frac{b_{\left(3\right)}}{\left(d\right)^2} 
	+ \frac{3237}{2048}\, \frac{c_{\left(2\right)}}{\left(d\right)^4} 
	- \frac{969}{1024}\, \frac{c_{\left(4\right)}}{\left(d\right)^2} 
	+ \frac{395}{256}\,c_{\left(6\right)} 
	- \frac{369}{128} \left(d\right)^2 c_{\left(8\right)} 
	+ \frac{15}{16}\left(d\right)^4 c_{\left(10\right)}  
	+ \frac{15525}{2048}\,\frac{d_{\left(3\right)}}{\left(d\right)^5}\,. 
	\nonumber\\ 
        \label{Z_13}
        \\
	{Z}_{\left(14\right)}\left(\ve{x}\right) &=& + \frac{8507}{512}\,\frac{b_{\left(2\right)}}{\left(d\right)^4}  
	-  \frac{1217}{256}\,\frac{b_{\left(4\right)}}{\left(d\right)^2} 
	+  \frac{393}{64}\,b_{\left(6\right)} 
	-  \frac{369}{32} \left(d\right)^2 b_{\left(8\right)} 
	+ \frac{15}{4} \left(d\right)^4 b_{\left(10\right)} 
	- 12\,\frac{c_{\left(1\right)}}{\left(d\right)^6} 
	+ 8\,\frac{c_{\left(3\right)}}{\left(d\right)^4} 
	- \frac{945}{512}\,\frac{d_{\left(4\right)}}{\left(d\right)^6}\,. 
	\label{Z_14}
	\\
	{Z}_{\left(15\right)}\left(\ve{x}\right) &=& - 16\,\frac{a_{\left(1\right)}}{\left(d\right)^8}  
	- \frac{2677}{1024}\, \frac{c_{\left(2\right)}}{\left(d\right)^6} 
	+ \frac{5515}{1536}\, \frac{c_{\left(4\right)}}{\left(d\right)^4} 
	+ \frac{335}{384}\, \frac{c_{\left(6\right)}}{\left(d\right)^2}  
	+ \frac{249}{64}\,c_{\left(8\right)}
	- \frac{15}{8} \left(d\right)^2 c_{\left(10\right)}
	+ \frac{5515}{1024}\,\frac{d_{\left(3\right)}}{\left(d\right)^7}\,. 
        \label{Z_15}
        \\
	{Z}_{\left(16\right)}\left(\ve{x}\right) &=& + 16\,\frac{a_{\left(1\right)}}{\left(d\right)^8} 
	- 24\,\frac{b_{\left(1\right)}}{\left(d\right)^6} 
	+ 16\,\frac{b_{\left(3\right)}}{\left(d\right)^4} 
	- \frac{10477}{512}\,\frac{c_{\left(2\right)}}{\left(d\right)^6} 
	+ \frac{5395}{768}\,\frac{c_{\left(4\right)}}{\left(d\right)^4} 
	+ \frac{311}{192}\,\frac{c_{\left(6\right)}}{\left(d\right)^2} 
	+ \frac{249}{32}\,c_{\left(8\right)} 
	- \frac{15}{4} \left(d\right)^2 c_{\left(10\right)}
        - \frac{2285}{512}\,\frac{d_{\left(3\right)}}{\left(d\right)^7}\,. 
	\nonumber\\ 
        \label{Z_16}
                \end{eqnarray}
        \end{widetext}

        \begin{widetext}
        \begin{eqnarray}
	{Z}_{\left(17\right)}\left(\ve{x}\right) &=& - \frac{7667}{512}\,\frac{b_{\left(2\right)}}{\left(d\right)^6} 
        + \frac{2153}{256}\,\frac{b_{\left(4\right)}}{\left(d\right)^4}
	+ \frac{143}{64}\,\frac{b_{\left(6\right)}}{\left(d\right)^2} 
	+ \frac{261}{32}\,b_{\left(8\right)} 
	- \frac{15}{4} \left(d\right)^2 b_{\left(10\right)} 
	- 8\,\frac{c_{\left(3\right)}}{\left(d\right)^6} 
	+ \frac{2205}{512}\,\frac{d_{\left(4\right)}}{\left(d\right)^8}\,.  
        \label{Z_17}
        \\
	{Z}_{\left(18\right)}\left(\ve{x}\right) &=& + \frac{2205}{2048}\,\frac{c_{\left(2\right)}}{\left(d\right)^8} 
	- \frac{3361}{1024}\,\frac{c_{\left(4\right)}}{\left(d\right)^6}  
	- \frac{365}{256}\,\frac{c_{\left(6\right)}}{\left(d\right)^4} 
	- \frac{129}{128}\,\frac{c_{\left(8\right)}}{\left(d\right)^2}  
	+ \frac{15}{16}\,c_{\left(10\right)} 
	+ \frac{2205}{2048}\,\frac{d_{\left(3\right)}}{\left(d\right)^9}\,. 
	\label{Z_18}
	\\ 
	{Z}_{\left(19\right)}\left(\ve{x}\right) &=& - \frac{5}{384}\,\frac{b_{\left(2\right)}}{\left(d\right)^4} 
        - \frac{1}{192}\,\frac{b_{\left(4\right)}}{\left(d\right)^2}
	+ \frac{5}{48}\,b_{\left(6\right)} 
	+ \frac{5}{128}\,\frac{d_{\left(4\right)}}{\left(d\right)^6} \,. 
        \label{Z_19}
        \\
	{Z}_{\left(20\right)}\left(\ve{x}\right) &=& - \frac{3997}{768}\,\frac{b_{\left(2\right)}}{\left(d\right)^4} 
	- \frac{569}{384}\,\frac{b_{\left(4\right)}}{\left(d\right)^2} 
	- \frac{95}{96}\,b_{\left(6\right)} 
	+ \frac{15}{16} \left(d\right)^2 b_{\left(8\right)} 
        + \frac{925}{256}\,\frac{d_{\left(4\right)}}{\left(d\right)^6}\,. 
        \label{Z_20}
        \\
	{Z}_{\left(21\right)}\left(\ve{x}\right) &=& - 48\,\frac{a_{\left(1\right)}}{\left(d\right)^8} 
	+ 40\,\frac{b_{\left(1\right)}}{\left(d\right)^6} 
	+ 8\,\frac{b_{\left(3\right)}}{\left(d\right)^4} 
	+ \frac{985}{128}\,\frac{c_{\left(2\right)}}{\left(d\right)^6} 
	+ \frac{601}{192}\,\frac{c_{\left(4\right)}}{\left(d\right)^4}  
	+ \frac{5}{48}\,\frac{c_{\left(6\right)}}{\left(d\right)^2}  
	- \frac{15}{8} c_{\left(8\right)} 
        + \frac{985}{128}\,\frac{d_{\left(3\right)}}{\left(d\right)^7}\,. 
        \label{Z_21}
        \\
	{Z}_{\left(22\right)}\left(\ve{x}\right) &=& + 64\,\frac{a_{\left(2\right)}}{\left(d\right)^{10}} 
	+ \frac{13039}{768}\,\frac{b_{\left(2\right)}}{\left(d\right)^6} 
	+ \frac{611}{384}\,\frac{b_{\left(4\right)}}{\left(d\right)^4} 
	+ \frac{5}{96}\,\frac{b_{\left(6\right)}}{\left(d\right)^2} 
	- \frac{15}{16}\,b_{\left(8\right)} 
	- \frac{48}{\left(d\right)^8}\,c_{\left(1\right)} 
	- \frac{6895}{256}\,\frac{d_{\left(4\right)}}{\left(d\right)^8}\,.  
        \label{Z_22}
        \\ 
	{Z}_{\left(23\right)}\left(\ve{x}\right) &=& + 12 \frac{a_{\left(2\right)}}{\left(d\right)^8} 
	+ \frac{31153}{2048} \frac{b_{\left(2\right)}}{\left(d\right)^4}
	- \frac{2371}{1024} \frac{b_{\left(4\right)}}{\left(d\right)^2}  
	- \frac{37}{256} b_{\left(6\right)} 
	- \frac{111}{128} \left(d\right)^2 b_{\left(8\right)}   
	+ \frac{15}{16} \left(d\right)^4 b_{\left(10\right)} 
	- 4 \frac{c_{\left(1\right)}}{\left(d\right)^6} 
	+ 8 \frac{c_{\left(3\right)}}{\left(d\right)^4} 
	- \frac{25875}{2048} \frac{d_{\left(4\right)}}{\left(d\right)^6} .
	\nonumber\\  
        \label{Z_23}
        \\
	{Z}_{\left(24\right)}\left(\ve{x}\right) &=& + 24\,\frac{a_{\left(1\right)}}{\left(d\right)^8} 
	- 36\,\frac{b_{\left(1\right)}}{\left(d\right)^6} 
	+ 16\,\frac{b_{\left(3\right)}}{\left(d\right)^4}
	- \frac{11343}{512}\,\frac{c_{\left(2\right)}}{\left(d\right)^6}  
	+ \frac{59}{256}\,\frac{c_{\left(4\right)}}{\left(d\right)^4} 
	- \frac{65}{64}\,\frac{c_{\left(6\right)}}{\left(d\right)^2} 
	- \frac{9}{32}\,c_{\left(8\right)} 
	   - \frac{15}{4} \left(d\right)^2 c_{\left(10\right)} 
	   + \frac{945}{512}\,\frac{d_{\left(3\right)}}{\left(d\right)^7}\,. 
	\nonumber\\  
        \label{Z_24}
        \\
	{Z}_{\left(25\right)}\left(\ve{x}\right) &=& + 64 \frac{a_{\left(2\right)}}{\left(d\right)^{10}} 
	+ \frac{93133}{3072} \frac{b_{\left(2\right)}}{\left(d\right)^6} 
	- \frac{11479}{1536} \frac{b_{\left(4\right)}}{\left(d\right)^4} 
	- \frac{433}{384} \frac{b_{\left(6\right)}}{\left(d\right)^2} 
	- \frac{9}{64} b_{\left(8\right)} 
	- \frac{15}{8} \left(d\right)^2 b_{\left(10\right)} 
	- 48 \frac{c_{\left(1\right)}}{\left(d\right)^8}
	+ 16 \frac{c_{\left(3\right)}}{\left(d\right)^6} 
	- \frac{19405}{1024} \frac{d_{\left(4\right)}}{\left(d\right)^8} .  
	\nonumber\\
        \label{Z_25}
        \\
	{Z}_{\left(26\right)}\left(\ve{x}\right) &=& - 64\,\frac{a_{\left(2\right)}}{\left(d\right)^{10}} 
        + \frac{5893}{1536}\,\frac{b_{\left(2\right)}}{\left(d\right)^6} 
        - \frac{5887}{768}\,\frac{b_{\left(4\right)}}{\left(d\right)^4} 
	- \frac{457}{192}\,\frac{b_{\left(6\right)}}{\left(d\right)^2} 
        - \frac{9}{32}\,b_{\left(8\right)} 
	- \frac{15}{4} \left(d\right)^2 b_{\left(10\right)} 
	+ 48\,\frac{c_{\left(1\right)}}{\left(d\right)^8} 
	+ \frac{6395}{512}\,\frac{d_{\left(4\right)}}{\left(d\right)^8} \,. 
        \label{Z_26}
        \\
	{Z}_{\left(27\right)}\left(\ve{x}\right) &=& - 48\,\frac{b_{\left(1\right)}}{\left(d\right)^8} 
	+ \frac{48}{\left(d\right)^6}\,b_{\left(3\right)} 
	- \frac{26781}{512}\,\frac{c_{\left(2\right)}}{\left(d\right)^8}  
	+ \frac{7457}{256}\,\frac{c_{\left(4\right)}}{\left(d\right)^6}  
	+ \frac{493}{64}\,\frac{c_{\left(6\right)}}{\left(d\right)^4}  
	+ \frac{129}{32}\,\frac{c_{\left(8\right)}}{\left(d\right)^2}  
	+ \frac{15}{4}\,c_{\left(10\right)}  
	- \frac{2205}{512}\,\frac{d_{\left(3\right)}}{\left(d\right)^9} \,.
        \label{Z_27}
        \\
	{Z}_{\left(28\right)}\left(\ve{x}\right) &=& - \frac{47575}{2048}\,\frac{b_{\left(2\right)}}{\left(d\right)^8} 
	+ \frac{10965}{1024}\,\frac{b_{\left(4\right)}}{\left(d\right)^6} 
	- \frac{445}{256}\,\frac{b_{\left(6\right)}}{\left(d\right)^4} 
	+ \frac{129}{128}\,\frac{b_{\left(8\right)}}{\left(d\right)^2} 
	+ \frac{15}{16}\,b_{\left(10\right)}
	- 24\,\frac{c_{\left(3\right)}}{\left(d\right)^8} 
	+ \frac{19845}{2048}\,\frac{d_{\left(4\right)}}{\left(d\right)^{10}}\,.  
        \label{Z_28}
\end{eqnarray}
\end{widetext} 
 
\section{Agreement of (\ref{S_I_1PN})-(\ref{S_I_2PN}) and (\ref{Second_Integration_2PN_A})-(\ref{Second_Integration_2PN_B})\label{Agreement_Second_Integration}}

In this Appendix some basic ideas are presented about how to get from the old representation (\ref{Second_Integration_Old}) with (\ref{S_I_1PN}) and
(\ref{S_I_2PN}), to the new representation (\ref{Second_Integration_New}) with ~(\ref{Second_Integration_2PN_A}) and (\ref{Second_Integration_2PN_B}).
For that demonstration one needs the following relations which are valid up to terms of the order ${\cal O}\left(c^{-4}\right)$:
\begin{eqnarray}
        \ve{x}_{\rm 1PN}\left(t\right) &=& \ve{x}_{\rm N}\left(t\right) + \Delta \ve{x}_{\rm 1PN}\left(t,t_0\right)\;,
        \label{Appendix_x_A}
        \\
        x_{\rm 1PN}\left(t\right) &=&
        x_{\rm N}\left(t\right) + \frac{\ve{x}_{\rm N}\left(t\right)  \cdot \Delta \ve{x}_{\rm 1PN}\left(t,t_0\right)}{x_{\rm N}\left(t\right) }\;,
        \label{Appendix_x_B}
        \\
        \frac{1}{\left(x_{\rm 1PN}\left(t\right)\right)^n} &=& \frac{1}{\left(x_{\rm N}\left(t\right)\right)^n}
        - \frac{n}{\left(x_{\rm N}\left(t\right)\right)^n} \,
        \frac{\ve{x}_{\rm N}\left(t\right) \cdot \Delta \ve{x}_{\rm 1PN}\left(t,t_0\right)}{\left(x_{\rm N}\left(t\right)\right)^2}\;.
\nonumber\\ 
	\label{Appendix_x_C}
\end{eqnarray}

\noindent
Let us notice here that 
\begin{eqnarray}
	\ve{x}_0 &=& \ve{x}_{\rm N}\left(t_0\right) = \ve{x}_{\rm 1PN}\left(t_0\right)
	\label{Replacement_4}  
\end{eqnarray}

\noindent
which follow from (\ref{Unperturbed_Lightray_2}) and (\ref{Introduction_4}). Furthermore, one encounters the following impact vector 
\begin{eqnarray}
	\hat{\ve{d}_{\sigma}} &=& \ve{\sigma} \times \left(\ve{x}_{\rm 1PN}\left(t\right) \times \ve{\sigma} \right)  
	\label{Appendix_impact_vector_x1PN}
\end{eqnarray}

\noindent
and its absolute value $\hat{d}_{\sigma} = | \hat{\ve{d}_{\sigma}} |$. This impact vector $\hat{\ve{d}_{\sigma}}$ in (\ref{Appendix_impact_vector_x1PN}) is 
related to the impact vector $\ve{d}_{\sigma}$ in (\ref{impact_vector_xN}) as follows (up to terms of the order ${\cal O}\left(c^{-4}\right)$): 
\begin{eqnarray}
        \hat{\ve{d}_{\sigma}} &=&
        \ve{d}_{\sigma} + \ve{\sigma} \times \left(\Delta \ve{x}_{\rm 1PN} \left(t,t_0\right)\times \ve{\sigma}\right),
        \label{Appendix_x_D}
        \\
        \hat{d}_{\sigma} &=&
        d_{\sigma} + \frac{\ve{d}_{\sigma} \cdot \Delta \ve{x}_{\rm 1PN}\left(t,t_0\right)}{d_{\sigma}}\;,
        \label{Appendix_x_E}
        \\
	\frac{1}{(\hat{d}_{\sigma})^n} &=&
        \frac{1}{\left(d_{\sigma}\right)^n} - \frac{n}{\left(d_{\sigma}\right)^n} \,
        \frac{\ve{d}_{\sigma} \cdot \Delta \ve{x}_{\rm 1PN}\left(t,t_0\right)}{\left(d_{\sigma}\right)^2}\;. 
        \label{Appendix_x_F}
\end{eqnarray}

\noindent
In relations (\ref{Appendix_x_A}) - (\ref{Appendix_x_C}) as well as (\ref{Appendix_x_D}) - (\ref{Appendix_x_F}) one needs the light ray perturbation 
in 1PN approximation, $\Delta \ve{x}_{\rm 1PN} \left(t,t_0\right)$, where it is advantageous to take  
Eqs.~(\ref{Appendix_Second_Integration_1PN}) and (\ref{Appendix_Second_Integration_1PN_A}). 

The entire procedure is separated into four steps: 

\noindent
{\bf First step:} The 1PN terms in Eq.~(\ref{S_I_1PN}) contain $6$ tensorial coefficients 
given in (\ref{coefficients_A3_N}) - (\ref{coefficients_D7_N}): 
\begin{eqnarray} 
	{\cal A}^i_{\left(3\right)}\left(\ve{x}_{\rm N}\right)\;,\; {\cal B}^i_{\left(3\right)}\left(\ve{x}_{\rm N}\right)\;,\;  
	{\cal C}^{i\,ab}_{\left(n\right)}\left(\ve{x}_{\rm N}\right)\;,\; {\cal D}^{i\,ab}_{\left(n\right)}\left(\ve{x}_{\rm N}\right).   
\label{Appendix_Tensors7}
\end{eqnarray}

\noindent 
These tensorial coefficients (\ref{Appendix_Tensors7}) consist of $10$ different tensors as given by (\ref{coefficient_U1}) - (\ref{coefficient_V8}) 
with the argument $\ve{x} = \ve{x}_{\rm N}$: 
\begin{eqnarray} 
	U^i_{\left(1\right)}\left(\ve{x}_{\rm N}\right)\;,\; U^i_{\left(2\right)}\left(\ve{x}_{\rm N}\right)\;,\;
	V^{i\,ab}_{\left(n\right)}\left(\ve{x}_{\rm N}\right).  
\label{Appendix_Tensors8}
\end{eqnarray} 

\noindent 
Accordingly, one may rewrite the 1PN terms in Eq.~(\ref{S_I_1PN}) in terms of these $10$ individual tensors: 
\begin{eqnarray} 
	\Delta {x}^{i}_{\rm 1PN}\left(t,t_0\right) &=&  \Delta {x}^{i}_{\rm 1PN}\left(t\right) - \Delta {x}^{i}_{\rm 1PN}\left(t_0\right) 
        \label{Appendix_Second_Integration_1PN}
\end{eqnarray}

\noindent
with
\begin{eqnarray} 
	\Delta {x}^{i}_{\rm 1PN}\left(t\right) &=& \frac{G M}{c^2} \sum\limits_{n=1}^2 \left({U}^i_{\left(n\right)}\,{F}_{\left(n\right)}\right)
        \left(\ve{x}_{\rm N}\right)  
\nonumber\\
	&& + \frac{G \hat{M}_{ab}}{c^2} \sum\limits_{n=1}^8 \left({V}^{i\,ab}_{\left(n\right)}\,{G}_{\left(n\right)}\right) 
        \left(\ve{x}_{\rm N}\right),    
        \label{Appendix_Second_Integration_1PN_A}
\end{eqnarray}

\noindent
where the scalar functions ${F}_{\left(n\right)}$ and ${G}_{\left(n\right)}$ are given by Eqs.~(\ref{F_1}) - (\ref{F_2}) and 
Eqs.~(\ref{G_1}) - (\ref{G_8}), respectively, where the argument $\ve{x} = \ve{x}_{\rm N}$. One may easily show that 
(\ref{Appendix_Second_Integration_1PN}) with (\ref{Appendix_Second_Integration_1PN_A}) is identical with (\ref{S_I_1PN}).  

\noindent
{\bf Second step:}
Similarly, the 2PN terms in Eq.~(\ref{S_I_2PN}) contain $51$ tensorial coefficients given by Eqs.~(E28) - (E39) 
and Eqs.~(E41) - (E65) as well as Eqs.~(E67) - (E87) in \cite{Zschocke_Quadrupole_1}: 
\begin{eqnarray} 
	&& {\cal E}^i_{\left(n\right)}\left(\ve{x}_{\rm N}\right), {\cal F}^i_{\left(n\right)}\left(\ve{x}_{\rm N}\right),  
	{\cal G}^i_{\left(5\right)}\left(\ve{x}_{\rm N}\right), {\cal H}^i_{\left(n\right)}\left(\ve{x}_{\rm N}\right),
	\nonumber\\ 
	&& {\cal K}^{i\,ab}_{\left(n\right)}\left(\ve{x}_{\rm N}\right), {\cal L}^{i\,ab}_{\left(n\right)}\left(\ve{x}_{\rm N}\right),   
        {\cal M}^{i\,ab}_{\left(n\right)}\left(\ve{x}_{\rm N}\right), {\cal N}^{i\,ab}_{\left(n\right)}\left(\ve{x}_{\rm N}\right),
	\nonumber\\ 
	&& {\cal P}^{i\,abcd}_{\left(n\right)}\left(\ve{x}_{\rm N}\right), {\cal Q}^{i\,abcd}_{\left(n\right)}\left(\ve{x}_{\rm N}\right).    
	\label{Appendix_Tensors5} 
\end{eqnarray}

\noindent
These $51$ tensorial coefficients (\ref{Appendix_Tensors5}) consist of $38$ different tensors
given by (\ref{coefficient_U1}) - (\ref{coefficient_V8}) and (\ref{coefficient_W1}) - (\ref{coefficient_W28}):
\begin{eqnarray} 
	\hspace{-1.0cm}  U^i_{\left(1\right)}\left(\ve{x}_{\rm N}\right),\; U^i_{\left(2\right)}\left(\ve{x}_{\rm N}\right),\; 
        V^{i\,ab}_{\left(n\right)}\left(\ve{x}_{\rm N}\right),\; W^{i\,abcd}_{\left(n\right)}\left(\ve{x}_{\rm N}\right).  
\label{Appendix_Tensors6}
\end{eqnarray}

\noindent
Accordingly, one may rewrite the 2PN terms in Eq.~(\ref{S_I_2PN}) in terms of these $38$ individual tensors:
\begin{eqnarray} 
        \Delta {x}^{i}_{\rm 2PN}\left(t,t_0\right) &=& \Delta {x}^{i}_{\rm 2PN}\left(t\right) - \Delta {x}^{i}_{\rm 2PN}\left(t_0\right) 
        \label{Appendix_Second_Integration_2PN}
\end{eqnarray}

\noindent
with
\begin{eqnarray}
	\Delta {x}^{i}_{\rm 2PN}\left(t\right) &=& 
	+ \frac{G^2 M^2}{c^4} \sum\limits_{n=1}^2 \left(U^i_{\left(n\right)}\,{\tilde{X}}_{\left(n\right)}\right)\left(\ve{x}_{\rm N}\right)
	\nonumber\\ 
	&& \hspace{-1.5cm} 
	+ \frac{G^2 M \hat{M}_{ab}}{c^4} \sum\limits_{n=1}^{8} \left(V^{i\,ab}_{\left(n\right)}\,{\tilde{Y}}_{\left(n\right)}\right)\left(\ve{x}_{\rm N}\right)
	\nonumber\\ 
	&& \hspace{-1.5cm} 
	+ \frac{G^2 \hat{M}_{ab} \hat{M}_{cd}}{c^4} \sum\limits_{n=1}^{28} \left(W^{i\,abcd}_{\left(n\right)}\,{\tilde{Z}}_{\left(n\right)}\right) 
	\left(\ve{x}_{\rm N}\right). 
        \label{Appendix_Second_Integration_2PN_B}
\end{eqnarray}

\noindent
The scalar functions in (\ref{Appendix_Second_Integration_2PN_B}) can be deduced just by inserting these $51$ tensorial 
coefficients (\ref{Appendix_Tensors5}) into (\ref{S_I_2PN}) and then combining all those scalar terms belonging to one and the same tensorial coefficient 
in (\ref{Appendix_Tensors6}). However, these scalar functions are an intermediate step and will not be given in their explicit form here, in order 
to simplify the representation. It is noticed again that (\ref{Appendix_Second_Integration_2PN}) and (\ref{Appendix_Second_Integration_2PN_B}) 
is identical with (\ref{S_I_2PN}). 

The form of (\ref{Appendix_Second_Integration_2PN}) and (\ref{Appendix_Second_Integration_2PN_B}) resembles already the structure of 
(\ref{Second_Integration_2PN_A}) and (\ref{Second_Integration_2PN_B}), respectively. However, the arguments in (\ref{Appendix_Second_Integration_2PN}) are   
the unperturbed light rays, $\ve{x}_{\rm N}$, while in (\ref{Second_Integration_2PN_A}) the arguments are the light rays in 1PN approximation, 
$\ve{x}_{\rm 1PN}$. 
Furthermore, the scalar functions ${\tilde{X}}_{\left(n\right)}$, ${\tilde{Y}}_{\left(n\right)}$, ${\tilde{Z}}_{\left(n\right)}$   
in (\ref{Appendix_Second_Integration_2PN_B}) are not identical with the scalar functions ${X}_{\left(n\right)}$, ${Y}_{\left(n\right)}$, 
${Z}_{\left(n\right)}$ in (\ref{Second_Integration_2PN_B}). 
In order to arrive at (\ref{Second_Integration_2PN_A}) and (\ref{Second_Integration_2PN_B}) two further steps are necessary.

\noindent
{\bf Third step:}
In order to arrive at (\ref{Second_Integration_2PN_A}) and (\ref{Second_Integration_2PN_B}) the argument in the tensorial coefficients as well as in the 
scalar functions in (\ref{Appendix_Second_Integration_1PN_A}) have to be replaced by the light ray in 1PN approximation. Then one obtains:
\begin{eqnarray}
        \Delta {x}^{i}_{\rm 1PN}\left(t\right) &=&
        \frac{G M}{c^2} \sum\limits_{n=1}^2 \left({U}^i_{\left(n\right)}\,{F}_{\left(n\right)}\right)
	\left(\ve{x}_{\rm 1PN}\right) 
	\nonumber\\ 
	&& \hspace{-1.75cm} + \frac{G \hat{M}_{ab}}{c^2} \sum\limits_{n=1}^8 \left({V}^{i\,ab}_{\left(n\right)}\,{G}_{\left(n\right)}\right)
        \left(\ve{x}_{\rm 1PN}\right) + \delta {x}^{i}_{\rm 2PN}\,, 
        \label{Appendix_Second_Integration_1PN_C}
\end{eqnarray}

\noindent 
where $\delta {x}^{i}_{\rm 2PN}$ is just the difference (\ref{Appendix_Second_Integration_1PN_A}) minus (\ref{Appendix_Second_Integration_1PN_C}): 
\begin{eqnarray}
	&& \delta {x}^{i}_{\rm 2PN} = 
	\nonumber\\ 
	&& + \frac{G M}{c^2} \sum\limits_{n=1}^2 \bigg[ \left({U}^i_{\left(n\right)}\,{F}_{\left(n\right)}\right) \left(\ve{x}_{\rm N}\right)  
	- \left({U}^i_{\left(n\right)}\,{F}_{\left(n\right)}\right)\left(\ve{x}_{\rm 1PN}\right) \bigg] 
        \nonumber\\
	&& + \frac{G \hat{M}_{ab}}{c^2} \bigg[\sum\limits_{n=1}^8 \left({V}^{i\,ab}_{\left(n\right)}\,{G}_{\left(n\right)}\right) 
        \left(\ve{x}_{\rm N}\right)
        - \left({V}^{i\,ab}_{\left(n\right)}\,{G}_{\left(n\right)}\right)
	\left(\ve{x}_{\rm 1PN}\right) \bigg]. 
	\nonumber\\ 
	\label{Appendix_Difference_2PN_C}  
\end{eqnarray}

\noindent
Eq.~(\ref{Appendix_Second_Integration_1PN_C}) is identical with (\ref{Appendix_Second_Integration_1PN_A}).  

\noindent
{\bf Fourth step:} 
In order to determine the expression in (\ref{Appendix_Difference_2PN_C}), one has to perform a series expansion of those terms in 
(\ref{Appendix_Difference_2PN_C}) having as argument the light ray in 1PN approximation. For that calculation one needs the same relations as given 
previously by Eqs.~(\ref{Appendix_x_A}) - (\ref{Appendix_x_C}) and Eqs.~(\ref{Appendix_x_D}) - (\ref{Appendix_x_F}). 

The determination of $\delta {x}^{i}_{\rm 2PN}$ in (\ref{Appendix_Difference_2PN_C}) has been assisted by the 
computer algebra system {\it Maple} \cite{Maple}. One finally arrives at the following form,  
\begin{eqnarray}
        \delta {x}^{i}_{\rm 2PN} &=&
        + \frac{G^2 M^2}{c^4} \sum\limits_{n=1}^2 \left(U^i_{\left(n\right)}\,{\hat{X}}_{\left(n\right)}\right)\left(\ve{x}_{\rm N}\right)
	\nonumber\\ 
	&& \hspace{-1.0cm} + \frac{G^2 M \hat{M}_{ab}}{c^4} \sum\limits_{n=1}^{8} \left(V^{i\,ab}_{\left(n\right)}\,{\hat{Y}}_{\left(n\right)}\right)
	\left(\ve{x}_{\rm N}\right) 
	\nonumber\\ 
	&& \hspace{-1.0cm} 
	+ \frac{G^2 \hat{M}_{ab} \hat{M}_{cd}}{c^4} \sum\limits_{n=1}^{28} \left(W^{i\,abcd}_{\left(n\right)}\,{\hat{Z}}_{\left(n\right)}\right)
        \left(\ve{x}_{\rm N}\right),  
	\label{Appendix_Difference_2PN_D}
\end{eqnarray}

\noindent
which is separated into three terms proportional to monopole-monopole, monopole-quadrupole and quadrupole-quadrupole. 
The tensorial coefficients are defined by (\ref{coefficient_U1}) - (\ref{coefficient_U2}), (\ref{coefficient_V1}) - (\ref{coefficient_V8}), 
and (\ref{coefficient_W1}) - (\ref{coefficient_W28}), respectively. The scalar functions in (\ref{Appendix_Difference_2PN_D}) are an intermediate step   
and will not be given in their explicit form here, in favor of a clear representation.

The term $\delta {x}^{i}_{\rm 2PN}$, defined by Eq.~(\ref{Appendix_Difference_2PN_C}) and determined by Eq.~(\ref{Appendix_Difference_2PN_D}),  
is obviously of second post-Newtonian order and should, therefore, be added to (\ref{Appendix_Second_Integration_2PN_B}) rather than 
(\ref{Appendix_Second_Integration_1PN_C}). Accordingly, the sum of (\ref{Appendix_Second_Integration_2PN_B}) and (\ref{Appendix_Second_Integration_1PN_C})  
can be written in the form 
\begin{eqnarray} 
        \ve{x}_{\rm 2PN}\left(t\right) &=& \ve{x}_0 + c \left(t - t_0 \right) \ve{\sigma}
	+ \Delta \ve{x}_{\rm 1PN}\left(t\right) -  \Delta \ve{x}_{\rm 1PN}\left(t_0\right) 
	\nonumber\\ 
	&& \hspace{-0.25cm} +\, \Delta \ve{x}_{\rm 2PN}\left(t\right) -  \Delta \ve{x}_{\rm 2PN}\left(t_0\right),  
        \label{Appendix_Second_Integration_New} 
        \\
        \Delta {x}^{i}_{\rm 1PN}\left(t\right) &=&
        \frac{G M}{c^2} \sum\limits_{n=1}^2 \left({U}^i_{\left(n\right)}\,{F}_{\left(n\right)}\right)
	\left(\ve{x}_{\rm 1PN}\right) 
	\nonumber\\ 
	&& \hspace{-0.25cm} + \frac{G \hat{M}_{ab}}{c^2} \sum\limits_{n=1}^8 \left({V}^{i\,ab}_{\left(n\right)}\,{G}_{\left(n\right)}\right)
        \left(\ve{x}_{\rm 1PN}\right), 
	\label{Appendix_Second_Integration_New_A}
        \\
        \Delta {x}^{i}_{\rm 2PN}\left(t\right) &=&
        \frac{G^2 M^2}{c^4} \sum\limits_{n=1}^2 \left(U^i_{\left(n\right)}\,{X}_{\left(n\right)}\right)\left(\ve{x}_{\rm N}\right)
	\nonumber\\ 
	&& \hspace{-0.25cm} + \frac{G^2 M \hat{M}_{ab}}{c^4} \sum\limits_{n=1}^{8} \left(V^{i\,ab}_{\left(n\right)}\,{Y}_{\left(n\right)}\right)
        \left(\ve{x}_{\rm N}\right)
	\nonumber\\ 
	&& \hspace{-0.25cm} + \frac{G^2 \hat{M}_{ab} \hat{M}_{cd}}{c^4} \sum\limits_{n=1}^{28} \left(W^{i\,abcd}_{\left(n\right)}\,{Z}_{\left(n\right)}\right)
        \left(\ve{x}_{\rm N}\right), 
	\label{Appendix_Second_Integration_New_B}
\end{eqnarray}

\noindent 
where, by taking account of (\ref{Appendix_Second_Integration_2PN_B}) and(\ref{Appendix_Difference_2PN_D}), the new scalar functions  
\begin{eqnarray}
	{X}_{\left(n\right)} &=& {\tilde{X}}_{\left(n\right)} + {\hat{X}}_{\left(n\right)}\;, 
	\label{Appendix_new_scalar_functions_X_S}
	\\
        {Y}_{\left(n\right)} &=& {\tilde{Y}}_{\left(n\right)} + {\hat{Y}}_{\left(n\right)}\;,
	\label{Appendix_new_scalar_functions_Y_S}
	\\ 
        {Z}_{\left(n\right)} &=& {\tilde{Z}}_{\left(n\right)} + {\hat{Z}}_{\left(n\right)}\;,
        \label{Appendix_new_scalar_functions_Z_S}
\end{eqnarray}

\noindent 
have been introduced. The solution (\ref{Appendix_Second_Integration_New}) with (\ref{Appendix_Second_Integration_New_A}) 
and (\ref{Appendix_Second_Integration_New_B}) agrees with expression (\ref{Second_Integration_New}) with (\ref{Second_Integration_2PN_A}) 
and (\ref{Second_Integration_2PN_B}), where the scalar functions (\ref{Appendix_new_scalar_functions_X_S}) - (\ref{Appendix_new_scalar_functions_Z_S}) are 
given by Eqs.~(\ref{X_1}) - (\ref{Z_28}) in their explicit form.

\section{Calculation of $\ve{k} \cdot \Delta \ve{x}_{\rm 2PN}$ in terms of $\ve{k}$}\label{Term_Shapiro_C}

In this Appendix we consider the term
\begin{eqnarray}
        \ve{k} \cdot \Delta\ve{x}_{{\rm 2PN}}\left(\ve{x}_1,\ve{x}_0\right) &=& \ve{k} \cdot \Delta\ve{x}_{{\rm 2PN}}\left(\ve{x}_1\right) 
        - \ve{k} \cdot \Delta\ve{x}_{{\rm 2PN}}\left(\ve{x}_0\right)
	\nonumber\\ 
\label{Term_Shapiro_C_5}
\end{eqnarray}

\noindent
in Eq.~(\ref{Shapiro_2PN}) which needs fully to be expressed in terms of vector $\ve{k}$. The expression of $\Delta\ve{x}_{{\rm 2PN}}\left(\ve{x}\right)$
is given by Eq.~(\ref{Second_Integration_2PN_B}), hence one obtains  
\begin{eqnarray}
        \ve{k} \cdot \Delta\ve{x}_{{\rm 2PN}}\left(\ve{x}\right) &=&
        + \frac{G^2 M^2}{c^4} \sum\limits_{n=1}^2\!\left(k^i U^i_{\left(n\right)}\,{X}_{\left(n\right)}\!\right)\left(\ve{x}\right)
	\nonumber\\ 
	&& \hspace{-1.25cm} + \frac{G^2 M \hat{M}_{ab}}{c^4} \sum\limits_{n=1}^{8}\!\left(k^i V^{i\,ab}_{\left(n\right)} {Y}_{\left(n\right)}\right)
        \left(\ve{x}\right)
	\nonumber\\ 
	&& \hspace{-1.25cm} + \frac{G^2 \hat{M}_{ab} \hat{M}_{cd}}{c^4} \sum\limits_{n=1}^{28}\!\left(k^i W^{i\,abcd}_{\left(n\right)} {Z}_{\left(n\right)}\right)
        \left(\ve{x}\right).  
\label{Term_Shapiro_C_10}
\end{eqnarray}

\noindent
The tensorial coefficients in (\ref{coefficient_U1}) - (\ref{coefficient_W28}) as well as the scalar functions in (\ref{X_1}) - (\ref{Z_28}) are given in terms
of vector $\ve{\sigma}$ rather than vector $\ve{k}$. But in view of relation (\ref{k_sigma_N}) we have $\ve{\sigma} = \ve{k} + {\cal O}\left(c^{-2}\right)$. 
Thus, a replacement $\ve{\sigma}$ by $\ve{k}$ in the tensorial coefficients as well as in the scalar functions in (\ref{Term_Shapiro_C_10}) would cause an error of 
the order ${\cal O}\left(c^{-6}\right)$ in line with the 2PN approximation. The tensorial coefficients in (\ref{Term_Shapiro_C_10}) are contracted with $k^i$. 
For instance one obtains up to terms of the order ${\cal O}\left(c^{-2}\right)$: $k^i U^i_{\left(1\right)} = 1$, $k^i U^i_{\left(2\right)} = 0$, 
$k^i V^{i\,ab}_{\left(1\right)} = k^a\,k^b$ , $\dots$, $k^i W^{i\,abcd}_{\left(28\right)} = 0$. After performing these contractions one may distinguish 
the following tensors:
\begin{eqnarray}
        S_{\left(1\right)}^{ab} &=& k^a\,k^b ,\; S_{\left(2\right)}^{ab} = k^a\,d_k^b ,\; S_{\left(3\right)}^{ab} = d_k^a\,d_k^b ,
        \label{Tensor_S}
        \\
        T_{\left(1\right)}^{abcd} &=& \delta^{ac}\,k^b\,k^d,\;T_{\left(2\right)}^{abcd} = \delta^{ac}\,k^b\,d_k^d ,\; 
        T_{\left(3\right)}^{abcd} = k^a\,k^b\,k^c\,k^d,
	\nonumber\\ 
	T_{\left(4\right)}^{abcd} &=& k^a\,k^b\,k^c\,d_k^d , \; 
        T_{\left(5\right)}^{abcd} =  k^a\,d_k^b\,k^c\,d_k^d , 
        \nonumber\\
	T_{\left(6\right)}^{abcd} &=& k^a\,k^b\,d_k^c\,d_k^d ,\; 
	T_{\left(7\right)}^{abcd} = k^a\,d_k^b\,d_k^c\,d_k^d ,\; 
	\nonumber\\ 
	T_{\left(8\right)}^{abcd} &=& \delta^{ac}\,\delta^{bd} , \; 
        T_{\left(9\right)}^{abcd} = \delta^{ac}\,d_k^b\,d_k^d ,\; 
	T_{\left(10\right)}^{abcd} = d_k^a\,d_k^b\,d_k^c\,d_k^d , 
	\nonumber\\ 
        \label{Tensor_T}
\end{eqnarray}

\noindent
where the symmetries $a \leftrightarrow b$ and $c \leftrightarrow d$ as well as $a \leftrightarrow c \land b \leftrightarrow d$ and
$a \leftrightarrow d \land b \leftrightarrow c$ have been taken into account, according to the corresponding symmetries of the quadrupole tensors 
in front of the individual terms in (\ref{Term_Shapiro_C_10}). 
As mentioned, in the scalar functions (\ref{X_1}) - (\ref{Z_28}) one may replace $\ve{\sigma}$ by $\ve{k}$. Then, one obtains the following expression:
\begin{eqnarray}
        \ve{k} \cdot \Delta\ve{x}_{{\rm 2PN}}\left(\ve{x}_1,\ve{x}_0\right) &=& \frac{G^2 M^2}{c^4}\;u_{\left(1\right)}\left(\ve{x}_1,\ve{x}_0\right)  
	\nonumber\\ 
	&& \hspace{-2.25cm} + \frac{G^2 M \hat{M}_{ab}}{c^4} \sum \limits_{n=1}^{3} S_{\left(n\right)}^{ab}\;v_{\left(n\right)}\left(\ve{x}_1,\ve{x}_0\right) 
	\nonumber\\ 
	&& \hspace{-2.25cm} 
	+ \frac{G^2 \hat{M}_{ab}\,\hat{M}_{cd}}{c^4} \sum \limits_{n=1}^{10} T_{\left(n\right)}^{abcd}\;w_{\left(n\right)}\left(\ve{x}_1,\ve{x}_0\right) 
        \label{Appendix_Term_Shapiro_C_Final_Form}
\end{eqnarray}

\noindent
where the scalar functions are given by 
\begin{widetext} 
\begin{eqnarray}
	u_{\left(1\right)}\left(\ve{x}_1,\ve{x}_0\right) &=& + \frac{4}{\left(d_k\right)^2}\,e_{\left(1\right)} 
	+ \frac{g_{\left(2\right)}}{4}- \frac{15}{4} \frac{h_{\left(1\right)}}{d_k}\,.  
	\label{u_1}
	\\
	v_{\left(1\right)}\left(\ve{x}_1,\ve{x}_0\right) &=& + \frac{4}{\left(d_k\right)^4}\,e_{\left(1\right)}
	- \frac{2}{\left(d_k\right)^2}\,f_{\left(1\right)} + 2\,f_{\left(3\right)}  
	- \frac{157}{64}\,\frac{g_{\left(2\right)}}{\left(d_k\right)^2} 
	+ \frac{97}{32}\,g_{\left(4\right)}
	- \frac{5}{8} \left(d_k\right)^2 g_{\left(6\right)}
	- \frac{285}{64}\, \frac{h_{\left(1\right)}}{\left(d_k\right)^3}\,.  
	\label{v_1}
	\\
	v_{\left(2\right)}\left(\ve{x}_1,\ve{x}_0\right) &=& + \frac{2}{\left(d_k\right)^2}\,f_{\left(2\right)} + 12\,f_{\left(4\right)} 
	- \frac{5}{4}\left(d_k\right)^2\,f_{\left(6\right)} 
	- \frac{4}{\left(d_k\right)^4}\,g_{\left(1\right)} - \frac{4}{\left(d_k\right)^2}\,g_{\left(3\right)}\;. 
        \label{v_2}
        \\
	v_{\left(3\right)}\left(\ve{x}_1,\ve{x}_0\right) &=& + \frac{8}{\left(d_k\right)^6}\,e_{\left(1\right)} 
	- \frac{4}{\left(d_k\right)^4}\,f_{\left(1\right)} - \frac{2}{\left(d_k\right)^2}\,f_{\left(3\right)} 
	- \frac{209}{64} \frac{g_{\left(2\right)}}{\left(d_k\right)^4}
	- \frac{91}{32} \frac{g_{\left(4\right)}}{\left(d_k\right)^2}
	+ \frac{5}{8} \, g_{\left(6\right)} - \frac{465}{64} \frac{h_{\left(1\right)}}{\left(d_k\right)^5}\,.
        \label{v_3}
        \\
	w_{\left(1\right)}\left(\ve{x}_1,\ve{x}_0\right) &=& + \frac{185}{256}\,\frac{g_{\left(2\right)}}{\left(d_k\right)^4}
	+ \frac{185}{384}\,\frac{g_{\left(4\right)}}{\left(d_k\right)^2} - \frac{59}{96}\,g_{\left(6\right)} 
	+ \frac{15}{16} \left(d_k\right)^2 g_{\left(8\right)} + \frac{185}{256}\,\frac{h_{\left(1\right)}}{\left(d_k\right)^5}\,.  
        \label{w_1}
        \\
	w_{\left(2\right)}\left(\ve{x}_1,\ve{x}_0\right) &=& + \frac{4}{\left(d_k\right)^4}\,f_{\left(2\right)} 
	+ \frac{f_{\left(4\right)}}{\left(d_k\right)^2} - \frac{3}{2}\,f_{\left(6\right)} + \frac{15}{8} \left(d_k\right)^2 f_{\left(8\right)} \,. 
        \label{w_2}
        \\
	w_{\left(3\right)}\left(\ve{x}_1,\ve{x}_0\right) &=& + \frac{14}{\left(d_k\right)^2}\,f_{\left(3\right)} 
	- \frac{5175}{2048}\,\frac{g_{\left(2\right)}}{\left(d_k\right)^4}
	+ \frac{835}{1024}\,\frac{g_{\left(4\right)}}{\left(d_k\right)^4}
	+ \frac{487}{256}\,g_{\left(6\right)} 
	- \frac{333}{128} \left(d_k\right)^2 g_{\left(8\right)} 
	+ \frac{15}{16} \left(d_k\right)^4 g_{\left(10\right)}
	- \frac{5175}{2048}\,\frac{h_{\left(1\right)}}{\left(d_k\right)^5}\,.
	\nonumber\\ 
	\label{w_3} 
	\\ 
	w_{\left(4\right)}\left(\ve{x}_1,\ve{x}_0\right) &=& - \frac{6}{\left(d_k\right)^4}\,f_{\left(2\right)}  
	+ \frac{3}{2}\,\frac{f_{\left(4\right)}}{\left(d_k\right)^2}
	+ \frac{449}{64}\,f_{\left(6\right)} - \frac{171}{16}\left(d_k\right)^2 f_{\left(8\right)} 
	+ \frac{15}{4} \left(d_k\right)^4 f_{\left(10\right)}\,.
        \label{w_4}
        \\
	w_{\left(5\right)}\left(\ve{x}_1,\ve{x}_0\right) &=& + \frac{2285}{512}\,\frac{g_{\left(2\right)}}{\left(d_k\right)^6}
        + \frac{749}{768}\,\frac{g_{\left(4\right)}}{\left(d_k\right)^4} + \frac{73}{192}\,\frac{g_{\left(6\right)}}{\left(d_k\right)^2}
	+ \frac{231}{32}\,g_{\left(8\right)} - \frac{15}{4} \left(d_k\right)^2 g_{\left(10\right)} 
	+ \frac{2285}{512}\,\frac{h_{\left(1\right)}}{\left(d_k\right)^7}\,. 
        \label{w_5}
        \\
	w_{\left(6\right)}\left(\ve{x}_1,\ve{x}_0\right) &=& + \frac{16}{\left(d_k\right)^6}\,f_{\left(1\right)}
	- \frac{5515}{1024}\,\frac{g_{\left(2\right)}}{\left(d_k\right)^6} 
	+ \frac{629}{1536}\,\frac{g_{\left(4\right)}}{\left(d_k\right)^4} + \frac{49}{384}\,\frac{g_{\left(6\right)}}{\left(d_k\right)^2} 
	+ \frac{231}{64}\,g_{\left(8\right)} - \frac{15}{8}\left(d_k\right)^2 g_{\left(10\right)} 
	- \frac{5515}{1024} \frac{h_{\left(1\right)}}{\left(d_k\right)^7}\,.
        \label{w_6}
        \\
	w_{\left(7\right)}\left(\ve{x}_1,\ve{x}_0\right) &=& - \frac{709}{128}\,\frac{f_{\left(2\right)}}{\left(d_k\right)^6} 
	+ \frac{319}{64}\,\frac{f_{\left(4\right)}}{\left(d_k\right)^4} +  \frac{25}{16}\,\frac{f_{\left(6\right)}}{\left(d_k\right)^2} 
	+ \frac{63}{8}\,f_{\left(8\right)} - \frac{15}{4} \left(d_k\right)^2 f_{\left(10\right)}\,. 
        \label{w_7}
        \\
	w_{\left(8\right)}\left(\ve{x}_1,\ve{x}_0\right) &=& + \frac{1}{128}\,\frac{g_{\left(2\right)}}{\left(d_k\right)^4}
	+ \frac{1}{192}\,\frac{g_{\left(4\right)}}{\left(d_k\right)^2} + \frac{5}{48}\,g_{\left(6\right)} 
	+ \frac{1}{128}\,\frac{h_{\left(1\right)}}{\left(d_k\right)^5}\,.
        \label{w_8}
        \\
        w_{\left(9\right)}\left(\ve{x}_1,\ve{x}_0\right) &=& - \frac{985}{256}\,\frac{g_{\left(2\right)}}{\left(d_k\right)^6}
        - \frac{217}{384}\,\frac{g_{\left(4\right)}}{\left(d_k\right)^4} - \frac{5}{96}\,\frac{g_{\left(6\right)}}{\left(d_k\right)^2} 
	- \frac{15}{16}\,g_{\left(8\right)} - \frac{985}{256}\,\frac{h_{\left(1\right)}}{\left(d_k\right)^7}\,. 
        \label{w_9}
        \\
        w_{\left(10\right)}\left(\ve{x}_1,\ve{x}_0\right) &=& + \frac{2205}{2048}\,\frac{g_{\left(2\right)}}{\left(d_k\right)^8}
        - \frac{3361}{1024}\,\frac{g_{\left(4\right)}}{\left(d_k\right)^6} - \frac{365}{256}\,\frac{g_{\left(6\right)}}{\left(d_k\right)^4}
        - \frac{129}{128}\,\frac{g_{\left(8\right)}}{\left(d_k\right)^2} + \frac{15}{16}\,g_{\left(10\right)} 
	+ \frac{2205}{2048}\,\frac{h_{\left(1\right)}}{\left(d_k\right)^9}\,,  
	\nonumber\\ 
        \label{w_10}
\end{eqnarray}
\end{widetext} 

\noindent
where the abbreviations  
\begin{eqnarray}
	e_{\left(n\right)}\left(\ve{x}_1,\ve{x}_0\right) &=& \left(x_1 + \ve{k} \cdot \ve{x}_1\right)^n - \left(x_0 + \ve{k} \cdot \ve{x}_0\right)^n \;, 
	\label{e_n}
	\\ 
	f_{\left(n\right)}\left(\ve{x}_1,\ve{x}_0\right) &=& \frac{1}{\left(x_1\right)^n} - \frac{1}{\left(x_0\right)^n} \;, 
	\label{f_n}
	\\
	g_{\left(n\right)}\left(\ve{x}_1,\ve{x}_0\right) &=& \frac{\ve{k} \cdot \ve{x}_1}{\left(x_1\right)^n} - \frac{\ve{k} \cdot \ve{x}_0}{\left(x_0\right)^n} \;, 
        \label{g_n}
	\\
	h_{\left(1\right)}\left(\ve{x}_1,\ve{x}_0\right) &=& \arctan \frac{\ve{k} \cdot \ve{x}_1}{d_k} - \arctan \frac{\ve{k} \cdot \ve{x}_0}{d_k} \;, 
	\label{h_1}
	\\ 
	h_{\left(2\right)}\left(\ve{x}_1,\ve{x}_0\right) &=& 
	+ \frac{\ve{k} \cdot \ve{x}_1}{d_k} \left(\arctan \frac{\ve{k} \cdot \ve{x}_1}{d_k} + \frac{\pi}{2}\right) 
	\nonumber\\ 
	&& - \frac{\ve{k} \cdot \ve{x}_0}{d_k} \left(\arctan \frac{\ve{k} \cdot \ve{x}_0}{d_k} + \frac{\pi}{2}\right),  
	\label{h_2}
\end{eqnarray}

\noindent 
have been introduced.

\section{Calculation of $\ve{k} \cdot \Delta \ve{x}_{\rm 1PN}$ in terms of vector $\ve{k}$}\label{Term_Shapiro_A} 

In this Appendix we consider the term 
\begin{eqnarray}
	\ve{k} \cdot \Delta\ve{x}_{{\rm 1PN}}\left(\ve{x}_1,\ve{x}_0\right) &=& \ve{k} \cdot \Delta\ve{x}_{{\rm 1PN}}\left(\ve{x}_1\right) 
	- \ve{k} \cdot \Delta\ve{x}_{{\rm 1PN}}\left(\ve{x}_0\right)
	\nonumber\\ 
\label{Term_Shapiro_A_5}
\end{eqnarray}

\noindent 
in Eq.~(\ref{Shapiro_2PN}) which needs fully to be expressed in terms of vector $\ve{k}$. The expression of $\Delta\ve{x}_{{\rm 1PN}}\left(\ve{x}\right)$ 
is given by Eq.~(\ref{Second_Integration_2PN_A}). One obtains 
\begin{eqnarray}
        \ve{k} \cdot \Delta\ve{x}_{{\rm 1PN}}\left(\ve{x}\right) &=& 
        + \frac{G M}{c^2} \sum\limits_{n=1}^2 \left(k^i {U}^i_{\left(n\right)}\,{F}_{\left(n\right)}\right)
	\left(\ve{x}\right)
	\nonumber\\ 
	&& + \frac{G \hat{M}_{ab}}{c^2} \sum\limits_{n=1}^8 \left(k^i {V}^{i\,ab}_{\left(n\right)}\,{G}_{\left(n\right)}\right)
        \left(\ve{x}\right)  
\label{Term_Shapiro_10}
\end{eqnarray}

\noindent
where the spatial variable $\ve{x}$ can either be $\ve{x}_1$ or $\ve{x}_0$. The tensorial coefficients in Eqs.~(\ref{coefficient_U1}) - (\ref{coefficient_V8}) 
and the scalar functions in Eqs.~(\ref{F_1}) - (\ref{G_8}) are given in terms of vector $\ve{\sigma}$ and need to be expressed in terms of vector $\ve{k}$. 

The boundary value problem is defined by Eqs.~(\ref{boundary_0}) and (\ref{boundary_1}), that means the spatial position of the source, $\ve{x}_0$, and the 
spatial position of the observer, $\ve{x}_1$. Hence, in (\ref{Term_Shapiro_A_5}) - (\ref{Term_Shapiro_10}) one naturally encounters both impact vectors 
\begin{eqnarray}
	\ve{d}_{\sigma} &=& \ve{\sigma} \times \left(\ve{x}_0 \times \ve{\sigma} \right),
\label{appendix_impact_vector_sigma_0}
\\
	\hat{\ve{d}_{\sigma}} &=& \ve{\sigma} \times \left(\ve{x}_1 \times \ve{\sigma} \right). 
\label{appendix_impact_vector_sigma_1}
\end{eqnarray}

\noindent 
For the treatment of the boundary value problem a further impact vector in terms of $\ve{k}$ is needed, defined by 
\begin{eqnarray}
\ve{d}_k &=& \ve{k} \times \left(\ve{x}_0 \times \ve{k} \right) = \ve{k} \times \left(\ve{x}_1 \times \ve{k} \right). 
\label{appendix_impact_vector_k}
\end{eqnarray}

\noindent 
In order to rewrite (\ref{Term_Shapiro_A_5}) fully in terms of vector $\ve{k}$ one needs a relation between the impact vector  
(\ref{appendix_impact_vector_sigma_0}) and (\ref{appendix_impact_vector_k}) and between the impact vector 
(\ref{appendix_impact_vector_sigma_1}) and (\ref{appendix_impact_vector_k}). These relations can be obtained by 
inserting (\ref{k_sigma_1PN}) into Eqs.~(\ref{appendix_impact_vector_sigma_0}) and (\ref{appendix_impact_vector_sigma_1}):   
\begin{eqnarray}
	\ve{d}_{\sigma} &=& \ve{d}_k + \frac{\ve{d}_k \cdot \Delta\ve{x}_{\rm 1PN}}{R}\,\ve{k}  
	+ \frac{\ve{k} \cdot \ve{x}_0}{R}\,\ve{k} \times \left(\Delta\ve{x}_{\rm 1PN} \times \ve{k}\right), 
	\nonumber\\ 
\label{Term_Shapiro_15A}
\\
	\hat{\ve{d}_{\sigma}} &=& \ve{d}_k + \frac{\ve{d}_k \cdot \Delta\ve{x}_{\rm 1PN}}{R}\,\ve{k}
	+ \frac{\ve{k} \cdot \ve{x}_1}{R}\,\ve{k} \times \left(\Delta\ve{x}_{\rm 1PN} \times \ve{k}\right), 
	\nonumber\\ 
\label{Term_Shapiro_15B}
\end{eqnarray}

\noindent
where $\Delta\ve{x}_{\rm 1PN} = \Delta\ve{x}_{\rm 1PN}\left(\ve{x}_1, \ve{x}_0\right)$. These relations are valid up to terms of the 
order ${\cal O}\left(c^{-4}\right)$. The subsequent relations will be applied which are valid up to terms of the order ${\cal O}\left(c^{-4}\right)$: 
\begin{eqnarray}
	\ve{k} \cdot \ve{\sigma} &=& 1 \,,
\label{Term_Shapiro_20}
	\\
	\ve{k} \cdot \ve{d}_{\sigma} &=& \frac{\ve{d}_k \cdot \Delta\ve{x}_{\rm 1PN}}{R} \,, 
\label{Term_Shapiro_25A} 
        \\
		\ve{k} \cdot \hat{\ve{d}_{\sigma}} &=& \frac{\ve{d}_k \cdot \Delta\ve{x}_{\rm 1PN}}{R} \,,
\label{Term_Shapiro_25B}
        \\
	\frac{1}{\left(d_{\sigma}\right)^n} &=& \frac{1}{\left(d_k\right)^n} 
	- \frac{n}{R}\,\frac{\left(\ve{k} \cdot \ve{x}_0\right) \,\left(\ve{d}_k \cdot \Delta \ve{x}_{\rm 1PN}\right) }{\left(d_k\right)^{n+2}}\,,
\label{Term_Shapiro_30A}
         \\
	 \frac{1}{(\hat{d}_{\sigma})^n} &=& \frac{1}{\left(d_k\right)^n}
        - \frac{n}{R}\,\frac{\left(\ve{k} \cdot \ve{x}_1\right) \,\left(\ve{d}_k \cdot \Delta \ve{x}_{\rm 1PN}\right) }{\left(d_k\right)^{n+2}}\,,
\label{Term_Shapiro_30B}
         \\
        \ve{\sigma} \cdot \ve{x}_0 &=& \ve{k} \cdot \ve{x}_0 - \frac{\ve{d}_k \cdot \Delta \ve{x}_{\rm 1PN}}{R} \,, 
\label{Term_Shapiro_35A}
         \\
	 \ve{\sigma} \cdot \ve{x}_1 &=& \ve{k} \cdot \ve{x}_1 - \frac{\ve{d}_k \cdot \Delta \ve{x}_{\rm 1PN}}{R} \,,  
\label{Term_Shapiro_35B}
\end{eqnarray}

\noindent
where $\Delta\ve{x}_{\rm 1PN} = \Delta\ve{x}_{\rm 1PN}\left(\ve{x}_1, \ve{x}_0\right)$. 
These relations follow from (\ref{k_sigma_1PN}) and (\ref{Term_Shapiro_15A}) and (\ref{Term_Shapiro_15B}). Here it useful to notice that 
$\left(\ve{k} \times \ve{x}\right) \cdot \left(\ve{k} \times \Delta \ve{x}_{\rm 1PN}\right) = \ve{d}_k \cdot \Delta \ve{x}_{\rm 1PN}$.  

Using (\ref{k_sigma_1PN}) and (\ref{Term_Shapiro_15A}) - (\ref{Term_Shapiro_25B}) 
one obtains for the tensorial coefficients in (\ref{Term_Shapiro_10}) when expressed in terms of vector $\ve{k}$ the following expressions, 
which are valid up to terms of the order ${\cal O}\left(c^{-4}\right)$: 
\begin{eqnarray}
	k^i {U}^i_{\left(1\right)}\left(\ve{x}\right) &=& 1 \,,  
	\label{Term_U1}
	\\
	k^i {U}^i_{\left(2\right)}\left(\ve{x}\right) &=& \frac{1}{R} \left(\ve{d}_k \cdot \Delta \ve{x}_{\rm 1PN}\right), 
        \label{Term_U2}
	\\
	k^i {V}^{i\,ab}_{\left(1\right)}\left(\ve{x}\right) &=& k^a \,k^b 
	- \frac{k^b}{R}\,\Delta x^a_{\rm 1PN} + \frac{k^a k^b}{R}\,\left(\ve{k} \cdot \Delta \ve{x}_{\rm 1PN}\right), 
        \label{Term_V1}
	\\
	k^i {V}^{i\,ab}_{\left(2\right)}\left(\ve{x}\right) &=& d_k^a\,k^b + \frac{1}{R}  \left(\ve{k} \cdot \ve{x}\right)  \Delta x^a_{\rm 1PN}\,k^b 
	\nonumber\\ 
	&& \hspace{-1.5cm} + \frac{1}{R} \left(\ve{d}_k \cdot \Delta \ve{x}_{\rm 1PN}\right) k^a\,k^b
	- \frac{1}{R} \left(\ve{k} \cdot \ve{x}\right) \left(\ve{k} \cdot \Delta \ve{x}_{\rm 1PN}\right) k^a\,k^b \,,  
	\nonumber\\ 
        \label{Term_V2}
	\\
	k^i {V}^{i\,ab}_{\left(3\right)}\left(\ve{x}\right) &=& k^a \,k^b - \frac{2}{R}\,k^{(a}\,\Delta x^{b)}_{\rm 1PN} 
	+ 2\,\frac{k^a k^b}{R}\,\left(\ve{k} \cdot \Delta \ve{x}_{\rm 1PN}\right),
	\nonumber\\ 
        \label{Term_V3}
	\\
	k^i {V}^{i\,ab}_{\left(4\right)}\left(\ve{x}\right) &=& k^a\,d_k^b + \frac{1}{R} \left(\ve{k} \cdot \ve{x}\right) k^a\,\Delta x^b_{\rm 1PN}  
	\nonumber\\ 
	&& \hspace{-1.5cm} + \frac{1}{R} \left(\ve{d}_k \cdot \Delta \ve{x}_{\rm 1PN}\right) k^a\,k^b
	- \frac{1}{R} \left(\ve{k} \cdot \ve{x}\right) \left(\ve{k} \cdot \Delta \ve{x}_{\rm 1PN}\right) k^a\,k^b  
	\nonumber\\ 
	&& \hspace{-1.5cm} - \frac{1}{R}\,\Delta x^a_{\rm 1PN}\,d_k^b 
	+ \frac{1}{R}\,\left(\ve{k} \cdot \Delta\ve{x}_{\rm 1PN}\right) k^a\,d_k^b \,, 
        \label{Term_V4}
	\\
	k^i {V}^{i\,ab}_{\left(5\right)}\left(\ve{x}\right) &=& d_k^a\,d_k^b + \frac{2}{R} \left(\ve{k} \cdot \ve{x}\right) d_k^{(a}\,\Delta x^{b)}_{\rm 1PN} 
	\nonumber\\ 
	&& \hspace{-1.90cm} + \frac{2}{R} \left(\ve{d}_k \cdot \Delta \ve{x}_{\rm 1PN}\right) d_k^{(a}\,k^{b)} 
	- \frac{2}{R} \left(\ve{k} \cdot \ve{x}\right) \left(\ve{k} \cdot \Delta \ve{x}_{\rm 1PN}\right) d_k^{(a}\,k^{b)} \,,  
	\nonumber\\ 
        \label{Term_V5}
	\\
        k^i {V}^{i\,ab}_{\left(6\right)}\left(\ve{x}\right) &=& 
	\frac{d_k^a\,d_k^b}{R} \left(\ve{d}_k \cdot \Delta \ve{x}_{\rm 1PN}\right), 
        \label{Term_V6}
	\\
        k^i {V}^{i\,ab}_{\left(7\right)}\left(\ve{x}\right) &=& 
	\frac{k^a\,k^b}{R} \left(\ve{d}_k \cdot \Delta \ve{x}_{\rm 1PN}\right), 
        \label{Term_V7}
	\\
        k^i {V}^{i\,ab}_{\left(8\right)}\left(\ve{x}\right) &=& 
	\frac{k^a\,d_k^b}{R} \left(\ve{d}_k \cdot \Delta \ve{x}_{\rm 1PN}\right), 
        \label{Term_V8}
\end{eqnarray}

\noindent
where in (\ref{Term_U2}) - (\ref{Term_V8}) the abbreviation 
$\Delta \ve{x}_{\rm 1PN} = \Delta \ve{x}_{\rm 1PN}\left(\ve{x}_1 , \ve{x}_0\right)$ is used, and $A^{(a}\,B^{b)} = \left(A^a\,B^b + A^b\,B^a\right)/2$ 
denotes symmetrization. Similarly, using (\ref{Term_Shapiro_30A}) - (\ref{Term_Shapiro_35B}) one obtains for the scalar functions in (\ref{Term_Shapiro_10}) 
when expressed in terms of vector $\ve{k}$ the following expressions: 
\begin{widetext} 
\begin{eqnarray}
        {F}_{\left(1\right)}\left(\ve{x}\right) &=& + 2\,\ln \left(x - \ve{k} \cdot \ve{x}\right) 
	+ \frac{2}{\left(d_k\right)^2} \left(x + \ve{k} \cdot \ve{x}\right)
	\frac{\ve{d}_k \cdot \Delta \ve{x}_{\rm 1PN}}{R} + {\cal O}\left(c^{-4}\right), 
        \label{F_1_k}
        \\
        {F}_{\left(2\right)}\left(\ve{x}\right) &=& - \frac{2}{\left(d_k\right)^2} \left(x + \ve{k} \cdot \ve{x}\right) 
	+ {\cal O}\left(c^{-2}\right), 
        \label{F_2_k}
        \\ 
        {G}_{\left(1\right)}\left(\ve{x}\right) &=& - \frac{2}{\left(d_k\right)^2}\,\frac{\ve{k} \cdot \ve{x}}{x} 
	+ \frac{4}{\left(d_k\right)^4} \, \frac{\left(\ve{k} \cdot \ve{x}\right)^2 \left(\ve{d}_k \cdot \Delta \ve{x}_{\rm 1PN}\right)}{R\,x} 
	+ \frac{2}{\left(d_k\right)^2}\,\frac{\ve{d}_k \cdot  \Delta \ve{x}_{\rm 1PN}}{R\,x}
	+ {\cal O}\left(c^{-4}\right), 
        \label{G_1_k}
        \\
        {G}_{\left(2\right)}\left(\ve{x}\right) &=& + \frac{4}{\left(d_k\right)^4} \left(x + \ve{k} \cdot \ve{x}\right)
        - \frac{2}{\left(d_k\right)^2}\,\frac{1}{x} 
	- \frac{4}{\left(d_k\right)^4}\,\frac{\ve{d}_k \cdot \Delta \ve{x}_{\rm 1PN}}{R} 
	\left(1 - \frac{\ve{k} \cdot \ve{x}}{x} + 4\,\frac{\ve{k} \cdot \ve{x}}{\left(d_k\right)^2} \left(x + \ve{k} \cdot \ve{x}\right) \right) 
	+ {\cal O}\left(c^{-4}\right),
        \label{G_2_k}
        \\
	{G}_{\left(3\right)}\left(\ve{x}\right) &=& + \frac{1}{\left(d_k\right)^2} \frac{\ve{k} \cdot \ve{x}}{x} + \frac{\ve{k} \cdot \ve{x}}{\left(x\right)^3} 
	- \frac{\ve{d}_k \cdot \Delta \ve{x}_{\rm 1PN}}{R\,x} 
	\left(\frac{1}{\left(d_k\right)^2} + \frac{1}{\left(x\right)^2} + 2\,\frac{\left(\ve{k} \cdot \ve{x}\right)^2}{\left(d_k\right)^4}\right) 
	+ {\cal O}\left(c^{-4}\right),
        \label{G_3_k}
        \\
        {G}_{\left(4\right)}\left(\ve{x}\right) &=& - \frac{4}{\left(d_k\right)^4} \left(x + \ve{k} \cdot \ve{x}\right)
        + \frac{2}{\left(d_k\right)^2}\,\frac{1}{x} + \frac{2}{\left(x\right)^3}
	+ \frac{4}{\left(d_k\right)^4}\,\frac{\ve{d}_k \cdot \Delta \ve{x}_{\rm 1PN}}{R} 
        \left(1 - \frac{\ve{k} \cdot \ve{x}}{x} + 4\,\frac{\ve{k} \cdot \ve{x}}{\left(d_k\right)^2} \left(x + \ve{k} \cdot \ve{x}\right) \right) 
	+ {\cal O}\left(c^{-4}\right),
	\nonumber\\ 
        \label{G_4_k}
        \\
        {G}_{\left(5\right)}\left(\ve{x}\right) &=& - \frac{2}{\left(d_k\right)^4}\,\frac{\ve{k} \cdot \ve{x}}{x}
        - \frac{1}{\left(d_k\right)^2}\,\frac{\ve{k} \cdot \ve{x}}{\left(x\right)^3} 
	- \frac{1}{\left(d_k\right)^2}\,\frac{\ve{d}_k \cdot \Delta \ve{x}_{\rm 1PN}}{R\,x}
	\left(\frac{4}{\left(d_k\right)^2} + \frac{1}{\left(x\right)^2} - 8\,\frac{\left(x\right)^2}{\left(d_k\right)^4}\right)
	+ {\cal O}\left(c^{-4}\right), 
        \label{G_5_k}
        \\
        {G}_{\left(6\right)}\left(\ve{x}\right) &=& - \frac{8}{\left(d_k\right)^6}\left(x + \ve{k} \cdot \ve{x}\right)
        + \frac{4}{\left(d_k\right)^4}\,\frac{1}{x} + \frac{1}{\left(d_k\right)^2}\,\frac{1}{\left(x\right)^3} 
	+ {\cal O}\left(c^{-2}\right),  
        \label{G_6_k}
        \\
        {G}_{\left(7\right)}\left(\ve{x}\right) &=& - \frac{2}{\left(d_k\right)^4} \left(x + \ve{k} \cdot \ve{x} \right)
        + \frac{1}{\left(d_k\right)^2}\,\frac{1}{x} - \frac{1}{\left(x\right)^3} + {\cal O}\left(c^{-2}\right), 
        \label{G_7_k}
        \\
        {G}_{\left(8\right)}\left(\ve{x}\right) &=& + \frac{4}{\left(d_k\right)^4}\,\frac{\ve{k} \cdot \ve{x}}{x}
        + \frac{2}{\left(d_k\right)^2}\,\frac{ \ve{k} \cdot \ve{x}}{\left(x\right)^3} + {\cal O}\left(c^{-2}\right), 
        \label{G_8_k}
        \end{eqnarray}
\end{widetext} 

\noindent
where the functions in (\ref{F_2_k}) and (\ref{G_6_k}) - (\ref{G_8_k}) need to be calculated up to terms of the order ${\cal O}\left(c^{-2}\right)$ because their 
corresponding tensorial coefficients in (\ref{Term_U2}) and (\ref{Term_V6}) - (\ref{Term_V8}) contain only terms of the order ${\cal O}\left(c^{-2}\right)$. By 
inserting (\ref{Term_U1}) - (\ref{Term_V8}) and (\ref{F_1_k}) - (\ref{G_8_k}) into (\ref{Term_Shapiro_10}) one obtains: 
\begin{widetext} 
\begin{eqnarray}
	\ve{k} \cdot \Delta\ve{x}_{{\rm 1PN}}\left(\ve{x}\right) &=& 
	+ \frac{2\,G\,M}{c^2} \,\ln \left(x - \ve{k} \cdot \ve{x}\right) 
	\nonumber\\
	&& \hspace{-2.5cm} - \frac{2\,G \hat{M}_{ab}}{c^2}
        \left[
	\frac{1}{\left(d_k\right)^4}\,\frac{\ve{k}\cdot \ve{x}}{x} \, d_k^a\,d_k^b  
	+ \frac{1}{2}\,\frac{1}{\left(d_k\right)^2}\,\frac{\ve{k}\cdot \ve{x}}{\left(x\right)^3}\, d_k^a\,d_k^b 
	+  \frac{1}{2}\,\frac{1}{\left(d_k\right)^2}\,\frac{\ve{k}\cdot \ve{x}}{x}\, k^a k^b 
	- \frac{1}{2}\,\frac{\ve{k}\cdot \ve{x}}{\left(x\right)^3}\,k^a k^b 
	- \frac{1}{\left(x\right)^3}\,d_k^a\,k^b
        \right]
        \nonumber\\
	&& \hspace{-2.5cm} - \frac{2\,G \hat{M}_{ab}}{c^2} \, \frac{1}{\left(d_k\right)^4}\,\frac{\ve{k}\cdot \ve{x}}{x} 
        \left[
	\frac{4}{\left(d_k\right)^2} \left(\ve{d}_k \cdot \Delta \ve{x}_{\rm 1PN}\right) d_k^a\,d_k^b 
	+ \left(\ve{d}_k \cdot \Delta \ve{x}_{\rm 1PN}\right) k^a\,k^b 
	+ 2 \left(\ve{k} \cdot \Delta \ve{x}_{\rm 1PN}\right) d_k^a\,k^b 
	- 2\,d_k^a\,\Delta x^b_{\rm 1PN} 
        \right]
	+ {\cal O}\left(c^{-6}\right), 
	\nonumber\\ 
	\label{Appendix_Term_Shapiro_A}
\end{eqnarray}
\end{widetext} 

\noindent
where $\Delta \ve{x}_{\rm 1PN} = \Delta \ve{x}_{\rm 1PN}\left(\ve{x}_1, \ve{x}_0\right)$ and the spatial argument $\ve{x}$ in (\ref{Appendix_Term_Shapiro_A}) 
can either be $\ve{x}_1$ or $\ve{x}_0$. By inserting (\ref{Delta_1PN}) with (\ref{Second_Integration_2PN_A}) into (\ref{Appendix_Term_Shapiro_A}) and 
taking account of (\ref{Term_Shapiro_A_5}), one finally arrives at
\begin{widetext} 
\begin{eqnarray}
        \ve{k} \cdot \Delta\ve{x}_{{\rm 1PN}}\left(\ve{x}_1, \ve{x}_0\right) &=& 
        + \frac{G\,M}{c^2}\,P_{\left(1\right)}\left(\ve{x}_1,\ve{x}_0\right)  
        + \frac{G \hat{M}_{ab}}{c^2} \sum\limits_{n=1}^{3} S^{ab}_{\left(n\right)} \,Q_{\left(n\right)}\left(\ve{x}_1,\ve{x}_0\right)
        \nonumber\\
        && \hspace{-3.0cm} 
	+ \frac{G^2 M^2}{c^4} \, r_{\left(1\right)}\left(\ve{x}_1,\ve{x}_0\right)
        + \frac{G^2 M\,\hat{M}_{ab}}{c^4} \sum\limits_{n=1}^{3} S^{ab}_{\left(n\right)} \, s_{\left(n\right)}\left(\ve{x}_1,\ve{x}_0\right)
	+ \frac{G^2 \hat{M}_{ab} \hat{M}_{cd}}{c^4} \sum\limits_{n=1}^{10} T^{abcd}_{\left(n\right)} \, t_{\left(n\right)}\left(\ve{x}_1,\ve{x}_0\right) 
	+ {\cal O}\left(c^{-6}\right), 
        \label{Appendix_Term_Shapiro_A_Final_Form}
\end{eqnarray}
\end{widetext}

%

\noindent
where the tensors have been defined by Eqs.~(\ref{Tensor_S}) - (\ref{Tensor_T}) and the scalar functions are 
\begin{eqnarray}
	P_{\left(1\right)}\left(\ve{x}_1,\ve{x}_0\right) &=& + 2\,\ln \frac{x_1 - \ve{k} \cdot \ve{x}_1}{x_0 - \ve{k} \cdot \ve{x}_0} 
	= - 2\,\ln \frac{x_1 + \ve{k} \cdot \ve{x}_1}{x_0 + \ve{k} \cdot \ve{x}_0}\;. 
	\nonumber\\ 
	\label{P_1} 
	\\
	Q_{\left(1\right)}\left(\ve{x}_1,\ve{x}_0\right) &=& - \frac{g_{\left(1\right)}}{\left(d_k\right)^2} + g_{\left(3\right)}\;.
        \label{Q_1}
	\\
        Q_{\left(2\right)}\left(\ve{x}_1,\ve{x}_0\right) &=& + 2\,f_{\left(3\right)}\;.  
        \label{Q_2}
	\\
        Q_{\left(3\right)}\left(\ve{x}_1,\ve{x}_0\right) &=& - \frac{2}{\left(d_k\right)^4}\,g_{\left(1\right)} - \frac{g_{\left(3\right)}}{\left(d_k\right)^2}\;.
        \label{Q_3}
	\\
        r_{\left(1\right)}\left(\ve{x}_1,\ve{x}_0\right) &=& 0\;.
        \label{r_1}
	\\
        s_{\left(1\right)}\left(\ve{x}_1,\ve{x}_0\right) &=& + \frac{4}{\left(d_k\right)^4}\,e_{\left(1\right)}\,g_{\left(1\right)}\;. 
        \label{s_1}
	\\
        s_{\left(2\right)}\left(\ve{x}_1,\ve{x}_0\right) &=& 0 \;.
        \label{s_2}
	\\
        s_{\left(3\right)}\left(\ve{x}_1,\ve{x}_0\right) &=& + \frac{8}{\left(d_k\right)^6}\,e_{\left(1\right)}\,g_{\left(1\right)}\;.
        \label{s_3}
	\\
        t_{\left(1\right)}\left(\ve{x}_1,\ve{x}_0\right) &=& 0\;. 
        \label{t_1}
	\\
	t_{\left(2\right)}\left(\ve{x}_1,\ve{x}_0\right) &=& - \frac{8}{\left(d_k\right)^6}\,g_{\left(1\right)}\,g_{\left(1\right)}\;.
        \label{t_2}
	\\
        t_{\left(3\right)}\left(\ve{x}_1,\ve{x}_0\right) &=& + \frac{4}{\left(d_k\right)^6}\,e_{\left(1\right)}\,g_{\left(1\right)} 
	- \frac{2}{\left(d_k\right)^4}\,f_{\left(1\right)}\,g_{\left(1\right)} 
	\nonumber\\ 
	&& + \frac{2}{\left(d_k\right)^2}\,f_{\left(3\right)}\,g_{\left(1\right)}\;. 
	\label{t_3}
	\\
        t_{\left(4\right)}\left(\ve{x}_1,\ve{x}_0\right) &=& + \frac{4}{\left(d_k\right)^6}\,g_{\left(1\right)}\,g_{\left(1\right)} 
	- \frac{4}{\left(d_k\right)^4}\,g_{\left(1\right)}\,g_{\left(3\right)}\;. 
        \label{t_4}
	\\
        t_{\left(5\right)}\left(\ve{x}_1,\ve{x}_0\right) &=& - \frac{16}{\left(d_k\right)^8}\,e_{\left(1\right)}\,g_{\left(1\right)} 
	+ \frac{8}{\left(d_k\right)^6}\,f_{\left(1\right)}\,g_{\left(1\right)}\;.
        \label{t_5}
	\\
        t_{\left(6\right)}\left(\ve{x}_1,\ve{x}_0\right) &=& + \frac{16}{\left(d_k\right)^8}\,e_{\left(1\right)}\,g_{\left(1\right)} 
        - \frac{8}{\left(d_k\right)^6}\,f_{\left(1\right)}\,g_{\left(1\right)} 
	\nonumber\\ 
	&& - \frac{4}{\left(d_k\right)^4}\,f_{\left(3\right)}\,g_{\left(1\right)}\;. 
        \label{t_6}
	\\
        t_{\left(7\right)}\left(\ve{x}_1,\ve{x}_0\right) &=& - \frac{8}{\left(d_k\right)^6}\,g_{\left(1\right)}\,g_{\left(3\right)}\;. 
        \label{t_7}
	\\
        t_{\left(8\right)}\left(\ve{x}_1,\ve{x}_0\right) &=& 0\;. 
        \label{t_8}
	\\
        t_{\left(9\right)}\left(\ve{x}_1,\ve{x}_0\right) &=& + \frac{16}{\left(d_k\right)^8}\,e_{\left(1\right)}\,g_{\left(1\right)} 
	- \frac{8}{\left(d_k\right)^6}\,f_{\left(1\right)}\,g_{\left(1\right)}\;.
        \label{t_9}
	\\
        t_{\left(10\right)}\left(\ve{x}_1,\ve{x}_0\right) &=& - \frac{4}{\left(d_k\right)^6}\,f_{\left(3\right)}\,g_{\left(1\right)}\;.
        \label{t_10}
\end{eqnarray}

\noindent 
The scalar functions $e_{\left(n\right)}$, $f_{\left(n\right)}$, $g_{\left(n\right)}$, $h_{\left(n\right)}$ were introduced by Eqs.~(\ref{e_n}) - (\ref{h_2}).

\section{Calculation of $\left|\ve{k} \times \Delta \ve{x}_{\rm 1PN}\right|^2$}\label{Term_Shapiro_B}

In this Appendix we consider the term
\begin{eqnarray}
        \left| \ve{k} \times \Delta\ve{x}_{{\rm 1PN}}\left(\ve{x}_1,\ve{x}_0\right) \right|^2 &=& 
	\Delta\ve{x}_{{\rm 1PN}}\left(\ve{x}_1,\ve{x}_0\right) \cdot \Delta\ve{x}_{{\rm 1PN}}\left(\ve{x}_1,\ve{x}_0\right)
	\nonumber\\ 
	&& - \left(\ve{k} \cdot \Delta\ve{x}_{{\rm 1PN}}\left(\ve{x}_1,\ve{x}_0\right)\right)^2 
\label{Term_Shapiro_B_5}
\end{eqnarray}

\noindent
in Eq.~(\ref{Shapiro_2PN}). The calculation of (\ref{Term_Shapiro_B_5}) can considerably be simplified by omitting all terms proportional to vector $\ve{k}$ 
in $\Delta\ve{x}_{{\rm 1PN}}$. Then, by inspection of (\ref{Second_Integration_2PN_A}) one obtains 
\begin{eqnarray}
        \left| \ve{k} \times \Delta\ve{x}_{{\rm 1PN}}\left(\ve{x}_1,\ve{x}_0\right) \right|^2 &=&
	\frac{G^2 M^2}{c^4} \, x_{\left(1\right)}\left(\ve{x}_1,\ve{x}_0\right) 
	\nonumber\\ 
	&& \hspace{-2.5cm} + \frac{G^2 M\,\hat{M}_{ab}}{c^4} \sum\limits_{n=1}^{3} S^{ab}_{\left(n\right)} \, y_{\left(n\right)}\left(\ve{x}_1,\ve{x}_0\right)
	\nonumber\\ 
	&& \hspace{-2.5cm} 
	+ \frac{G^2 \hat{M}_{ab} \hat{M}_{cd}}{c^4} \sum\limits_{n=1}^{10} T^{abcd}_{\left(n\right)} \, z_{\left(n\right)}\left(\ve{x}_1,\ve{x}_0\right)
	\label{Appendix_Term_Shapiro_B_Final_Form}
\end{eqnarray}

\noindent
where the tensors have been defined by Eqs.~(\ref{Tensor_S}) - (\ref{Tensor_T}) and scalar functions are 
\begin{widetext} 
\begin{eqnarray}
	x_{\left(1\right)}\left(\ve{x}_1,\ve{x}_0\right) &=& + \frac{4}{\left(d_k\right)^2}\,e_{\left(1\right)}\,e_{\left(1\right)}\;. 
	\label{x_1}
	\\
        y_{\left(1\right)}\left(\ve{x}_1,\ve{x}_0\right) &=& + \frac{8}{\left(d_k\right)^4} e_{\left(1\right)} e_{\left(1\right)} 
	- \frac{4}{\left(d_k\right)^2} e_{\left(1\right)} f_{\left(1\right)} + 4 e_{\left(1\right)} f_{\left(3\right)}\,.
        \label{y_1}
        \\
        y_{\left(2\right)}\left(\ve{x}_1,\ve{x}_0\right) &=& - \frac{8}{\left(d_k\right)^4}\,e_{\left(1\right)}\,g_{\left(1\right)} 
	- \frac{8}{\left(d_k\right)^2}\,e_{\left(1\right)}\,g_{\left(3\right)}\;. 
        \label{y_2}
        \\
        y_{\left(3\right)}\left(\ve{x}_1,\ve{x}_0\right) &=& + \frac{16}{\left(d_k\right)^6}\,e_{\left(1\right)}\,e_{\left(1\right)} 
	- \frac{8}{\left(d_k\right)^4}\,e_{\left(1\right)}\,f_{\left(1\right)} 
	- \frac{4}{\left(d_k\right)^2}\,e_{\left(1\right)}\,f_{\left(3\right)}\;. 
        \label{y_3}
        \\
        z_{\left(1\right)}\left(\ve{x}_1,\ve{x}_0\right) &=& + \frac{4}{\left(d_k\right)^4}\,g_{\left(1\right)}\,g_{\left(1\right)}\;. 
        \label{z_1}
        \\
        z_{\left(2\right)}\left(\ve{x}_1,\ve{x}_0\right) &=& - \frac{16}{\left(d_k\right)^6}\,e_{\left(1\right)}\,g_{\left(1\right)} 
	+ \frac{8}{\left(d_k\right)^4}\,f_{\left(1\right)}\,g_{\left(1\right)}\;. 
        \label{z_2}
        \\
        z_{\left(3\right)}\left(\ve{x}_1,\ve{x}_0\right) &=& + \frac{4}{\left(d_k\right)^6}\,e_{\left(1\right)}\,e_{\left(1\right)} 
	- \frac{4}{\left(d_k\right)^4}\,g_{\left(1\right)}\,g_{\left(1\right)} 
	- \frac{4}{\left(d_k\right)^4}\,e_{\left(1\right)}\,f_{\left(1\right)} 
	+ \frac{1}{\left(d_k\right)^2}\,f_{\left(1\right)}\,f_{\left(1\right)} 
	+ \frac{4}{\left(d_k\right)^2}\,e_{\left(1\right)}\,f_{\left(3\right)} 
	- 2\,f_{\left(1\right)}\,f_{\left(3\right)} 
	\nonumber\\ 
	&& + \left(d_k\right)^2 f_{\left(3\right)}\,f_{\left(3\right)}\;.
        \label{z_3}
        \\
        z_{\left(4\right)}\left(\ve{x}_1,\ve{x}_0\right) &=& + \frac{8}{\left(d_k\right)^6} e_{\left(1\right)}\,g_{\left(1\right)} 
	- \frac{4}{\left(d_k\right)^4} f_{\left(1\right)}\,g_{\left(1\right)} - \frac{8}{\left(d_k\right)^4} e_{\left(1\right)}\,g_{\left(3\right)} 
	- \frac{4}{\left(d_k\right)^2} f_{\left(3\right)}\,g_{\left(1\right)} + \frac{4}{\left(d_k\right)^2} f_{\left(1\right)}\,g_{\left(3\right)} 
	- 4\,f_{\left(3\right)}\,g_{\left(3\right)} .
        \label{z_4}
        \\
        z_{\left(5\right)}\left(\ve{x}_1,\ve{x}_0\right) &=& - \frac{16}{\left(d_k\right)^8}\,e_{\left(1\right)}\,e_{\left(1\right)} 
        + \frac{16}{\left(d_k\right)^6}\,e_{\left(1\right)}\,f_{\left(1\right)} + \frac{8}{\left(d_k\right)^4}\,g_{\left(1\right)}\,g_{\left(3\right)} 
        - \frac{4}{\left(d_k\right)^4}\,f_{\left(1\right)}\,f_{\left(1\right)} + \frac{4}{\left(d_k\right)^2}\,g_{\left(3\right)}\,g_{\left(3\right)} \;.  
        \label{z_5}
        \\
        z_{\left(6\right)}\left(\ve{x}_1,\ve{x}_0\right) &=& + \frac{16}{\left(d_k\right)^8} e_{\left(1\right)}\,e_{\left(1\right)} 
        - \frac{16}{\left(d_k\right)^6} e_{\left(1\right)}\,f_{\left(1\right)} + \frac{4}{\left(d_k\right)^4} e_{\left(1\right)}\,f_{\left(3\right)} 
        + \frac{4}{\left(d_k\right)^4} f_{\left(1\right)}\,f_{\left(1\right)} - \frac{2}{\left(d_k\right)^2} f_{\left(1\right)}\,f_{\left(3\right)} 
        - 2\,f_{\left(3\right)}\,f_{\left(3\right)} . 
        \label{z_6}
        \\
        z_{\left(7\right)}\left(\ve{x}_1,\ve{x}_0\right) &=& - \frac{16}{\left(d_k\right)^6}\,e_{\left(1\right)}\,g_{\left(3\right)} 
	- \frac{8}{\left(d_k\right)^4}\,f_{\left(3\right)}\,g_{\left(1\right)} + \frac{8}{\left(d_k\right)^4}\,f_{\left(1\right)}\,g_{\left(3\right)}\;.
        \label{z_7}
        \\
        z_{\left(8\right)}\left(\ve{x}_1,\ve{x}_0\right) &=& 0\;. 
        \label{z_8}
        \\
        z_{\left(9\right)}\left(\ve{x}_1,\ve{x}_0\right) &=& + \frac{16}{\left(d_k\right)^8}\,e_{\left(1\right)}\,e_{\left(1\right)} 
        - \frac{16}{\left(d_k\right)^6}\,e_{\left(1\right)}\,f_{\left(1\right)} + \frac{4}{\left(d_k\right)^4}\,f_{\left(1\right)}\,f_{\left(1\right)}\;. 
	\label{z_9}
        \\
        z_{\left(10\right)}\left(\ve{x}_1,\ve{x}_0\right) &=& - \frac{8}{\left(d_k\right)^6}\,e_{\left(1\right)}\,f_{\left(3\right)} 
	+ \frac{4}{\left(d_k\right)^4}\,f_{\left(1\right)}\,f_{\left(3\right)} + \frac{1}{\left(d_k\right)^2}\,f_{\left(3\right)}\,f_{\left(3\right)}\;.
        \label{z_10}
\end{eqnarray}
\end{widetext} 

\noindent
The scalar functions $e_{\left(n\right)}$, $f_{\left(n\right)}$, $g_{\left(n\right)}$, $h_{\left(n\right)}$ are given by Eqs.~(\ref{e_n}) - (\ref{h_2}).

\section{Estimation of Shapiro time-delay}\label{Estimation_Shapiro}  

\subsection{The expression of the Shapiro time-delay} 

According to Eq.~(\ref{Shapiro_2PN}) the time delay of a light signal in the field of one body
at rest, where its monopole and quadrupole structure is taken into account, is given by 
\begin{eqnarray}
c \left(t_1 - t_0 \right) &=& R - \ve{k} \cdot \Delta\ve{x}_{{\rm 1PN}}\left(\ve{x}_1,\ve{x}_0\right) 
	- \ve{k} \cdot \Delta \ve{x}_{{\rm 2PN}}\left(\ve{x}_1,\ve{x}_0\right) 
	\nonumber\\ 
	&& + \,\frac{1}{2\,R} \left| \ve{k} \times \Delta \ve{x}_{{\rm 1PN}}\left(\ve{x}_1,\ve{x}_0\right)\right|^2 + {\cal O}\left(c^{-6}\right). 
\label{Appendix_Shapiro_2PN}
\end{eqnarray}

\noindent
The term $\ve{k} \cdot \Delta\ve{x}_{\rm 2PN}$ has been given by Eq.~(\ref{Appendix_Term_Shapiro_C_Final_Form}) in Appendix~\ref{Term_Shapiro_C}. 
The term $\ve{k} \cdot \Delta\ve{x}_{\rm 1PN}$ has been given by Eq.~(\ref{Appendix_Term_Shapiro_A_Final_Form}) in Appendix~\ref{Term_Shapiro_A}. 
The term $\left|\ve{k} \times \Delta\ve{x}_{\rm 1PN}\right|^2$ has been given by Eq.~(\ref{Appendix_Term_Shapiro_B_Final_Form}) in Appendix~\ref{Term_Shapiro_B}. 
According to these results the Shapiro time delay in 2PN approximation in the gravitational field of one body at rest with monopole and quadrupole structure is
given as follows (cf. Eq.~(\ref{Shapiro_2PN_Final})):
\begin{eqnarray}
	c \left(t_1 - t_0 \right) &=& R + \Delta c \tau^{M}_{\rm 1PN} + \Delta c \tau^{M_{ab}}_{\rm 1PN}
	\nonumber\\ 
	&& \hspace{-1.5cm} + \,\Delta c \tau^{M \times M}_{\rm 2PN} + \Delta c \tau^{M \times M_{ab}}_{\rm 2PN} + \Delta c \tau^{M_{ab} \times M_{cd}}_{\rm 2PN}\,,  
	\label{Appendix_Shapiro_2PN_Final}
\end{eqnarray}

\noindent 
up to terms of the order ${\cal O}\left(c^{-6}\right)$ and where the individual terms are 
\begin{eqnarray}
	\Delta c \tau^{M}_{\rm 1PN} &=& - \frac{G\,M}{c^2}\,P_{\left(1\right)}\left(\ve{x}_1,\ve{x}_0\right),  
\label{Appendix_Shapiro_2PN_Final_M}
	\\
	\Delta c \tau^{M_{ab}}_{\rm 1PN} &=& 
	- \frac{G \hat{M}_{ab}}{c^2} \sum\limits_{n=1}^{3} S^{ab}_{\left(n\right)} \,Q_{\left(n\right)}\left(\ve{x}_1,\ve{x}_0\right),
\label{Appendix_Shapiro_2PN_Final_Q}
       \end{eqnarray}

\begin{eqnarray} 
	\Delta c \tau^{M \times M}_{\rm 2PN} &=& + \frac{G^2 M^2}{c^4} \, R_{\left(1\right)}\left(\ve{x}_1,\ve{x}_0\right),
\label{Appendix_Shapiro_2PN_Final_MM}
	\\
	\Delta c \tau^{M \times M_{ab}}_{\rm 2PN} &=& 
	+ \frac{G^2 M\,\hat{M}_{ab}}{c^4} \sum\limits_{n=1}^{3} S^{ab}_{\left(n\right)} \, S_{\left(n\right)}\left(\ve{x}_1,\ve{x}_0\right),
\label{Appendix_Shapiro_2PN_Final_MQ}
	\\
	\Delta c \tau^{M_{ab} \times M_{cd}}_{\rm 2PN} &=& 
	+ \frac{G^2 \hat{M}_{ab} \hat{M}_{cd}}{c^4} \sum\limits_{n=1}^{10} T^{abcd}_{\left(n\right)} \, T_{\left(n\right)}\left(\ve{x}_1,\ve{x}_0\right).
	\nonumber\\ 
\label{Appendix_Shapiro_2PN_Final_QQ}
\end{eqnarray}

\noindent
The tensors $S^{ab}_{\left(n\right)}$ and $T_{\left(n\right)}^{abcd}$ are defined by Eqs.~(\ref{Tensor_S}) and (\ref{Tensor_T}) and the scalar functions
are introduced:
\begin{eqnarray}
   R_{\left(1\right)} &=& - r_{\left(1\right)} - u_{\left(1\right)} + \frac{1}{2\,R}\,x_{\left(1\right)} \;,
\label{Shapiro_2PN_Final_Function_R}
\\
   S_{\left(n\right)} &=& - s_{\left(n\right)} - v_{\left(n\right)} + \frac{1}{2\,R}\,y_{\left(n\right)} \;,
\label{Shapiro_2PN_Final_Function_S}
\\
   T_{\left(n\right)} &=& - t_{\left(n\right)} - w_{\left(n\right)} + \frac{1}{2\,R}\,z_{\left(n\right)} \;. 
\label{Shapiro_2PN_Final_Function_T}
\end{eqnarray}

\noindent
The functions in (\ref{Shapiro_2PN_Final_Function_R}) are defined by Eqs.~(\ref{r_1}) and (\ref{u_1}) and (\ref{x_1}). 
The functions in (\ref{Shapiro_2PN_Final_Function_S}) are defined by Eqs.~(\ref{s_1}) - (\ref{s_3}) and (\ref{v_1}) - (\ref{v_3}) and (\ref{y_1}) - (\ref{y_3}). 
The functions in (\ref{Shapiro_2PN_Final_Function_T}) are defined by Eqs.~(\ref{t_1}) - (\ref{t_10}) and (\ref{w_1}) - (\ref{w_10}) and (\ref{z_1}) - (\ref{z_10}). 
In these functions the abbreviations as given by Eqs.~(\ref{e_n}) - (\ref{h_2}) have been used. 

In this Appendix we will determine the upper limits of the individual terms in Shapiro time delay formula (\ref{Appendix_Shapiro_2PN_Final}). 
One may distinguish two scenarios of Shapiro time delay measurements: one-way and two-way scenario. 
In the one-way scenario a signal is emitted from the celestial object (e.g. spacecraft, pulsar) and received by the observer. 
In the two-way scenario a signal is emitted from the observer, then reflected off the celestial object (e.g. planet or spacecraft), 
and finally received back by the observer. If one assumes that the gravitating body as well as observer and celestial object are at rest, 
then both these scenarios just differ by a factor $2$. Here the upper limits are given for the one-way Shapiro effect.

\subsection{Estimation of 2PN monopole-monopole term}

The 2PN monopole-monopole term in (\ref{Appendix_Shapiro_2PN_Final}) reads
\begin{eqnarray}
	\Delta c \tau_{\rm 2PN}^{M \times M} &=& 
        \frac{G^2 M^2}{c^4}\,R_{\left(1\right)}\left(\ve{x}_1,\ve{x}_0\right),  
\label{Estimation_Shapiro_2PN_M_M_5}
\end{eqnarray}

\noindent
where the scalar function $R_{\left(1\right)}$ has been defined by Eq.~(\ref{Shapiro_2PN_Final_Function_R}). 
Eq.~(\ref{Estimation_Shapiro_2PN_M_M_5}) agrees with the 2PN term in Eq.~(3.2.51) in \cite{Brumberg1991} as well as Eq.~(69) in \cite{Article_Zschocke1}
(for PPN parameter the values of GR, $\gamma = 1$, must be chosen); note that $R\,d_k = \left|\ve{x}_0 \times \ve{x}_1\right|$. 
Inserting the abbreviations (\ref{r_1}) and (\ref{u_1}) and (\ref{x_1}) into (\ref{Shapiro_2PN_Final_Function_R}) one obtains for the 
function $R_{\left(1\right)}$: 
\begin{eqnarray}
        R_{\left(1\right)} &=& + \frac{2}{\left(d_k\right)^2} \frac{\left(x_1 - x_0\right)^2 - R^2}{R}   
	- \frac{1}{4} \left(\frac{\ve{k} \cdot \ve{x}_1}{\left(x_1\right)^2} - \frac{\ve{k} \cdot \ve{x}_0}{\left(x_0\right)^2} \right)
	\nonumber\\ 
	&& + \, \frac{15}{4}\, \frac{1}{d_k} \left(\arctan \frac{\ve{k} \cdot \ve{x}_1}{d_k} - \arctan \frac{\ve{k} \cdot \ve{x}_0}{d_k} \right).  
\label{appendix_R1A}
\end{eqnarray}

\noindent 
In order to determine the upper limit of (\ref{appendix_R1A}), the relations for the angle $\beta_0 = \delta\left(\ve{k},\ve{x}_0\right)$ 
and $\beta_1 = \delta\left(\ve{k},\ve{x}_1\right)$ are very useful: 
\begin{eqnarray}
	\cos \beta_0 &=& \frac{\ve{k} \cdot \ve{x}_0}{x_0} = \frac{\left(x_1\right)^2 - \left(x_0\right)^2 - R^2}{2 R x_0}\;, 
        \label{Relation_x0}
        \\ 
        \cos \beta_1 &=& \frac{\ve{k} \cdot \ve{x}_1}{x_1} = \frac{\left(x_1\right)^2 - \left(x_0\right)^2 + R^2}{2 R x_1}\;. 
        \label{Relation_x1}
\end{eqnarray}

\noindent
These relations are exactly valid and can be shown by using (\ref{Tangent_Vector1}).  
The impact parameters are $d_k = x_0\,\sin \beta_0 = x_1\,\sin \beta_1$. Then, the expression in (\ref{appendix_R1A}) can be 
rewritten in terms of variable $z$ in (\ref{parameter_z}) as well as angle $\alpha = \delta\left(\ve{x}_0,\ve{x}_1\right)$ in (\ref{parameter_beta0_beta1}). 
By using the computer algebra system {\it Maple} \cite{Maple} one obtains for the upper of (\ref{Estimation_Shapiro_2PN_M_M_5}) 
\begin{eqnarray} 
	\left| \Delta c \tau_{\rm 2PN}^{M \times M} \right| 
	&\le& \frac{8}{\left(d_k\right)^2}\,x_1\, \frac{G^2 M^2}{c^4}\;.
	\label{Estimation_Shapiro_2PN_M_M_15}
\end{eqnarray}

\noindent
Numerical values of (\ref{Estimation_Shapiro_2PN_M_M_15}) are presented in Table~\ref{Table3} for the Sun and giant planets. 
If one implements the inequality 
\begin{eqnarray}
	R\,\frac{x_1\,x_0}{\left(x_1 + x_0\right)^2} &\le& x_1 
\end{eqnarray}

\noindent 
into the first term on the r.h.s. of Eq.~(70) in \cite{Article_Zschocke1}, one verifies that the estimation in (\ref{Estimation_Shapiro_2PN_M_M_15}) is in 
agreement with that estimation in our article \cite{Article_Zschocke1}. Here we note that the term which was estimated by Eq.~(71) in \cite{Article_Zschocke1} has 
been absorbed in our upper limit given in (\ref{Estimation_Shapiro_2PN_M_M_15}).

\subsection{Estimation of 2PN monopole-quadrupole term}

The 2PN monopole-quadrupole term in (\ref{Appendix_Shapiro_2PN_Final}) reads
\begin{eqnarray}
	\hspace{-0.5cm}	\Delta c \tau_{\rm 2PN}^{M \times M_{ab}} = 
	\frac{G^2 M\,\hat{M}_{ab}}{c^4} \sum\limits_{n=1}^{3} S^{ab}_{\left(n\right)} \, S_{\left(n\right)}\left(\ve{x}_1,\ve{x}_0\right), 
\label{Estimation_Shapiro_2PN_M_Q_5}
\end{eqnarray}
 
\noindent
where the tensorial coefficients $S^{ab}_{\left(n\right)}$ are given by Eqs.~(\ref{Tensor_S}) and the scalar functions $S_{\left(n\right)}$ have been defined 
by Eq.~(\ref{Shapiro_2PN_Final_Function_S}). Their explicit form is obtained by inserting the abbreviations (\ref{e_n}) - (\ref{h_2}) into the scalar functions 
$s_{\left(n\right)}$ (given by Eqs.~(\ref{s_1}) - (\ref{s_3})) and $v_{\left(n\right)}$ (given by Eqs.~(\ref{v_1}) - (\ref{v_3})) and $y_{\left(n\right)}$ 
(given by Eqs.~(\ref{y_1}) - (\ref{y_3})) into (\ref{Shapiro_2PN_Final_Function_S}).  
In order to estimate the upper limit of the individual terms in (\ref{Estimation_Shapiro_2PN_M_Q_5}) 
the assumption is adopted that to a good approximation the giant planets can be considered as axially symmetric bodies, 
that means the STF quadrupole tensor in the following form is used (cf. Eq.~(\ref{M_ab}))
\begin{eqnarray}
        \hat{M}_{ab} = M\,J_2\,P^2 \left(\frac{1}{3}\,\delta_{ab} - \delta_{a3}\,\delta_{b3} \right), 
        \label{Quadrupole_Tensor_in_z_axis}
\end{eqnarray}

\noindent 
where it is assumed that the $x^3$-axis of the coordinate system is aligned with the symmetry axis $\ve{e}_3$ of the massive body. The parameter 
in (\ref{Quadrupole_Tensor_in_z_axis}), that means $M$ (mass of the body) $J_2$ (actual second zonal harmonic coefficient), $P$ (equatorial radius of the body) 
are given in Table~\ref{Table1} for the Sun and giant planets of the solar system. It is advisable to apply relations (\ref{Relation_x0}) - (\ref{Relation_x1}) 
as well as the parameter (\ref{parameter_z}) - (\ref{parameter_beta0_beta1}), which considerably simplify the expressions 
in (\ref{Estimation_Shapiro_2PN_M_Q_5}). Then, the estimation proceeds in very similar way as for (\ref{Estimation_Shapiro_2PN_M_M_5})  
and one finds, by means of the computer algebra system {\it Maple} \cite{Maple} the following upper limit:  
\begin{eqnarray}
	\left|\Delta c \tau_{\rm 2PN}^{M \times M_{ab}} \right| 
	&\le& \frac{12}{\left(d_k\right)^2} \,x_1 \frac{G^2 M^2}{c^4} \,\frac{P^2}{\left(d_k\right)^2} \left|J_2\right| . 
\label{Estimation_Shapiro_2PN_M_Q_10}
\end{eqnarray}

\noindent 
Numerical values of (\ref{Estimation_Shapiro_2PN_M_Q_10}) are presented in Table~\ref{Table3} for the Sun and giant planets.
In order to get correct upper limits one has to take into account that $\ve{k}$ and $\ve{d}_k$ are perpendicular to each other, which restricts their 
possible values and angles with $\ve{e}_3$ (see also endnote [99] in \cite{Zschocke_Quadrupole_1}).

\subsection{Estimation of 2PN quadrupole-quadrupole term}

The 2PN quadrupole-quadrupole term in (\ref{Appendix_Shapiro_2PN_Final}) reads
\begin{eqnarray}
    \Delta c \tau_{\rm 2PN}^{M_{ab} \times M_{cd}} &=& 
	\frac{G^2 \hat{M}_{ab} \hat{M}_{cd}}{c^4} \sum\limits_{n=1}^{10} T^{abcd}_{\left(n\right)} \, T_{\left(n\right)}\left(\ve{x}_1,\ve{x}_0\right),
	\nonumber\\ 
\label{Estimation_Shapiro_2PN_Q_Q_5}
\end{eqnarray}

\noindent
where the tensorial coefficients $T^{abcd}_{\left(n\right)}$ are given by Eqs.~(\ref{Tensor_T}) and the scalar functions $T_{\left(n\right)}$ have been defined 
by Eq.~(\ref{Shapiro_2PN_Final_Function_T}). In order to estimate the upper limit of the individual terms in (\ref{Estimation_Shapiro_2PN_Q_Q_5}) the assumption 
is adopted that to a good approximation the giant planets can be considered as axially symmetric bodies, that means for the product of two mass-quadrupole tensors  
(cf. Eq.~(\ref{M_ab})) the following expression is used 
\begin{eqnarray}
	\hat{M}_{ab}\,\hat{M}_{cd} &=& M^2\,\left|J_2\right|^2\,P^4 
	\nonumber\\ 
	&& \hspace{-1.75cm} \times \left(\frac{1}{9} \delta_{ab} \delta_{cd} - \frac{1}{3} \delta_{ab} \delta_{c3} \delta_{d3} 
        - \frac{1}{3} \delta_{a3} \delta_{b3} \delta_{cd} + \delta_{a3} \delta_{b3} \delta_{c3} \delta_{d3} \right), 
	\nonumber\\ 
        \label{Two_Quadrupole_Tensor_in_z_axis}
\end{eqnarray}

\noindent
where it is assumed that the $x^3$-axis of coordinate system is aligned with the symmetry axis $\ve{e}_3$ of the massive body. It is advisable to introduce 
the parameter (\ref{parameter_z}) - \ref{parameter_beta0_beta1}) as well as relations \ref{Relation_x0}) - (\ref{Relation_x1}), which considerably simplify 
the expressions in (\ref{Estimation_Shapiro_2PN_Q_Q_5}). Then, the estimation proceeds in very similar way as for (\ref{Estimation_Shapiro_2PN_M_M_5})
and one finds, by means of the computer algebra system {\it Maple} \cite{Maple} the following upper limit:
\begin{eqnarray}
	\left|\Delta c \tau_{\rm 2PN}^{M_{ab} \times M_{cd}} \right| 
        &\le& \frac{8}{\left(d_k\right)^2} \,x_1 \frac{G^2 M^2}{c^4} \,\frac{P^4}{\left(d_k\right)^4} \left|J_2\right|^2 .  
\label{Estimation_Shapiro_2PN_Q_Q_10}
\end{eqnarray}

\noindent
Numerical values of (\ref{Estimation_Shapiro_2PN_Q_Q_10}) are presented in Table~\ref{Table3} for the Sun and giant planets. 
In order to get correct upper limits one has to take into account that $\ve{k}$ and $\ve{d}_k$ are perpendicular to each other, which restricts their possible 
values and angles with $\ve{e}_3$ (see also endnote [99] in \cite{Zschocke_Quadrupole_1}).


\end{document}